\documentclass[11pt,a4paper]{article}
\usepackage{jcappub} 
\usepackage[english]{babel}
\usepackage[T1]{fontenc}
\usepackage[utf8]{inputenc}

\usepackage{amsmath, amsfonts, amssymb, cancel, graphicx, hyperref, color, xcolor, float, bm}

\graphicspath{ {./Figures/} }
	
\newcommand{\vk} {{\boldsymbol k}}
\newcommand{\vq} {{\boldsymbol q}}
\newcommand{\vp} {{\boldsymbol p}}

\newcommand{\beq} {\begin{equation}}
\newcommand{\eeq} {\end{equation}}
\newcommand{\bal} {\begin{aligned}}
\newcommand{\eal} {\end{aligned}}

\newcommand{\bsp} {\begin{split}}
\newcommand{\esp} {\end{split}}

\newcommand{\Nmodes} {N_\text{modes}}

\definecolor{darkgreen}{RGB}{235,118,56}

\renewcommand{\vec}{\bm}
\newcommand{\derd}{\text{d}}
\newcommand{\hMpc}{h^{-1}\text{Mpc}}
\newcommand{\ihMpc}{h\text{Mpc}^{-1}}
\newcommand{\confh}{\mathcal{H}}

\date{\today}

\author[\dagger]{Tobias Baldauf,}
\author[*]{Emmanuel Schaan}
\author[\dagger]{and Matias Zaldarriaga}
\affiliation[\dagger]{School of Natural Sciences, Institute for Advanced Study, Princeton, NJ 08540, U.S.A.}
\affiliation[*]{Department of Astrophysical Sciences, Princeton University, Princeton, NJ 08540, U.S.A.}

\emailAdd{baldauf@ias.edu}
\emailAdd{eschaan@astro.princeton.edu}
\emailAdd{matiasz@ias.edu}

\title{On the reach of perturbative descriptions for dark matter displacement fields}

\abstract{
We study Lagrangian Perturbation Theory (LPT) and its regularization in the Effective Field Theory (EFT) approach. 
We evaluate the LPT displacement with the same phases as a corresponding $N$-body simulation, which allows us to compare perturbation theory to the non-linear simulation with significantly reduced cosmic variance, and provides a more stringent test than simply comparing power spectra. 
We reliably detect a non-vanishing leading order EFT coefficient and a stochastic displacement term, uncorrelated with the LPT terms. This stochastic term is expected in the EFT framework, and, to the best of our understanding, is not an artifact of numerical errors or transients in our simulations. 
This term constitutes a limit to the accuracy of perturbative descriptions of the displacement field and its phases, corresponding to a $1\%$ error on the non-linear power spectrum at $k=0.2 \ihMpc$ at $z=0$. Predicting the displacement power spectrum to higher accuracy or larger wavenumbers thus requires a model for the stochastic displacement.
}

\begin{document}

%%%%%%%%%%%%%%%%%%%%%%%%%%%%%%%%%%%%%%%%%%%%%%%%%%%%%%%%%
\maketitle

%%%%%%%%%%%%%%%%%%%%%%%%%%%%%%%%%%%%%%%%%%%%%%%%%%%%%%%%%

%%%%%%%%%%%%%%%%%%%%%%%%%%%%%%%%%%%%%%%%%%%%%%%%%%%%%%%%%
%%%%%%%%%%%%%%%%%%%%%%%%%%%%%%%%%%%%%%%%%%%%%%%%%%%%%%%%%
\section{Introduction}
%%%%%%%%%%%%%%%%%%%%%%%%%%%%%%%%%%%%%%%%%%%%%%%%%%%%%%%%%
%%%%%%%%%%%%%%%%%%%%%%%%%%%%%%%%%%%%%%%%%%%%%%%%%%%%%%%%%

% Importance of LSS in the future

The detection of the acoustic peaks in the temperature and polarization power spectrum of the Cosmic Microwave Background (CMB) anisotropies imply that the seeds for structure formation were put in place before the hot big bang, perhaps during a period of inflation. So far these primordial seeds are our only fossil record from this early period.

The measurements of the CMB anisotropies recently released by the {\it Planck} collaboration \cite{Planck:2015params, Planck:2015nonGauss} have extracted most of the information available in the CMB about the primordial seeds. Further improvements will have to come from measurements of structure at later times in the history of the Universe. Gravitational instability makes the primordial fluctuations grow with time, making them easier to detect at late epochs. However, the evolution is no longer linear and thus on small scales much of the information about the initial conditions is, for practical purposes, lost. 

The CMB measurements have produced superb constraints on departures from Gaussianity of the initial seeds. Although very impressive these constraints fail to reach the clear theoretical targets (e.g. \cite{Alvarez:2014vva}), and measurements of LSS are a natural candidate to try to get closer to the theoretical thresholds. This has sparked renewed interest in modeling accurately  the effects of the non-linear evolution of structure in the late Universe, as well as new ideas for potential surveys that could reach the required accuracy \cite{Dore:2014cca}. Accurate modeling of the development of structure is also required to extract information about the dark components of the matter budget  (such as neutrinos) from LSS observations. 

% History of PT

One of the avenues to study the growth of structure is using perturbative techniques. 
Perturbation theory for LSS dates back to the very early days of modern Cosmology {\it e.g.} \cite{Zeldovich:1969sb, Peebles:LSS}. It is extremely successful at calculating correlators at the lowest order or tree level (for a complete review of perturbation theory results see \cite{Bernardeau:2001qr}). On the other hand, results for the first nontrivial correction to tree level results, the ``loop corrections'', are less satisfactory. 
The reason for the failure at the loop level is clear. Perturbation theory cannot be used to describe the small scales because the series simply does not converge in that regime. Unfortunately the errors in the small scales pollute large scale results. This behaviour becomes worse, as higher loops are considered, which involve integrals over the spurious perturbative predictions on small scales.

% Recent developments

Recent years have seen a quick development of perturbation theory with the introduction of the so-called 
Effective Field Theory (EFT) of Large Scale Structure for both Eulerian and Lagrangian descriptions \cite{Baumann:2010tm,Carrasco:2012cv,Porto:2013qua}. The EFT framework explicitly keeps track of the effects of the small scales by keeping explicit counterterms that both fix the mistakes introduced by perturbation theory and replace them by the correct level of ``leakage'' from the small scale dynamics. Although the amplitude of these corrections is not known a-priori, their functional form is. Thus the EFT of large scale structure generalizes the standard perturbative calculation adding new parameters that need to be fitted to simulations or observations. 

In recent years the EFT approach has been used to compute the power spectrum at one loop and two loops in a variety of cosmologies, as well as the bispectrum \cite{Carrasco:2012cv,Carrasco:2013mua,Pajer:2013jj,Carrasco:2013sva,Baldauf:2014qfa,Angulo:2014tfa}. So far, the free parameters inherent to the EFT were fitted to simulations, trying  to match the power spectrum or bispectrum as accurately as possible.  This approach is potentially problematic as the EFT works best on very large scales where the non-linear corrections are small and the sample variance in the simulations is large. So in practice the fitting of these parameters was done in the mildly non-linear regime, the same regime were the scheme was being tested, allowing the possibility of overfitting.

In this paper we introduce a novel approach, where we use the EFT to predict the displacement field in simulations on a realization by realization basis. Thus we predict the actual field in the simulation and not just its power spectrum. By doing so we sidestep the cosmic variance issue and directly measure the EFT parameters on very large scales. We can then see how these new terms affect the mildly non-linear regime without having to worry about overfitting. This approach is closely related to \cite{Tassev:2012cq}, where transfer functions were introduced to quantify the relation between the perturbative predictions and the results of simulations. The EFT coefficients are just describing a specific parametrized form for these transfer functions. We will generalize \cite{Tassev:2012cq}  by going to higher order in perturbation theory and including terms with different spatial structure which are predicted by the EFT.

% This paper is organized as follows:

This paper is organized as follows: in Sec.~\ref{sec:review} we review the Lagrangian perturbation theory to establish our notation. 
We present our simulation suite in Sec.~\ref{sec:1loop}, and compare the one-loop and higher-loop predictions with simulations in Sec.~\ref{sec:1loop} and Sec.~\ref{sec:tf}.
We present the evidence for a stochastic displacement in Sec.~\ref{sec:stoch}, which constitutes a limit to the accuracy of perturbative treatments of the displacement field. We summarize our results in Sec.~\ref{sec:conclusions}.
The extensive Appendix gives more details on systematic effects in the simulations, the derivation of recursion relations for the perturbative displacement fields, cosmic variance, higher order counter terms and transient effects arising from simulation initial conditions.

%%%%%%%%%%%%%%%%%%%%%%%%%%%%%%%%%%%%%%%%%%%%%%%%%%%%%%%%%%%%%%%
%%%%%%%%%%%%%%%%%%%%%%%%%%%%%%%%%%%%%%%%%%%%%%%%%%%%%%%%%%%%%%%
\section{Review of LPT and EFT: towards a well-defined perturbation theory}
\label{sec:review}
%%%%%%%%%%%%%%%%%%%%%%%%%%%%%%%%%%%%%%%%%%%%%%%%%%%%%%%%%%%%%%%
%%%%%%%%%%%%%%%%%%%%%%%%%%%%%%%%%%%%%%%%%%%%%%%%%%%%%%%%%%%%%%%

%%%%%%%%%%%%%%%%%%%%%%%%%%%%%%%%%%%%%%%%%%%%%%%%%%%%%%%%%%%%%%%
\subsection{LPT expansion}
%%%%%%%%%%%%%%%%%%%%%%%%%%%%%%%%%%%%%%%%%%%%%%%%%%%%%%%%%%%%%%%

The motion of dark matter particles is described by the Euler-Poisson system:
\beq
\left\{
\bal
&\frac{\derd \boldsymbol v}{\derd t} = - \nabla_{\boldsymbol r} \Phi \\
&\Delta_{\boldsymbol r} \Phi = 4\pi \mathcal{G} \rho.
\eal
\right.
\eeq
Here, $\boldsymbol v$ is the physical velocity $\boldsymbol v \equiv \frac{\derd \boldsymbol r}{\derd t}$, $\boldsymbol r$ is the physical position ($\boldsymbol r = a \boldsymbol x$, with $\boldsymbol x$ comoving position) and $t$ the cosmic time ($\derd t = a \derd \tau$, with $\tau$ conformal time). In the Lagrangian picture, we follow the motion of a dark matter particle initially at position $\boldsymbol q$ and introduce the displacement $\boldsymbol \psi$ such that $\boldsymbol x = \boldsymbol q + \boldsymbol \psi$. Subtracting the Hubble flow leads to:
\beq
\left\{
\bal
&\ddot{\boldsymbol \psi} + \mathcal{H} \dot{\boldsymbol \psi} = -\nabla_{\boldsymbol x} \varphi \\
&\Delta_{\boldsymbol x} \varphi = \frac{3}{2} \mathcal{H}^2 \Omega_\text{m} \delta = \frac{3}{2} \mathcal{H}^2 \Omega_\text{m} \left[ \frac{1}{\text{Det} \left( \delta_{ij} + \psi_{i,j} \right)} - 1 \right]   .
\eal
\right.
\label{eq:eom_lpt}
\eeq
Here ` $\dot{}$ ' denotes the derivative with respect to conformal time, while ` $_{,i}$ ' denotes the derivative with respect to the comoving Lagrangian coordinate $q^i$, and $\mathcal{H}$ is the conformal Hubble parameter. The above equation breaks at shell crossing, when the determinant vanishes.
The conformal Hubble parameter is determined by the Friedmann equation:
\beq
\mathcal{H}^2 = a^2 \mathcal{H}_0^2 \left[ \frac{\Omega_\text{m}^0}{a^3} + \Omega_\Lambda^0 \right] .
\eeq

We proceed with a Helmholtz decomposition of the displacement $\boldsymbol \psi = \vec{\nabla} \phi + \vec{\nabla} \times \boldsymbol \omega$. We shall call $\phi$ the scalar displacement or displacement potential, and $\boldsymbol \omega$ the vector or transverse component of the displacement. We make an ansatz for perturbative solutions of the form $\phi = \sum_{n\geq 1} \phi^{(n)}$ and $\boldsymbol \omega = \sum_{n\geq 1} {\boldsymbol \omega}^{(n)}$, with:
\beq
\left\{
\bal
&\phi^{(n)} \equiv \frac{D^n}{n!} \int \frac{\derd^3\boldsymbol p_1 ... \derd^3\boldsymbol p_n}{(2\pi)^{3n}} 
(2\pi)^3  \delta^D(\boldsymbol p_{1..n} - \boldsymbol k) 
L_{n}(\boldsymbol p_1, ..., \boldsymbol p_n) 
\delta_0(\boldsymbol p_1) ...  \delta_0(\boldsymbol p_n) \\
&{\boldsymbol\omega}^{(n)} \equiv \frac{D^n}{n!} \int \frac{\derd^3\boldsymbol p_1 ... \derd^3\boldsymbol p_n}{(2\pi)^{3n}}
(2\pi)^3  \delta^D(\boldsymbol p_{1..n} - \boldsymbol k) 
{\boldsymbol T}_{n}(\boldsymbol p_1, ..., \boldsymbol p_n) 
\delta_0(\boldsymbol p_1) ...  \delta_0(\boldsymbol p_n) .\\
\eal
\right.
\label{eq:convolution}
\eeq
Here $\delta_0$ is the linear density field, $D$ is the linear growth factor and $L_{n}$ and ${\boldsymbol T}_{n}$ are the LPT kernels, which can be computed order by order as we will show shortly \cite{Bouchet:1995, Catelan:1995, Matsubara:2008re,Rampf:2012xa}. Such an expansion is exact in the case of an Einstein-de Sitter (EdS) universe, with $D$ reducing to the linear growth factor. It constitutes a good approximation for a $\Lambda$CDM universe \cite{Bernardeau:2001qr, Bouchet:1995}.

The goal of this study is to compare the predictions of LPT and EFT for the displacement to the non-linear displacement from $N$-body simulations. 
However, the EFT expansion is only valid for low values of $k$, i.e. on large scales, where the cosmic variance in the simulation is largest. This cosmic variance introduces relative errors of up to $10\%$ on the power spectrum at the largest scales of our simulations ($k \sim 5\times 10^{-3}$ h/Mpc), which would make it almost impossible to measure the EFT parameters from the power spectrum (see App.~\ref{sec:appendix_cosmic_variance}).
A simple way to reduce this scatter to less than $0.1\%$ is to consider ratios of power spectra (dividing the non-linear power spectrum by the linear power spectrum measured from the same realization).
But an even better way to proceed is to compute the LPT displacement directly on the simulation grid, with the same exact initial condition as the one used in the N-body simulation. This completely suppresses the cosmic variance. Computing the LPT displacement on the simulation grid with the same initial conditions as the simulation will also allow us to perform a more stringent test of the EFT, by comparing it to the simulation at the level of the displacement field itself, and not only at the level of the power spectrum.

Since computing the convolutions of \eqref{eq:convolution} on the simulation grid would be too computationally expensive, we will use a real-space formulation of the LPT expansion that is easier to evaluate.
From the equations of motion presented above, one can solve for the $n$th order displacement $\psi^{(n)}$ recursively (see App.~\ref{sec:appendix_lpt} for the full derivation, following \cite{Bouchet:1995, Zheligovsky:2013eca}). For the scalar displacement, this yields:
\beq
\left\{
\bal
&\psi^{(2)}_{i,i} = -\frac{3}{7} \mathcal{L}^{1,1} \\
&\psi^{(3)}_{i,i} = -\frac{10}{9} \mathcal{L}^{2,1} - \frac{1}{3} \mathcal{M}^{1,1,1} \\
&\psi^{(4)}_{i,i} = -\frac{17}{33} \mathcal{L}^{2,2} - \frac{14}{11} \mathcal{L}^{3,1} - \frac{13}{33} \left( \mathcal{M}^{1,1,2} +\text{ perm.} \right) \\
&\psi^{(5)}_{i,i} = -\frac{18}{13} \mathcal{L}^{4,1} - \frac{14}{13} \mathcal{L}^{3,2} 
- \frac{6}{13} \left( \mathcal{M}^{1,1,3} +\text{ perm.} \right)
- \frac{5}{13} \left( \mathcal{M}^{1,2,2} +\text{ perm.} \right) \\
&... \\
&\psi^{(n)}_{i,i} = - \sum_{n_1+n_2 = n \atop {n1, n2 > 0}} \frac{2\left( n_1^2 + n_2^2 \right) + n - 3}{\left( 2n+3 \right)\left( n-1 \right)} \,\, \mathcal{L}^{n_1, n_2}
- \sum_{n_1+n_2+n_3 = n \atop {n_1,n_2,n_3>0}}  \frac{2\left( n_1^2 + n_2^2 + n_3^2 \right) + n - 3}{\left( 2n+3 \right)\left( n-1 \right)} \,\, \mathcal{M}^{n_1, n_2, n_3} ,
\eal
\right.
\label{eq:lpt_grid_scalar}
\eeq
where we have introduced
\beq
\left\{
\bal
&\mathcal{L}^{m,n} = \frac{1}{2} \sum_{i,j} \left[   \psi^{(m)}_{i\,,i} \psi^{(n)}_{j\,,j} - \psi^{(m)}_{i\,,j} \psi^{(n)}_{j\,,i}   \right] \\
&\mathcal{M}^{l,m,n} = \frac{1}{3!} \epsilon_{i_1 i_2 i_3} \epsilon_{j_1 j_2 j_3}  \,\, \psi^{(l)}_{i_1\,,j_1}  \psi^{(m)}_{i_2\, ,j_2}  \psi^{(n)}_{i_3\,,j_3}  .\\
\eal
\right.
\eeq
and $\epsilon_{i_1 i_2 i_3}$ is the Levi-Civita tensor. Notice how the scalar part of the displacement on the l.h.s. of \eqref{eq:lpt_grid_scalar} is in principle sourced by both the scalar and vector displacements on the r.h.s., through the $\mathcal{L}$ and $\mathcal{M}$ terms. 

For the vector part of the displacement, we find:
\beq
\left\{
\bal
&\nabla \times {\boldsymbol\psi}^{(3)} = \frac{1}{3} {\boldsymbol\Pi}^{1,2} \\
&... \\
&\nabla \times {\boldsymbol\psi}^{(n)} = \sum_{p=1}^{n-1} \frac{n-2p}{2n}  \,\, {\boldsymbol\Pi}^{p, n-p} \\
\eal
\right.
\label{eq:lpt_grid_vector}
\eeq
with:
\beq
{\boldsymbol\Pi}^{m,n} = \sum_{i}  \mathbf{\nabla}\left(\psi^{(m)}_{i}\right) \times  \mathbf{\nabla}\left(\psi^{(n)}_{i}\right) .
\eeq
Here again, the vector part of the displacement on the l.h.s. of \eqref{eq:lpt_grid_vector} is sourced by both the scalar and vector displacements on the r.h.s..

Eqs.~\eqref{eq:lpt_grid_scalar} and \eqref{eq:lpt_grid_vector} allow to compute recursively the scalar and vector parts of the displacement on the simulation grid to arbitrary order in LPT. Notice that the scalar and vector displacements are not decoupled: in particular, the vector displacement at third order is sourced by the scalar displacements at first and second order, and contributes in return to the scalar displacement starting at fourth order. 

From Eqs.~\eqref{eq:lpt_grid_scalar} and \eqref{eq:lpt_grid_vector}, one can recursively derive analytical expressions for $L_{n}$ and $\vec T_{n}$.
Focusing on the equations for the scalar displacement \eqref{eq:lpt_grid_scalar}, we see that there are two fundamental vertices corresponding to $\mathcal{L}$ and $\mathcal{M}$. In Fourier space, these correspond to kernels which we shall denote by $\kappa_2$ and $\kappa_3$ respectively:
\beq
\kappa_2(\vec p_1,\vec p_2)=1-\left(\frac{\vec p_1\cdot \vec p_2}{p_1 p_2}\right)^2,
\eeq
and
\beq
\bal
\kappa_3(\vec p_1,\vec p_2,\vec p_3) &=
\frac{\text{Det}\left[ \vp_1, \vp_2, \vp_3 \right]^2}{p_1^2 p_2^2 p_3^2} \\
&= 1-\left(\frac{\vec p_1\cdot \vec p_2}{p_1 p_2}\right)^2-\left(\frac{\vec p_2\cdot \vec p_3}{p_2 p_3}\right)^2-\left(\frac{\vec p_3\cdot \vec p_1}{p_3 p_1}\right)^2+2\frac{(\vec p_1\cdot \vec p_2)( \vec p_2\cdot \vec p_3) (\vec p_3\cdot \vec p_1)}{p_1^2 p_2^2 p_3^2}. \\
\eal
\eeq
They are the building blocks of higher order vertices.
They obey  $\kappa_2(\vec p,-\vec p)=0$, $\kappa_3(\vec k,\vec p,-\vec p)=0$. For fixed external momentum $\vec k$ and large loop momentum $p\to \infty$ they scale as
\beq
\bal
\kappa_2(\vec p,\vec k-\vec p) &\xrightarrow{p/k\to \infty}\frac{k^2}{p^2} \\
\kappa_3(\vp_1, \vp_2, \vk-\vp_1-\vp_2) &\xrightarrow{p_1/k \to \infty} \frac{k^2}{p_1^2}\kappa_2(\vec k,\vec p_2) \\
\eal
\eeq

From Eq.~\eqref{eq:lpt_grid_scalar}, we derive the LPT kernels $L_{n}$ for the scalar potential in terms of these fundamental vertices $\kappa_2$ and $\kappa_3$:
\beq
\left\{
\bal
&L_{1}(\vp_1) = \frac{1}{p_1^2} \\
&L_{2}(\vp_1, \vp_2)  = \frac{3}{7 p_{12}^2} \kappa_2(\vp_1,\vp_2) \\
&L_{3a}^{( \text{asym})}(\vp_1, \vp_2, \vp_3)  = -\frac{1}{3 p_{123}^2} \kappa_3(\vp_1,\vp_2,\vp_3) \\
&L_{3b}^{(\text{asym})}(\vp_1, \vp_2, \vp_3)  = \frac{5}{7 p_{123}^2} \kappa_2(\vp_1,\vp_2)
\kappa_2(\vp_{12},\vp_3)
\eal
\right.
\label{eq:lptkernels123}
\eeq
where $p_{i_1\ldots i_n} \equiv p_{i1} + \ldots +p_{i_n}$.

Similarly, from Eq.~\eqref{eq:lpt_grid_vector}, we obtain the LPT kernels ${\boldsymbol T}_{n}$ for the vector component of the displacement:
\beq
\left\{
\bal
&{\boldsymbol T}_{1}(\vp_1) = {\boldsymbol T}_{2}(\vp_1, \vp_2) = 0\\
&{\boldsymbol T}_{3}(\vp_1, \vp_2, \vp_3)  = \frac{1}{p_{123}^2} L_{1}(\vp_1) L_{2}(\vp_2, \vp_3) \left( \vp_1\cdot\vp_{23}\right) \left( \vp_1\times\vp_{23}\right) \; . \\
\eal
\right.
\eeq

\begin{figure}[H]
\centering
\includegraphics[width=0.8\textwidth]{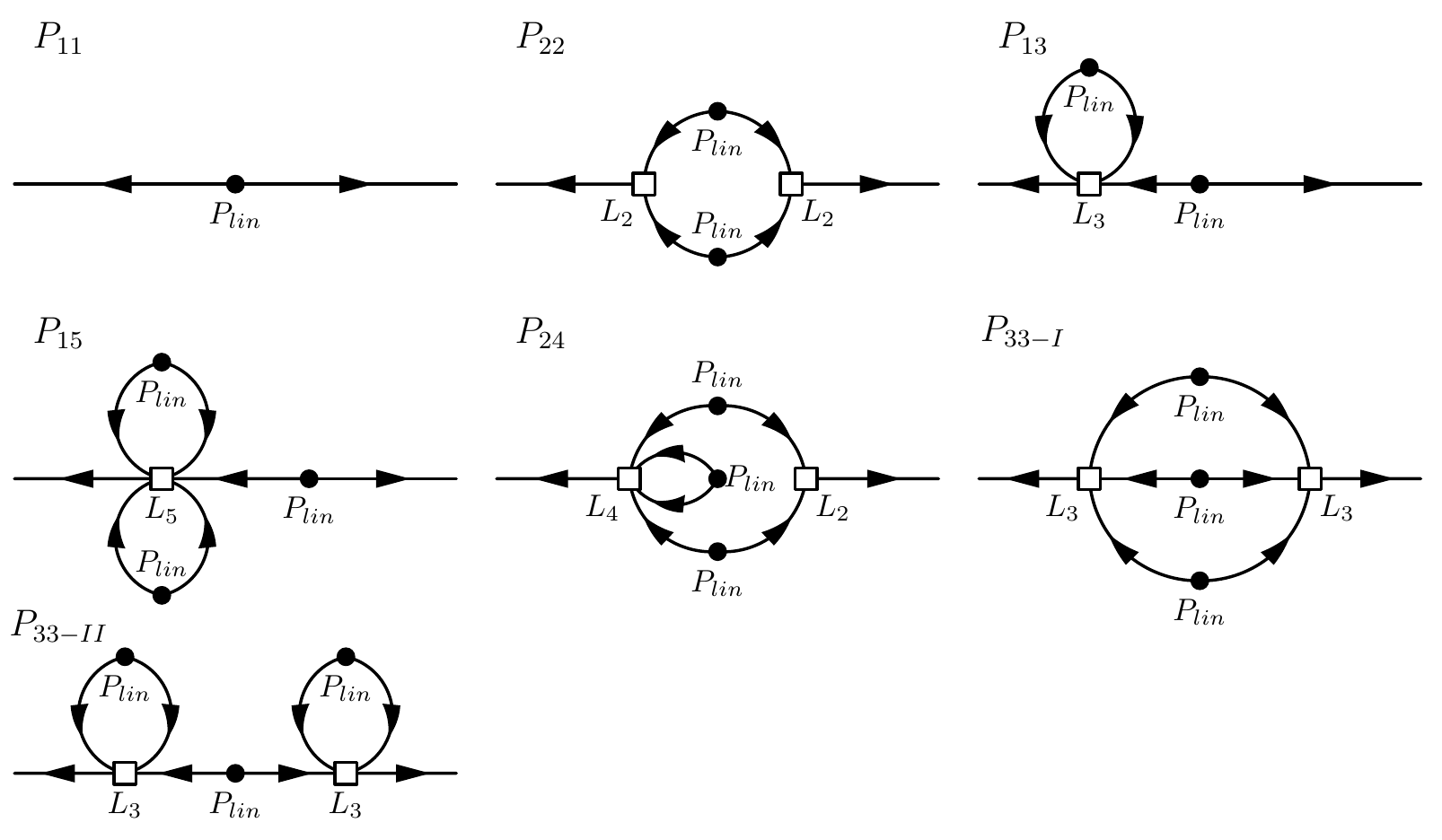}
\caption{Diagrams for the one-loop (first row) and two-loop (second and third row) contributions to the power spectrum of the scalar displacement $\phi$. The empty squares symbolize the LPT kernels $L_n$, while the linear matter power spectrum is represented by the black dots. The explicit formulae for the expressions are given in App.~\ref{sec:appendix_lpt}.}
\label{fig:diagrams_1loop_2loop}
\end{figure}

\begin{figure}[H]
\centering
\includegraphics[width=0.49\textwidth]{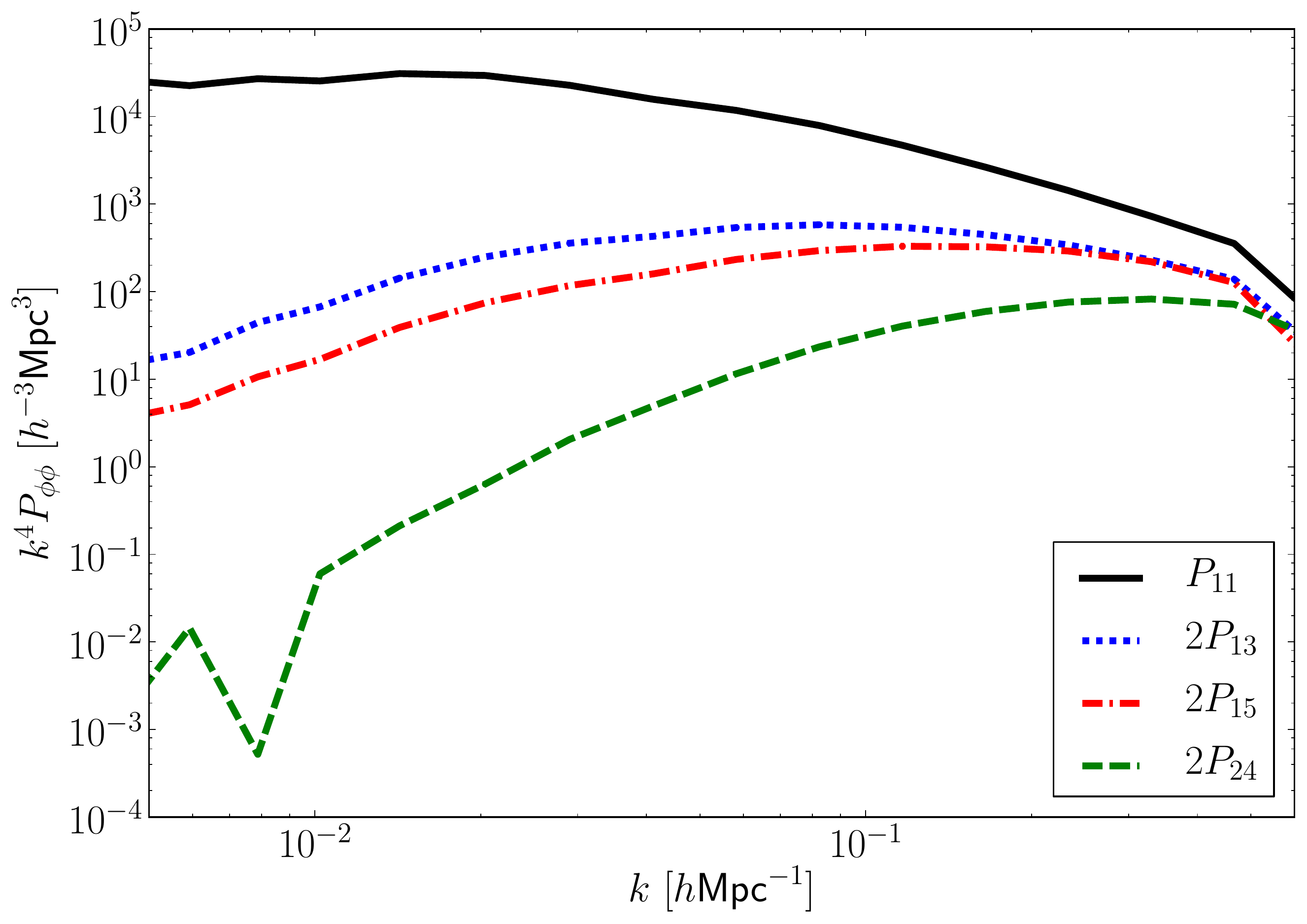}
\includegraphics[width=0.49\textwidth]{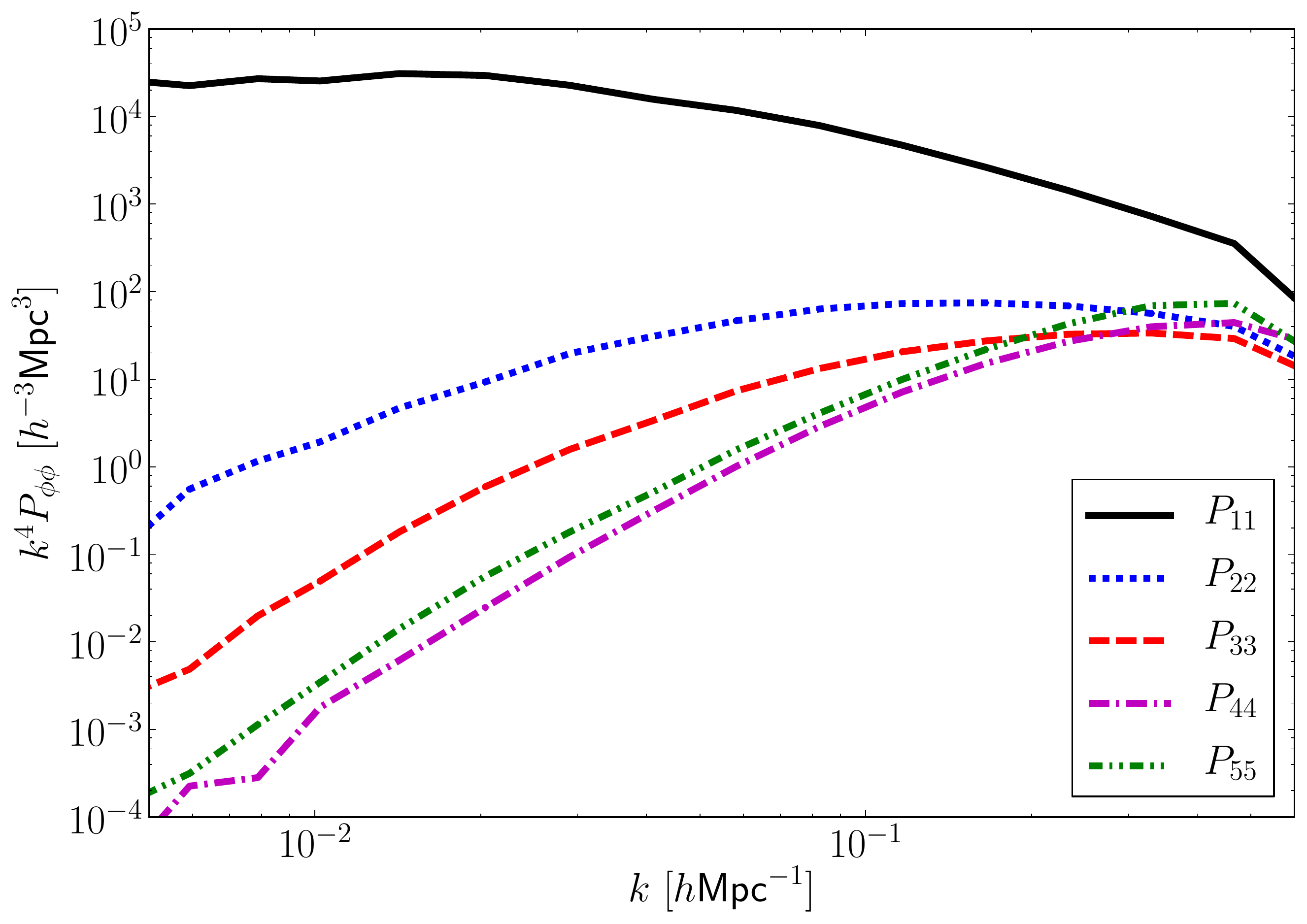}
\caption{Power spectra of the LPT displacements up to 5th order, for one particular realization of the linear density field $\delta_0$ with cutoff $k_\text{max}=0.6 \ihMpc$.}
\label{fig:lpt_terms_2loop}
\end{figure}

Eventually, we define the power spectrum $P_{ij}(k)$ of two random fields $\phi^{(i)}$ and $\phi^{(j)}$ as:
\beq
\left\langle \phi^{(i)}(\vec k)\phi^{(j)} (\vec k')\right\rangle = (2\pi)^3\delta^\text{(D)}(\vec k+\vec k')P_{ij}(k)
\eeq
The power spectrum $P_{ij}$ between $i$th and $j$th order LPT displacements can be computed in terms of the linear density power spectrum $P_\text{lin}$, defined analogously. The 1-loop and 2-loop power spectra for the the scalar displacement $\phi$ are represented diagrammatically in Fig.~\ref{fig:diagrams_1loop_2loop}, computed in App.~\ref{sec:appendix_lpt}, and shown in Fig.~\ref{fig:lpt_terms_2loop}. In what follows, unless otherwise indicated, we focus on the scalar displacement and call ``nLPT'' the scalar displacement up to order $n$ ($\phi_{n\text{LPT}} = \phi^{(1)} +... + \phi^{(n)}$).

The reader might be more familiar with density power spectra than displacement power spectra. For this reason we will often show $k^4 P_{\phi}(k)$ instead of $P_\phi(k)$, since the former is equal to the density power spectrum to lowest order ($\delta^{(1)} = i \vec k \cdot \vec \psi^{(1)}=-k^2 \phi^{(1)}$).

Note that the LPT arising from the kernels in Eq.~\eqref{eq:lptkernels123} are IR safe by themselves, i.e., there is no cancellation of IR contributions as it is the case in Eulerian perturbation theory \cite{Carrasco:2013sva}. Thus no IR safe integrand needs to be employed, and the respective integrals can be evaluated independently. However, as we will show in detail, the loops are UV sensitive, meaning that their value depends on the UV cutoff used, even when this cutoff is already in the non-linear regime. This cutoff dependence is unphysical, and makes the LPT predictions somewhat ill-defined, as we will show in the following sections.

%>>>>>>>>>>>>>>>>>>>>>>>>>>>>>>>>>>>>>>>>>>>>>>>>>>>>>>>>>>>>>>>>>>>>>>>>>>>>>>>>>>>>>>
%>>>>>>>>>>>>>>>>>>>>>>>>>>>>>>>>>>>>>>>>>>>>>>>>>>>>>>>>>>>>>>>>>>>>>>>>>>>>>>>>>>>>>>
%
%>>>>>>>>>>>>>>>>>>>>>>>>>>>>>>>>>>>>>>>>>>>>>>>>>>>>>>>>>>>>>>>>>>>>>>>>>>>>>>>>>>>>>>
%>>>>>>>>>>>>>>>>>>>>>>>>>>>>>>>>>>>>>>>>>>>>>>>>>>>>>>>>>>>>>>>>>>>>>>>>>>>>>>>>>>>>>>

%%%%%%%%%%%%%%%%%%%%%%%%%%%%%%%%%%%%%%%%%%%%%%%%%%%%%%%%%%%%%%%

%%%%%%%%%%%%%%%%%%%%%%%%%%%%%%%%%%%%%%%%%%%%%%%%%%%%%%%%%%%%%%%
\subsection{The EFT counterterms}
\label{subsec:eft_counterterms}
%%%%%%%%%%%%%%%%%%%%%%%%%%%%%%%%%%%%%%%%%%%%%%%%%%%%%%%%%%%%%%%

In the EFT approach, one acknowledges that the mistakes perturbation theory makes on small scales will affect the dynamics of the particles on large scales. This can be expressed as follows:
\beq
\ddot{\boldsymbol \psi} + \mathcal{H} \dot{\boldsymbol \psi} =
-\nabla_{\boldsymbol x} \varphi_\text{LPT}
+{\boldsymbol a}_\text{ct}+{\boldsymbol a}_\text{stoch}.
\label{EFT}
\eeq
This equation is identical to its LPT counterpart Eq.~\eqref{eq:eom_lpt}, except for ${\boldsymbol a}_\text{ct}$, the sum of the so-called counterterms, and ${\boldsymbol a}_\text{stoch}$, the stochastic acceleration.

The counterterms in ${\boldsymbol a}_\text{ct}$ model the effect of the unknown small scale dynamics on the large scales that are being tracked by perturbation theory.
Once a few normalization coefficients (the EFT parameters) are fixed, the Fourier amplitude and phase of ${\boldsymbol a}_\text{ct}$ are completely determined by the perturbative solution (i.e. ultimately by the linear displacement field) on a realization by realization basis.
The terms in ${\boldsymbol a}_\text{ct}$ have the correct spatial structure to absorb potential divergencies encountered in the loop calculations.

On the other hand ${\boldsymbol a}_\text{stoch}$ is a stochastic contribution due to the small-scale detail of the particular realization considered. Its exact Fourier amplitude and phase cannot be predicted, but at the order considered in this paper, only the power spectrum of ${\boldsymbol a}_\text{stoch}$ will be relevant, and it will be regarded as realization-independent.

In \cite{Porto:2013qua} the form of the counterterms in the r.h.s. of Eq~\eqref{EFT} was motivated by considering the theory for the displacements once smoothed on a sufficiently large scale so that perturbation theory is valid.  The equation of motion in the EFT can be interpreted as the equation of motion of the center of mass of a set of point-like particles, each following the LPT equation of motion~\eqref{eq:eom_lpt}. 
Regardless of the physical origin of these terms, the response terms are expanded in powers of the displacement and its derivatives, and to lowest order, the resulting expression for the scalar displacement is very simple \cite{Porto:2013qua}:
\beq
\phi_{1-\text{loop EFT}} = \left( 1 + \alpha k^2 \right)\phi^{(1)} + \phi^{(2)} + \phi^{(3)} + \phi_\text{stoch},
\label{eq:def_alpha_phi}
\eeq
This equation differs from its LPT analog only by the addition of the EFT counterterm $\alpha k^2 \phi^{(1)}$ (where $\alpha$ is some unknown function of time), and the stochastic term $\phi_\text{stoch}$.
In particular, the corresponding 1-loop power spectrum is given by:
\beq
P_{1-\text{loop EFT}} = P_{1-\text{loop LPT}} + 2\alpha k^2 P_{11}
\eeq
Indeed, the power spectrum of the stochastic term is expected to be subdominant compared to the EFT term $\alpha k^2 P_{11}$, as shown in Fig.~\ref{fig:scalings_p} and explained in more details in the next section.
As we shall show shortly, the EFT counterterm has the right scale-dependence to correct the mistake on $P_{13}$ due to high-$k$ modes, shown in Eq.~\eqref{eq:uvmistake_p13_p22}. This way, the EFT term can absorb the UV mistake in the LPT power spectrum. Furthermore, the coefficient $\alpha$ will compensate the cutoff $k_\text{max}$ in the loop integrals, resulting in a cutoff-independent power spectrum.\\

Note that we can also understand $\alpha$ as a regulator of loop corrections to the displacement field itself (rather than loop corrections to the displacement power spectrum). Namely we can close one loop in the third order displacement field (i.e. contract two of the three linear density fields) to yield a modified linear displacement field 
\beq
\phi^{(3)}(\vec k)\supset \delta_0(\vec k) \int\frac{\derd^3 p}{(2\pi)^2} L_3(\vec p,-\vec p,\vec k) P_\text{lin}(p)
=\phi^{(1)}(\vec k) \frac{P_{13}(k)}{P_{11}(k)} 
\eeq
The loop is the same one as in $P_{13}$, i.e. it scales as $k^2$ for large loop momenta. It can be regularized by $\alpha k^2$. This also makes clear that the regularization of a $n$-th order field (here $\phi^{(3)}$) requires the leading order counterterm to be a $n-2$-th order field (here $\phi^{(1)}$), or equivalently that the lowest counterterm counts as a second order field in the power counting.

In what follows, we call ``EFT'' or ``3EFT'' the model $\phi_\text{3EFT} = (1 + \alpha k^2)\phi^{(1)} + \phi^{(2)} + \phi^{(3)}$, and we shall compare it to the ``nLPT'' models in terms of agreement with the simulation.

%%%%%%%%%%%%%%%%%%%%%%%%%%%%%%%%%%%%%%%%%%%%%%%%%%%%%%%%%%%%%%%
\subsection{Estimates for the sizes of non-linear corrections: EdS scalings vs. $\Lambda$CDM}
%%%%%%%%%%%%%%%%%%%%%%%%%%%%%%%%%%%%%%%%%%%%%%%%%%%%%%%%%%%%%%%

When performing a perturbative calculation, it is useful to know how many orders are required to reach a desired accuracy level. In the case of the EFT this is also necessary in order to understand how many counterterms have to be kept. In this section we summarize the standard estimates based on an Einstein-de Sitter (EdS) universe with power law initial power spectrum  and contrast them with what one gets in $\Lambda$CDM. Since the power law approximation is not particularly accurate in the case of a $\Lambda$CDM universe, we will discuss in more detail the situation at one-loop order in $\Lambda$CDM in the next subsection. 

\begin{figure}[t]
\centering
\includegraphics[width=0.48\textwidth]{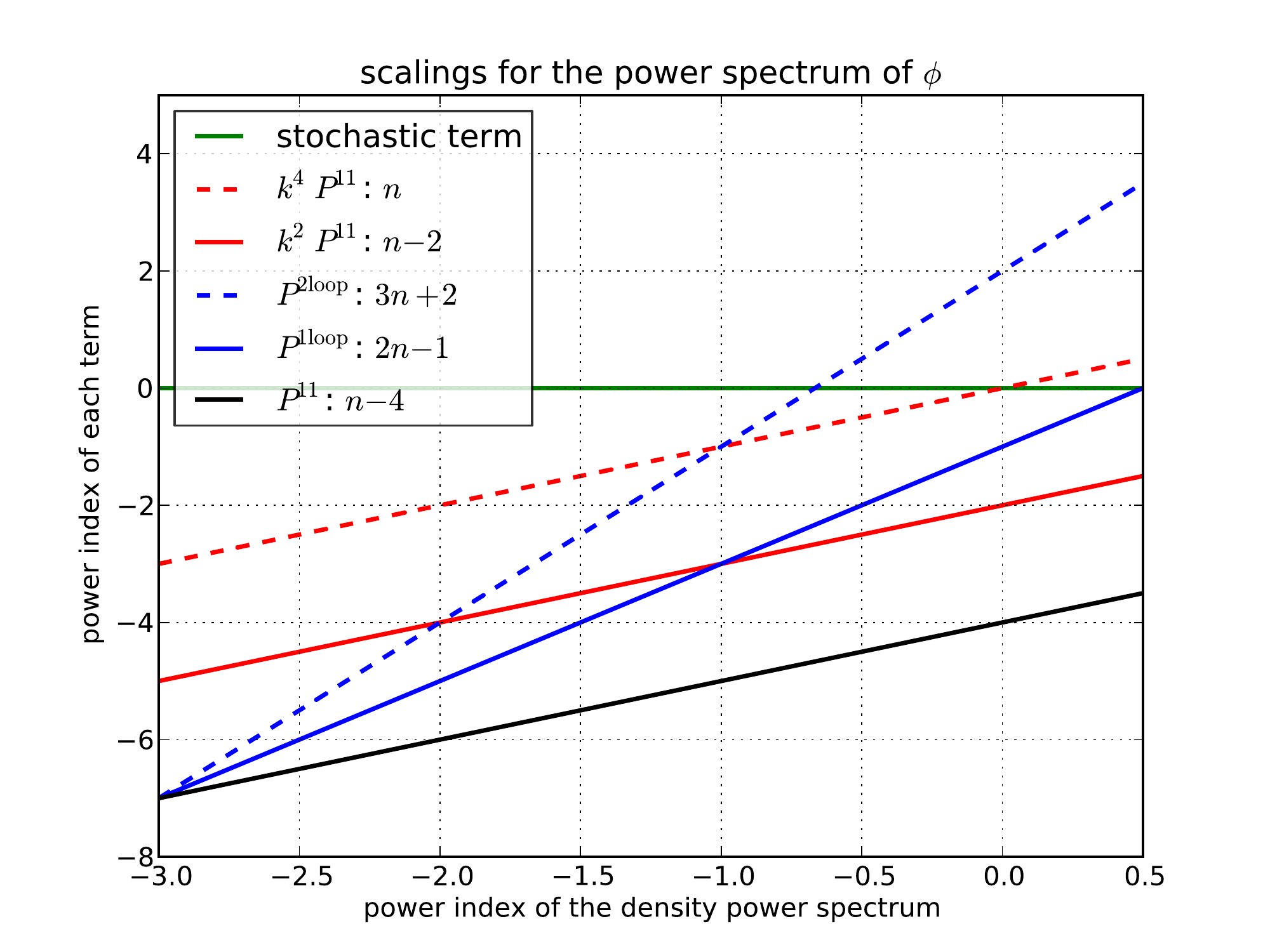}
\includegraphics[width=0.49\textwidth]{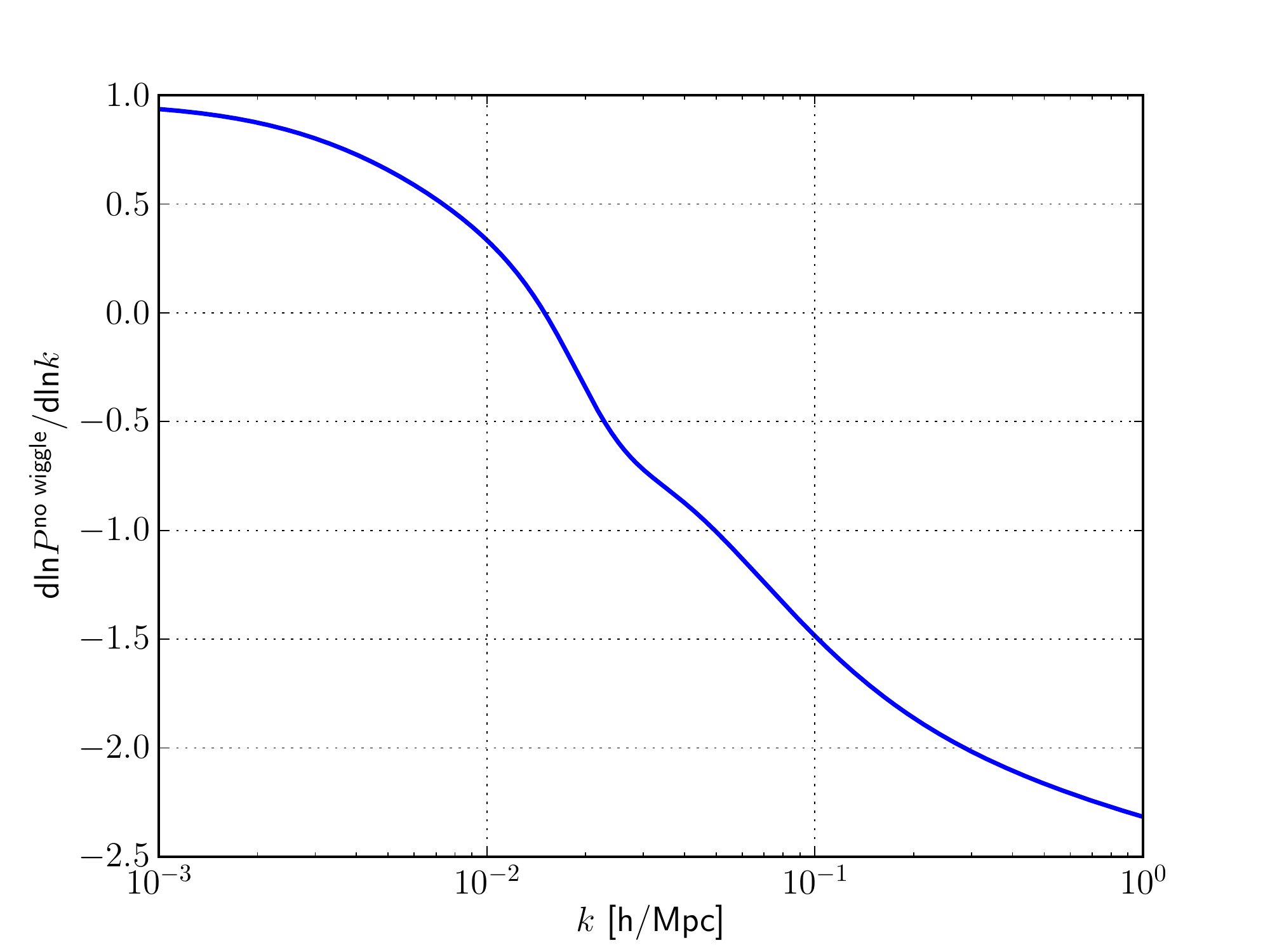}
\caption{\emph{Left panel:} Power index for each contribution to the power spectrum of $\phi$, as a function of the power index of the linear density power spectrum $P_\text{lin}$.
\emph{Right panel:} Power index of the no-wiggle power spectrum from Eisenstein \& Hu \cite{Eisenstein:1997ik}.}
\label{fig:scalings_p}
\end{figure}

We start by reviewing the simple case of an EdS universe, where the initial density power spectrum is a simple power law: $P_\text{lin}(k) = 2\pi^2 k_\text{nl}^{-3} \left( k/k_\text{nl} \right)^n$. In this case, each extra loop contributes a factor $\sim \int \derd \ln q \frac{q^3}{2\pi^2}P_\text{lin}(q) \sim \left( k/k_\text{nl} \right)^{n+3}$ (when the loop integrals become divergent, the situation is slightly more complicated and our discussion here applies to the answer after renormalization). One can therefore estimate the size of the 1-loop corrections as $\sim b(k/k_\text{nl})^{n+3} P_{11}$, and the size of the 2-loop corrections as $\sim b\left[ (k/k_\text{nl})^{n+3} \right]^2 P_{11}$, where $b$ is some unknown factor. Fig.~\ref{fig:scalings_p} shows the power index of the various LPT terms in the EdS universe, as a function of the power spectrum power index $n$.

\begin{figure}[H]
\centering
\includegraphics[width=10cm]{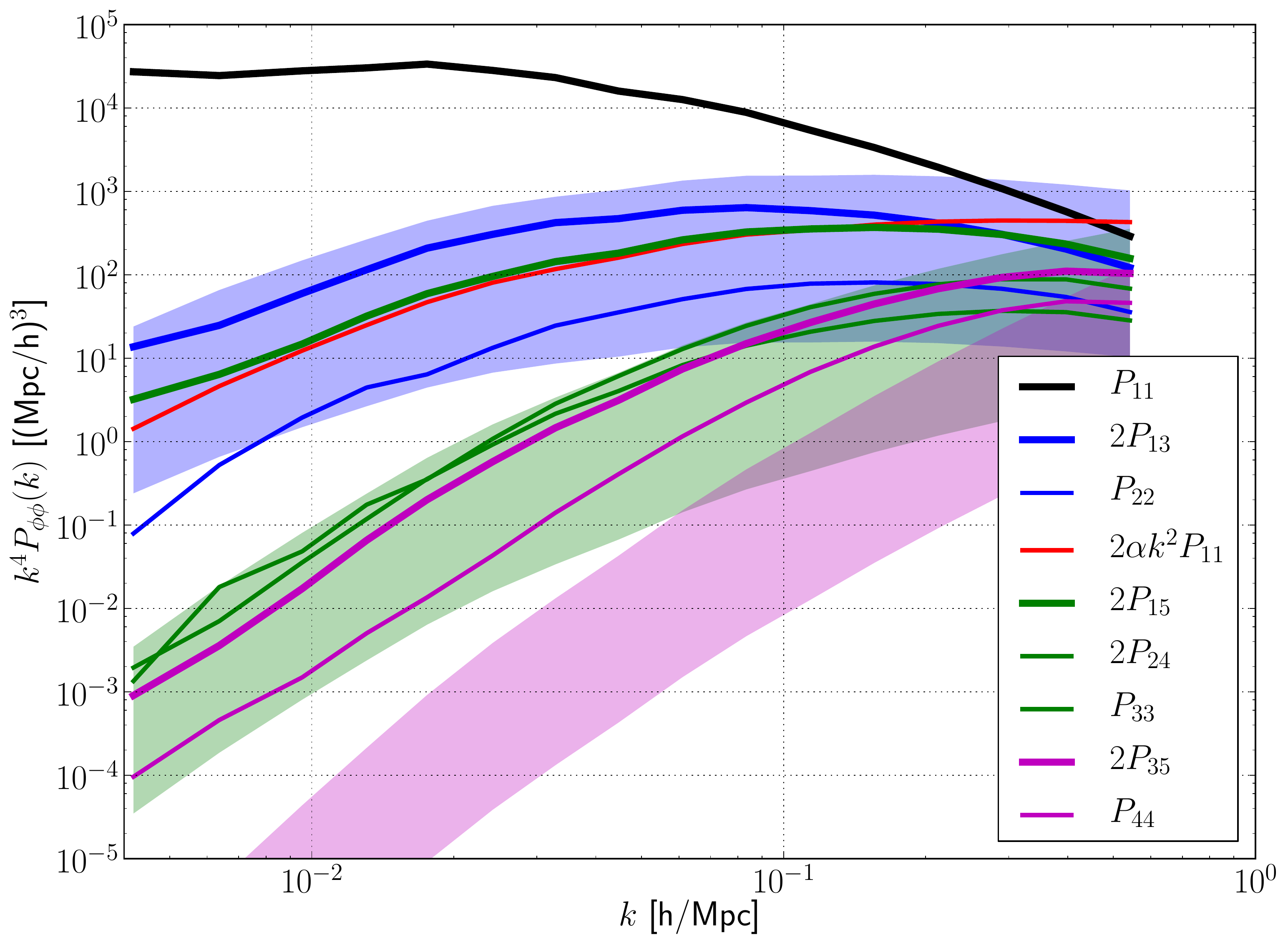}
\caption{Comparison between the scalings from EdS (shaded areas) and the actual loop contributions to the power spectrum (solid lines). The scalings correspond to $n = -1.4$, $k_\text{nl} = 1.05$ h/Mpc and $b$ floating between $0.1$ and $10$. These scalings approximately match the 1-loop terms, but fail at reproducing the higher order terms.}
\label{fig:p_test_scalings}
\end{figure}

We compare these EdS scalings to the actual $\Lambda$CDM LPT terms in Fig.~\ref{fig:p_test_scalings}, where we chose $n=-1.46$ and $k_\text{nl}=0.21 h/$Mpc to match the amplitude and the slope of the linear power spectrum at $k_0=0.1 h/$Mpc, and we let the factor $b$ float between $0.1$ and $10$ (colored bands). 
Fig.~\ref{fig:p_test_scalings} shows that this approach is too na\"\i ve. In particular, various LPT terms corresponding to the same loop order may differ in size by factors of hundreds on large scales, so much so that a 2-loop term like $P_{15}$ can be larger than a 1-loop term like $P_{22}$. It is also clear that as we approach $k\sim 0.2-0.3h/\text{Mpc}$ all terms have roughly the same value. Thus one might suspect that at those scales the perturbative expansion fails. 
We explain this in more detail in the case of the 1-loop power spectrum, in the next subsection.

%>>>>>>>>>>>>>>>>>>>>>>>>>>>>>>>>>>>>>>>>>>>>>>>>>>>>>>>>>>>>>>>>>>>>>>>>>>>>>>>>>>>>>>
%>>>>>>>>>>>>>>>>>>>>>>>>>>>>>>>>>>>>>>>>>>>>>>>>>>>>>>>>>>>>>>>>>>>>>>>>>>>>>>>>>>>>>>

%%%%%%%%%%%%%%%%%%%%%%%%%%%%%%%%%%%%%%%%%%%%%%%%%%%%%%%%%%%%%%%
\subsection{Size of the one-loop contributions: cutoff-dependence and UV mistake}
%%%%%%%%%%%%%%%%%%%%%%%%%%%%%%%%%%%%%%%%%%%%%%%%%%%%%%%%%%%%%%%

\begin{figure}[t]
\centering
\includegraphics[width=0.49\textwidth]{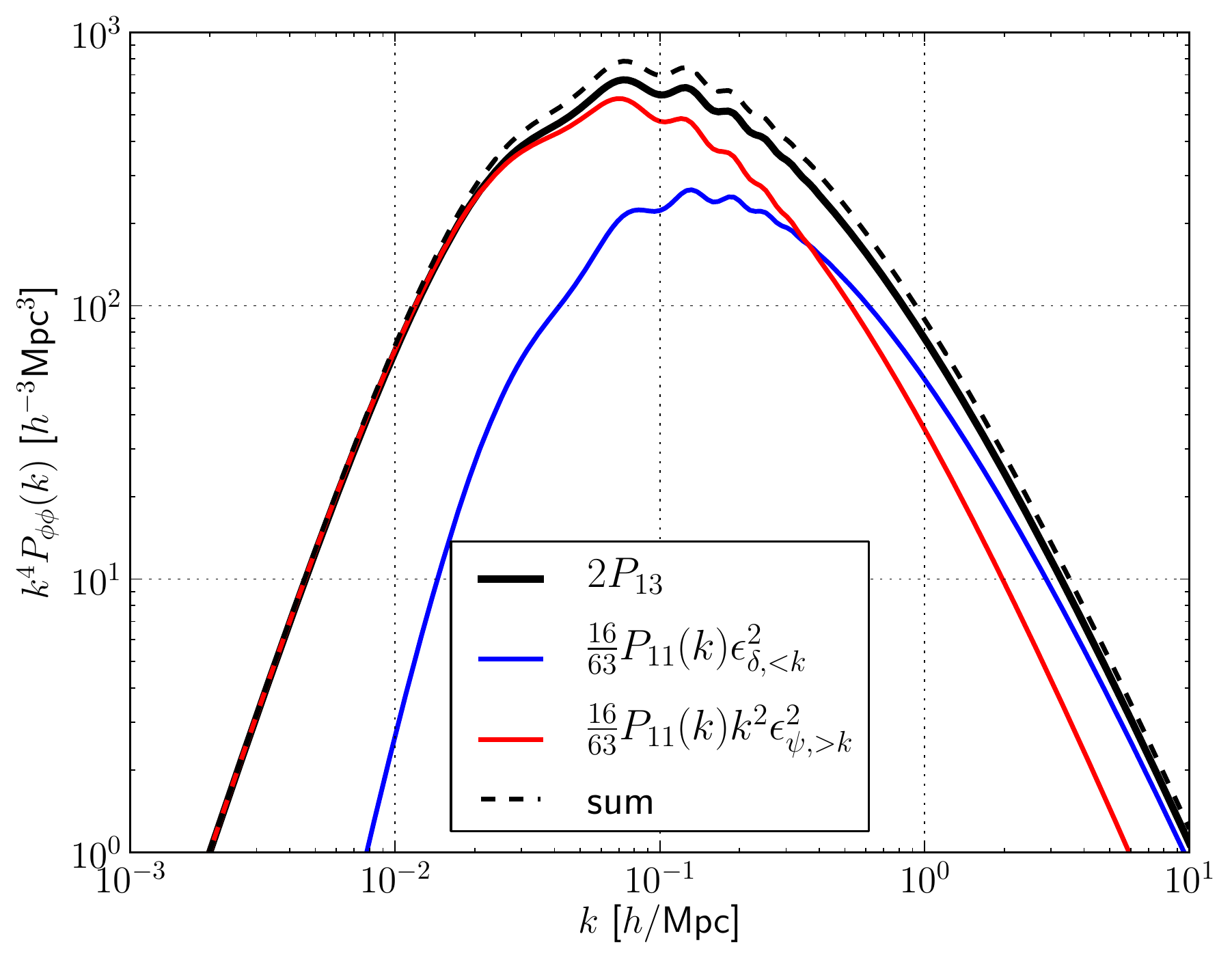}
\includegraphics[width=0.49\textwidth]{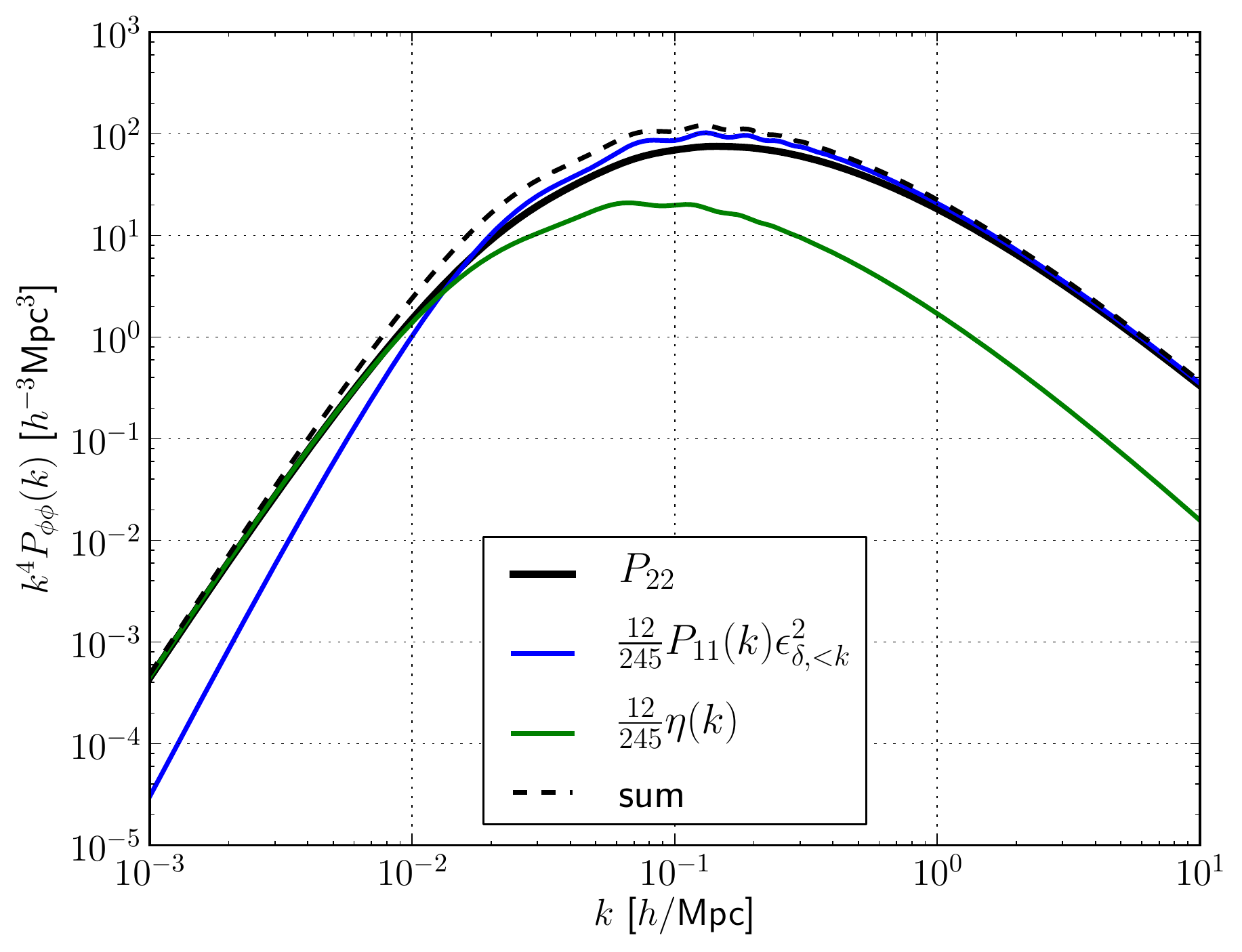}
\caption{\emph{Left panel:} contributions to $P_{13}$ from the large-scale density modes (red line) and small-scale displacement modes (blue line).
\emph{Right panel:} contributions to $P_{22}$ from the large-scale density modes (blue line) and the $\eta$ term.
The dominant contribution comes from the small-scale displacement modes, which only enter the expression of $P_{13}$ and not $P_{22}$.}
\label{fig:size_p13_p22}
\vspace{0.3cm}
\centering
\includegraphics[width=0.49\textwidth]{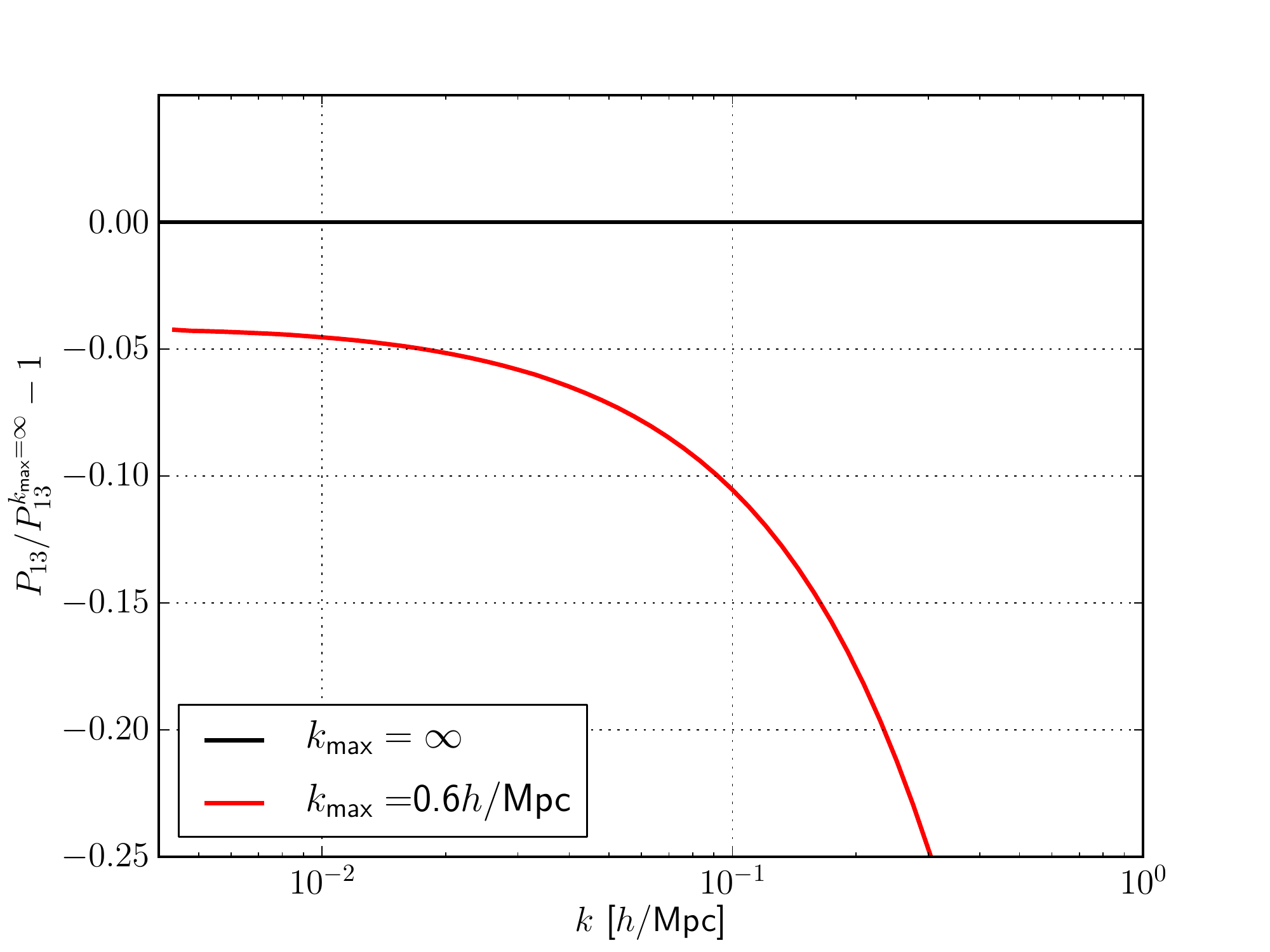}
\includegraphics[width=0.49\textwidth]{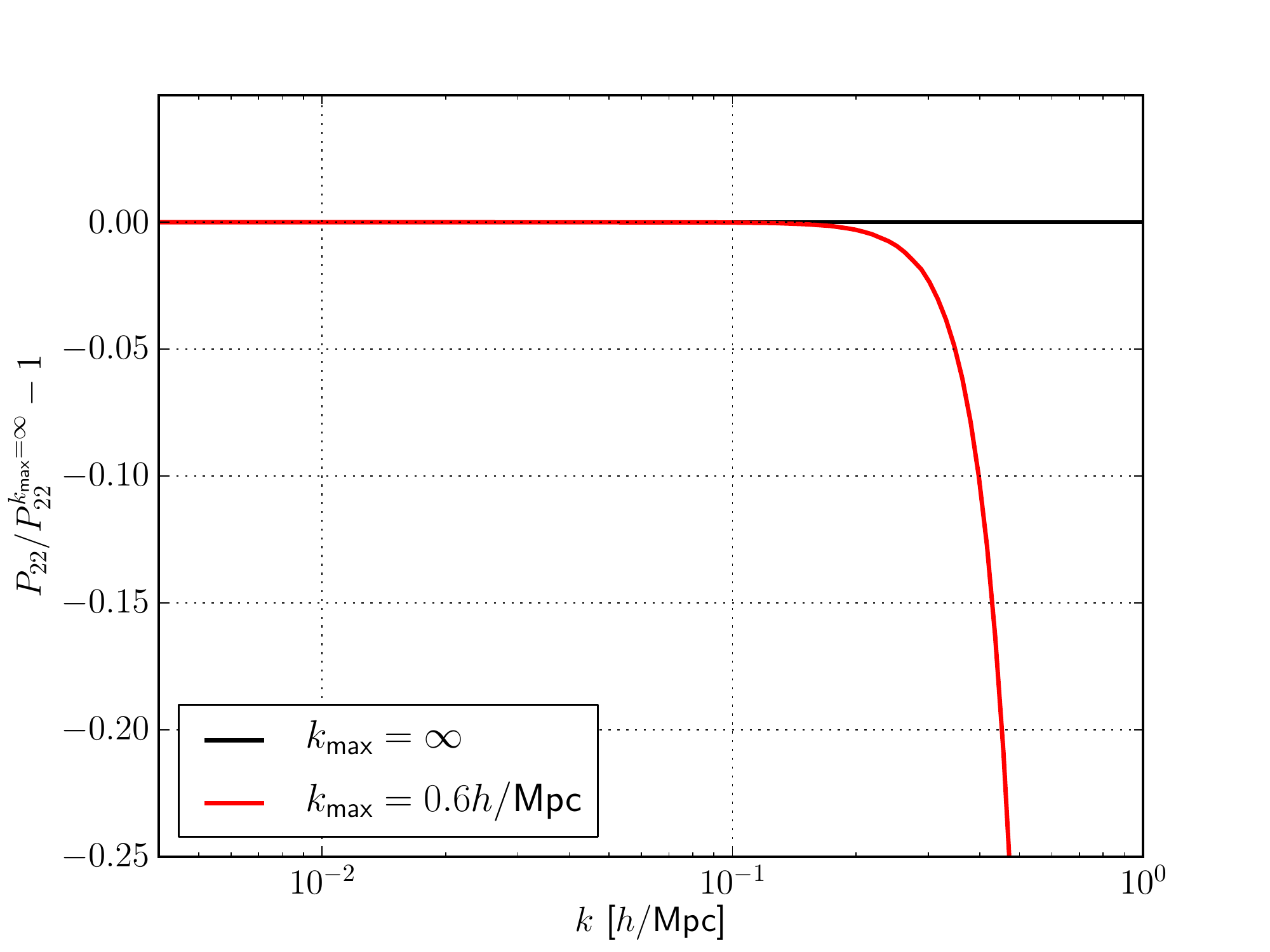}
\caption{Effect of the cutoff on the loop integral for $P_{13}$ (left) and $P_{22}$ (right, same scale). The cutoff-dependence makes $P_{13}$ ill-defined beyond $\sim 10\%$ at $k=0.1 h/$Mpc, whereas $P_{22}$ is very much cutoff-independent.}
\label{fig:cutoffdpdce_p13_p22}
\end{figure}

We wish to better understand the size of the 1-loop LPT contributions $P_{13}$ and $P_{22}$, and in particular to explain why $P_{13} \gg P_{22}$.
To do so, we decompose and approximate the loop integral as
\beq
\int_0^\infty\derd p\; f(k,p) = 
\int_0^k\derd p\; f(k,p) + \int_k^\infty \derd p\;f(k,p) 
\simeq \int_0^k\derd p\; f(k,p\ll k) + \int_k^\infty\derd p\; f(k,p \gg k),
\eeq
where in the last step we replace the integrand by its limit when $p\ll k$ and $p\gg k$ respectively. This procedure yields the following approximations to $P_{13}$ and $P_{22}$:
\beq
\bal
2P_{13}(k) &\simeq \frac{16}{63} P_{11}(k) \left[ \epsilon_{\delta, <k} + k^2 \epsilon_{\psi, >k} \right] \\
P_{22}(k) &\simeq \frac{12}{245} \left[2 P_{11}(k)\epsilon_{\delta, <k} + \eta(k) \right] ,
\eal
\label{eq:size_p13_p22}
\eeq
where we defined
\begin{align}
\epsilon_{\delta, <k} \equiv& \int_0^k \frac{\derd p}{2\pi^2}p^2P_\text{lin}(p)\propto\int_0^k \frac{\derd \ln p}{2\pi^2} p^{3+n}\\
\epsilon_{\psi, >k} \equiv& \int_k^\infty \frac{\derd p}{2\pi^2}P_\text{lin}(p)\propto\int_k^\infty \frac{\derd \ln p}{2\pi^2} p^{1+n}\\
\eta(k) \equiv& \int_k^\infty \frac{\derd p}{2\pi^2}\frac{P_\text{lin}(p)^2}{p^2}\propto\int_k^\infty \frac{\derd \ln p}{2\pi^2} p^{2n-1} \; .
\end{align}
The term $\epsilon_{\delta, <k}$ is the variance of the large-scale density modes with $p\leq k$, i.e. it encodes tides. The term $\epsilon_{\psi, >k}$ is the variance of the small scale displacement modes with $p\geq k$.
The contributions of these terms to $P_{22}$ and $P_{13}$ are shown in Fig.~\ref{fig:size_p13_p22}. Note that the integrand for $\eta$ peaks where the slope of the power spectrum is $n=1/2$, i.e., before the wavenumber of matter radiation equality at $k\approx 0.006 h/\text{Mpc}$. This integral is thus very insensitive to the UV cutoff, which is thus the case for $P_{22}$ as well.
It appears that the dominant term for $k \lesssim 0.4 h/$Mpc is $\epsilon_{\psi, >k}$, which is absent in $P_{22}$. This explains why $P_{13} \gg P_{22}$ for $k \lesssim 0.4 h/$Mpc. 
Besides, the common contribution from $\epsilon_{\delta, <k}$ to $P_{13}$ and $P_{22}$ is multiplied by a larger factor in the case of $P_{13}$. This is why $P_{13} \geq P_{22}$ for $k \gtrsim 0.4 h/$Mpc.

Fig.~\ref{fig:size_p13_p22} shows that the low-$k$ values of $P_{13}$ are dominated by $\epsilon_{\psi, >k}$, the contribution from high momenta $p$ in the loop. These high momenta cannot be correctly described by perturbation theory, and thus bring a mistake in the value of $P_{13}$.  These high momenta in the loop also make the result sensitive to the UV cutoff in the integral. 
On the other hand, $P_{22}$ is very convergent as long as $\Lambda\gg k_\text{eq}$, since $\eta(k)$ is rather insensitive to the slope in the UV.
Thus the cutoff-dependence of the 1-loop power spectrum mostly comes from $P_{13}$ (see Fig.~\ref{fig:cutoffdpdce_p13_p22}.). 
This cutoff-dependence is not physical, and means that the LPT power spectrum at 1-loop is at most defined up to $\sim 10\%$ at $k=0.1 h/$Mpc. This cutoff-dependence is even worse for some higher-loop terms. In particular, $P_{15}$ calculated with a cutoff of $\Lambda=0.6\ihMpc$ deviates from the calculation without a cutoff by $50\%$ at $k=0.1\ihMpc$. The contribution to $P_{33}$ that arises from the $P_{33-II}$ diagram in Fig.~\ref{fig:diagrams_1loop_2loop} inherits the cutoff dependence from $P_{13}$. The $P_{33-I}$ diagram has negligible corrections at $k=0.1\ihMpc$ and $P_{24}$ has $10\%$ correction at this wavenumber.

Eq.~\ref{eq:size_p13_p22} allows us to estimate the scale-dependence of the UV mistake introduced by the high momenta in the loop:
\beq
\bal
&P_{13}^\text{UV}(k) \propto k^2 P_{11}(k) \\
&P_{22}^\text{UV}(k) \propto \text{const.} \\
\eal
\label{eq:uvmistake_p13_p22}
\eeq
The EFT counterterms presented in Sec.~\ref{subsec:eft_counterterms} have the correct form to cancel these contributions, and to provide well-defined cutoff-independent predictions. In particular, the term $\alpha k^2 P_{11}$ corrects the cutoff-dependence of $P_{13}$, while the stochastic term $P_\text{stoch}$ corrects the (smaller) cutoff-dependence of $P_{22}$.

\subsection{Next to leading order EFT counterterms}\label{sec:nloctr}
So far we only discussed the leading order counterterm in the EFT. 
Let us now consider the higher order counterterms, that contribute to the two loop calculation.
The second order counterterm can be derived from a stress tensor $\tau_{ij}$, which is constructed from all possible second order terms with at least two derivatives of the gravitational potential and two free indices \cite{Baldauf:2014qfa}:
\begin{align}
\tau_{ij}=c_1\delta_{ij}^\text{(K)} (\partial^2 \phi^{(1)})^2+c_2 \partial_i\partial_j \phi_1 \partial^2 \phi^{(1)}+c_3\partial_i \partial_l \phi \partial_l\partial_j \phi^{(1)}+c_4\delta_{ij}^\text{(K)} \partial_l \partial_m \phi^{(1)}\partial_l \partial_m \phi^{(1)}.
\end{align}
From this we can derive the ansatz for the acceleration counterterm, where we realize that only three out of the four terms above are linearly independent. Another derivative yields the counterterm for the displacement divergence
\beq
\nabla_{\boldsymbol x} \cdot {\boldsymbol a}_\text{ct} =  \alpha \partial^2(\partial^2 \phi^{(1)}) + e_1 \partial^2 (\partial^2 \phi^{(1)})^2+ e_2 \partial^2 (\partial_i\partial_j \phi^{(1)})^2+ e_3 \partial_i (\partial_i\partial_j \phi^{(1)}\partial_j \partial^2 \phi^{(1)}).
\eeq
Reordering the second order counterterms, we define the three linearly-independent operators
\beq
\bal
E_{2,1}(\vec k_1,\vec k_2)=&1,\\
E_{2,2}(\vec k_1,\vec k_2)=& \kappa_2(\vec k_1,\vec k_2),\\
E_{2,3}(\vec k_1,\vec k_2)=&\frac{(k_1^2+k_2^2)}{(\vec k_1+\vec k_2)^2} \kappa_2(\vec k_1,\vec k_2),
\eal
\eeq
which define the fourth order counterterms by
\begin{equation}
\phi^{(\tilde 2,i)}(\vec k)=\epsilon_i \int \frac{\derd^3 p}{(2\pi)^3}E_{2,i}(\vec p,\vec k-\vec p)\delta_0(\vec p)\delta_0(\vec k-\vec p).
\end{equation}
These objects are in principle introduced at the level of the equation of motion and need to be integrated with the Green's function to generate a displacement field. These integrals only change the unknown prefactors and can thus be reabsorbed into new constants at the level of the displacement field.\\
Note that two of the three operators, namely $E_{2,2}$ and $E_{2,3}$, are also automatically generated by solving the equation of motion in the presence of the leading order counterterm $\alpha k^2 \phi^{(1)}$, as we explain in more details in App.~\ref{app:ctr}. In particular, these next-to-leading order counterterms are generated by either entering the leading order counterterm into $\mathcal{L}^{1,1}$, i.e.
\beq
\kappa_2(\vec k_1,\vec k_2)\frac{(k_1^2+k_2^2)}{(\vec k_1+\vec k_2)^2}\delta_0(\vec k_1)\delta_0(\vec k_2)
\eeq
or by replacing the linear potential in $k^4 \phi^{(1)}$ by the second order potential, i.e.
\beq
\kappa_2(\vec k_1,\vec k_1)\delta_0(\vec k_1)\delta_0(\vec k_2)
\eeq
Note that these two terms are generated in a fixed proportion, determined by the time dependence of the leading order counterterm $\alpha$. Note also that $E_1$ is not generated. We will see later that all three terms $E_{2,1}$, $E_{2,2}$ and $E_{2,3}$ are needed to cure potential divergencies of the displacement field at fourth order.

There are also new counterterms at third (and higher) order in the fields. We will denote them $\phi^{(\tilde 3,j)}(\vec k)$ and the corresponding vertices $E_{3,j}$ without explicitly deriving their functional form.

We showed above that the leading order counterterm $\alpha k^2 \phi$ can be regarded as a loop regulator for the displacement field itself (and not only its power spectrum). Since this counterterm is linear in $\phi$, it is a loop regulator for a diagram with a single external leg. The diagram for the displacement with one loop and one external leg is obtained by considering $\phi^{(3)}$, and closing a loop (i.e. contracting two of the three linear density fields). Indeed, we found that the $\alpha k^2 \phi$ counterterm had the right form to cure the cutoff-dependence of this diagram.

\begin{figure}[t]
\centering
\includegraphics[width=0.7\textwidth]{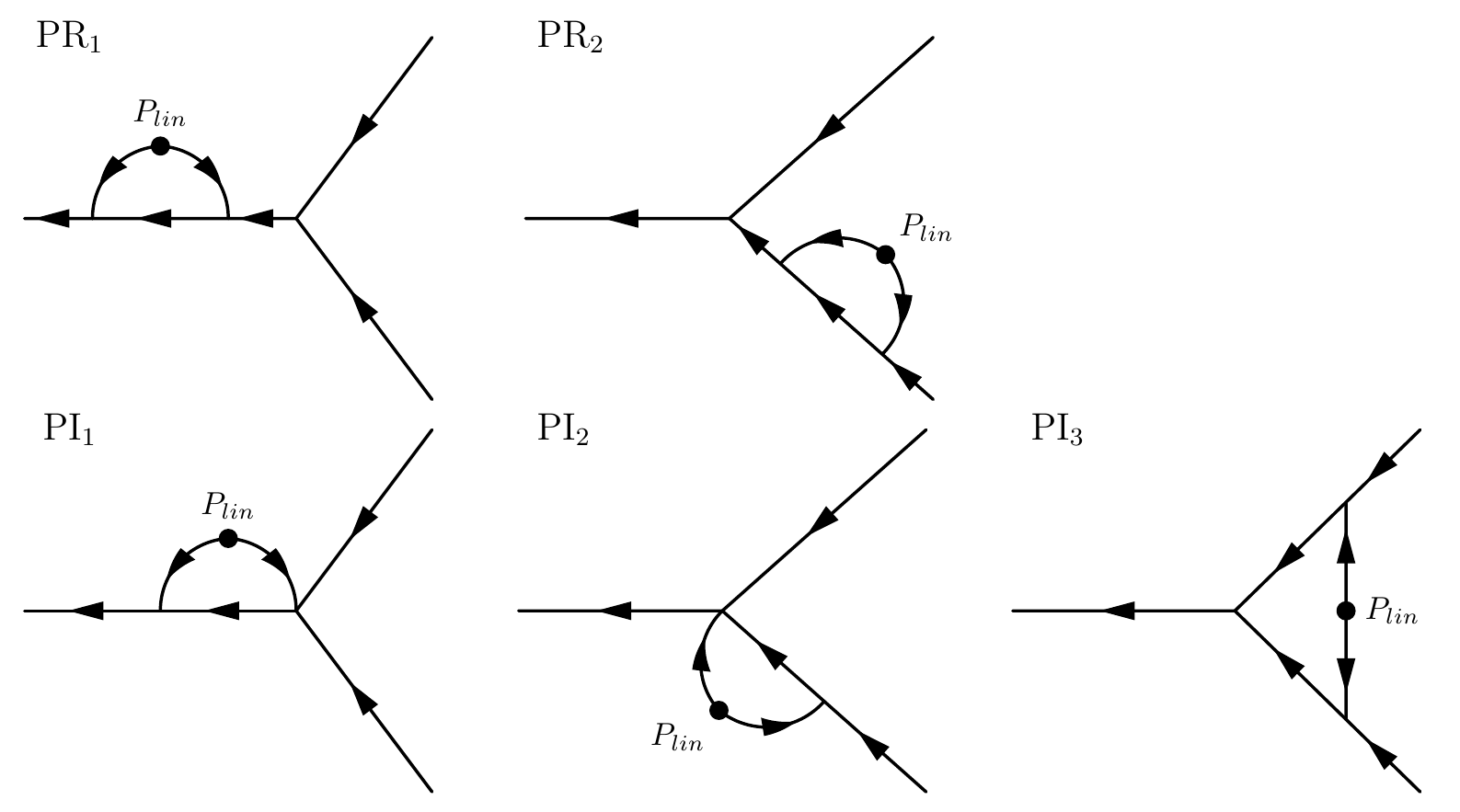}
\caption{Diagrams of the reducible and irreducible contributions to the fourth order displacement divergence. Note that the vertices in these diagrams are the fundamental LPT vertices $\kappa_2$ and $\kappa_3$ instead of the full LPT vertices $L_i$. The above diagrams arise from writing all possible diagrams for $\phi_4$ and then correlating intermediate fields, leaving two of the ingoing fields floating.}
\label{fig:div2nd}
\end{figure}

Similarly, the next-to-leading order operators $E_{2,1}$, $E_{2,2}$ and $E_{2,3}$ correspond to counterterms quadratic in $\phi$, and
can be seen as regulators for the loops of LPT displacements with two external legs (corresponding to terms quadratic in $\phi$).
In order to close a loop, we need to add two input fields, which leads us to consider the fourth order displacement:
\begin{align}
\phi^{(4)}(\vec k_1+\vec k_2)\supset \frac{1}{4}\delta_0(\vec k_1)\delta_0(\vec k_2) \int\frac{\derd^3 p}{(2\pi)^3} L_4(\vec p,-\vec p,\vec k_1,\vec k_2) P_\text{lin}(p )
\end{align}
The contributing diagrams, with one loop and two external legs, are shown in Fig.~\ref{fig:div2nd} in terms of the fundamental LPT vertices $\kappa_2$ and $\kappa_3$.

Some of them are trivial or reducible counterterms, which are directly related to the leading order counterterm that was introduced to cure $P_{13}$
\footnote{Here we introduce the short-hand notation $\int _{\vec{p}} \equiv \int \frac{ \derd^3 \vec{p}} {(2\pi)^3} $.}:
\beq
\text{PR}_1\propto \int_{\vec p}  \frac{\kappa_2(\vec k_1,\vec k_2)\kappa_2(\vec k_1+\vec k_2+\vec p,-\vec p)\kappa_2(\vec k_1+\vec k_2,\vec p)}{(\vec k_1+\vec k_2)^2}P_\text{lin}(p )\stackrel{p\gg k}{=}\frac{24}{5}\epsilon_{\psi, >k}  E_{2,2}(\vec k_1,\vec k_2),
\eeq
\beq
\text{PR}_2\propto\int_{\vec p}
\frac{\kappa_2(\vec k_1,\vec k_2)\bigl[\kappa_2(\vec k_1+\vec p,-\vec p)
\kappa_2(\vec k_1,\vec p)+(\vec k_1 \leftrightarrow \vec k_2)\bigr]}{(\vec k_1+\vec k_2)^2}
P_\text{lin}(p )\stackrel{p\gg k}{=}\frac{24}{5}\epsilon_{\psi, >k} E_{2,3}(\vec k_1,\vec k_2),
\eeq
Others have a slightly more complex vertex structure, that requires additional counterterms:
\beq
\text{PI}_1\propto\int_{\vec p} \frac{\kappa_3(\vec k_1, \vec k_2, \vec p) \kappa_2(\vec k_1 + \vec k_2 + \vec p, -\vec p)}{(\vec k_1+\vec k_2)^2}P_\text{lin}(p )\stackrel{p\gg k}{=}\frac{12}{5}\epsilon_{\psi, >k} E_{2,2}(\vec k_1,\vec k_2),
\eeq
\beq
\text{PI}_2\propto\int_{\vec p}  \frac{\bigl[\kappa_3(\vec k_1 + \vec p, 
  \vec k_2, -\vec p) \kappa_2(\vec k_1, \vec p) +(\vec k_1\leftrightarrow \vec k_2)\bigr]}{(\vec k_1+\vec k_2)^2}\stackrel{p\gg k}{=}\frac{12}{5}\epsilon_{\psi, >k} E_{2,3}(\vec k_1,\vec k_2),
\eeq
\begin{align}
\text{PI}_3 &\propto \int_{\vec p}  \frac{\kappa_2(\vec k_1+\vec p,\vec k_1-\vec p)\kappa_2(\vec k_1,\vec p)
\kappa_2(\vec k_2,-\vec p)}{(\vec k_1+\vec k_2)^2}\nonumber\\
&\stackrel{p \gg k}{=}3 \epsilon_{\psi, >k}\Bigl[\frac{48}{35}E_{2,1}(\vec k_1,\vec k_2) -\frac{8}{35}E_{2,2}(\vec k_1,\vec k_2) -\frac{8}{35}E_{2,3}(\vec k_1,\vec k_2)\Bigr].
\end{align}
We see that all the divergencies can be projected on the basis $E_{2,i}$, which is thus sufficient to cure the divergencies.

\begin{figure}[t]
\centering
\includegraphics[width=0.7\textwidth]{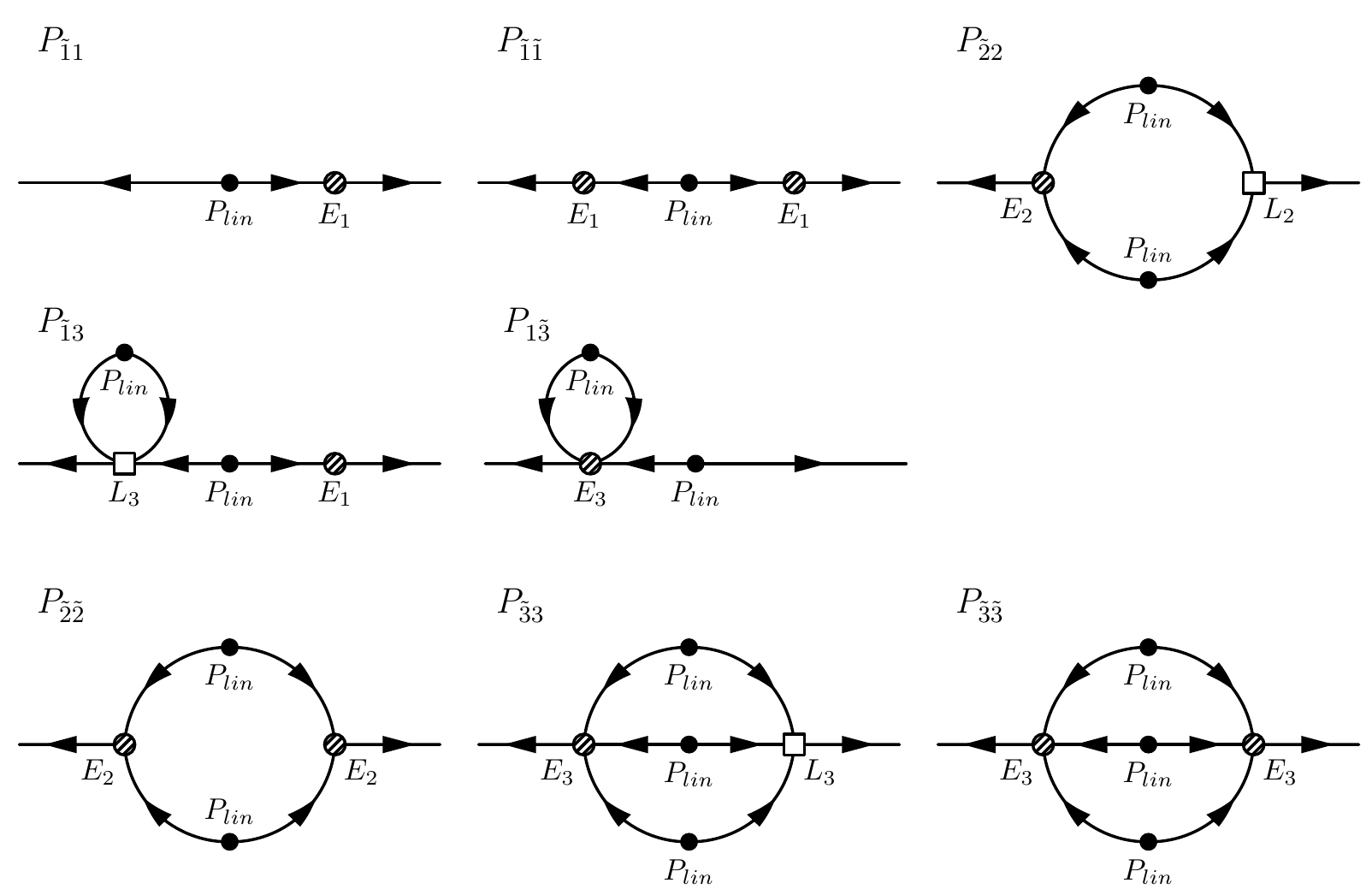}
\caption{Diagrams for the EFT corrections to the 2-loop power spectrum. Here $E_1(k) \equiv \alpha k^2$.}
\label{fig:diag_p_eft_2loop}
\end{figure}

Given the leading and next-to-leading counterterms we just derived, we wish to obtain the corresponding EFT corrections to the 2-loop power spectrum.
At the  2-loop level, the LPT power spectra receive corrections that arise from replacing the linear power spectra in the one loop expressions by the one loop counterterm $k^2 P_\text{lin}$, or by replacing the vertices with the EFT vertices $E_{i,j}$, as shown in Fig.~\ref{fig:diag_p_eft_2loop}.
This leads to the following terms (where the tilde denotes a counterterm):
\begin{align}
P_{\tilde 2 2,i}\propto &\int_{\vec{p}} P_\text{lin}(q) P_\text{lin}(|\vec k-\vec p|) E_{2}(\vec p,\vec k-\vec p)L_2(-\vec p,-\vec k+\vec p),\nonumber\\
P_{\tilde 2\tilde 2,ij}\propto &\int_{\vec{p}} P_\text{lin}(p)  P_\text{lin}(|\vec k-\vec p|) E_{2,i}(\vec p,\vec k-\vec p)E_{2,j}(-\vec p,-\vec k+\vec p),\nonumber\\
P_{1\tilde3,i}\propto & \; P_\text{lin}(k)\int_{\vec{p}} P_\text{lin}(p) E_{3,i}(\vec p,-\vec p,\vec k),\\
P_{\tilde 3\tilde 3,ij}\propto & \int_{\vec{p}} \int_{\vec{p'}} P_\text{lin}(p)P_\text{lin}(p')P_\text{lin}(|\vec k-\vec p-\vec p'|) E_{3,i}(\vec p,\vec p',\vec k-\vec p-\vec p')E_{3,j}(-\vec p,-\vec p',-\vec k+\vec p+\vec p'),\nonumber\\
P_{\tilde 33,i}\propto & \int_{\vec{p}}\int_{\vec{p'}} P_\text{lin}(p)P_\text{lin}(p')P_\text{lin}(|\vec k-\vec p-\vec p'|) E_{3,i}(\vec p,\vec p',\vec k-\vec p-\vec p') L_3(-\vec p,-\vec p',-\vec k+\vec p+\vec p').\nonumber
\end{align}
Note that at the two loop level, only correlators of the second order counterterms with LPT terms can contribute: these are the terms $P_{\tilde{2} 2,i}$ and $P_{1 \tilde{3},i}$. Thus going to two-loop order in the EFT requires including these counterterms.
The auto-power of the second order counterterms (i.e. the terms $P_{\tilde 2\tilde 2,ij}$) and the auto-power of the third order counterterms (i.e. $P_{\tilde 3\tilde 3,ij}$) correspond to at least three-loop diagrams.

In the tLPT models, which we shall define shortly, any cross spectrum with the LPT terms (such as $P_{\tilde{2} 2,i}$ and $P_{1 \tilde{3},i}$) is automatically accounted for by the transfer functions. Thus, there is no need to explicitly calculate the counterterms at the two-loop tLPT level.

In the low-$k$ regime we have $P_{\tilde 2 2,1}\propto k^4$, $P_{\tilde 2 2,2}\propto k^6$ and $P_{\tilde 2 2,3}\propto k^4$, i.e., two of the terms scale as $P_{22}$ while one of them is suppressed by an additional $k^2$. For the cubic counterterms we expect $P_{1\tilde{3}}\propto k^{2i} P,\ i=0,1,2\ldots$. These scalings will become relevant later, when we focus on the scale dependence of the transfer functions.

%%%%%%%%%%%%%%%%%%%%%%%%%%%%%%%%%%%%%%%%%%%%%%%%%%%%%%%%%
%%%%%%%%%%%%%%%%%%%%%%%%%%%%%%%%%%%%%%%%%%%%%%%%%%%%%%%%%
\section{Testing LPT and EFT at the field level: moving beyond cosmic variance}
\label{sec:1loop}
%%%%%%%%%%%%%%%%%%%%%%%%%%%%%%%%%%%%%%%%%%%%%%%%%%%%%%%%%
%%%%%%%%%%%%%%%%%%%%%%%%%%%%%%%%%%%%%%%%%%%%%%%%%%%%%%%%%

%%%%%%%%%%%%%%%%%%%%%%%%%%%%%%%%%%%%%%%%%%%%%%%%%%%%%%%%%
\subsection{Simulation Suite: measurements and systematic errors}
%%%%%%%%%%%%%%%%%%%%%%%%%%%%%%%%%%%%%%%%%%%%%%%%%%%%%%%%%

We use a suite of $N$-body simulations to study the fully non-linear evolution of the cosmic density field. The initial conditions for the $N_\text{p}=1024^3$ particles are set up at $z_\text{init}=99$ with the publicly available 2LPT code \cite{Crocce:2006ve}. We subsequently follow their trajectories using GADGET-II \cite{Springel2005}.
We use two different box sizes with the same number of particles, the L simulation (for ``large'') has a box length of $1500 \hMpc$ and the M simulation (for ``medium'') has a box length of $500 \hMpc$.
The cosmological parameters are based on the WMAP7 \cite{WMAP7} CMB analysis:
$\Omega_\text{m}=0.272$, $\sigma_8=0.81$, $n_\text{s}=0.967$.
We have 16 independent realizations for the L simulation and one for the M simulation.
Based on the particle IDs we can reconstruct their displacement from the uniform Lagrangian grid
\beq
\vec \psi=\vec x-\vec q.
\eeq
The displacement vector is then assigned to the Lagrangian grid position, the three Cartesian components of the displacement grid are transformed to Fourier space and the displacement potential is estimated.
The number of particles and the box size limit the calculation to wavenumbers smaller than the Nyquist wavenumber $k_\text{Ny}=\pi N_\text{c}/L_\text{box}$, where $N_\text{c}$ is the number of grid cells per dimension ($1024$ for both simulations) and $L_\text{box}$ is the size of the simulation box.
This corresponds to $k_\text{Ny,L}=2.14 \ihMpc$ for the L simulation and $k_\text{Ny,M}=6.4 \ihMpc$ for the M simulation.

We calculate the LPT displacements for the same initial conditions, i.e. the phases of the perturbative displacements agree with the ones that seeded the fully non-linear $N$-body simulation. This allows us to perform cross-correlations on a mode by mode basis, rather than averaging the perturbation theory and the simulations separately and comparing only their auto-power spectrum. This is clearly a much more stringent test of any perturbative treatment than just comparing the fairly smooth scale dependence of a power spectrum. Matching the power spectrum means that one tries to get $\mathcal{O}(10^2)$ numbers right, whereas we are trying to match a $\mathcal{O}(10^9)$ dimensional vector.
In practice, we measure the auto and cross-correlations of the various displacements $\phi_\text{nl}$, $\phi^{(1)}$, $\phi^{(2)}$, $\phi^{(3)}$, $\phi^{(4)}$ and $\phi^{(5)}$.

In order to compute the LPT displacements on the grid, we use the real-space formulation \eqref{eq:lpt_grid_scalar} and \eqref{eq:lpt_grid_vector}. This involves a sequence of products and Fast Fourier Transforms, which is computationally more efficient than the convolutions \eqref{eq:convolution}. In particular, inverse Laplacians and derivatives are calculated in Fourier space, but convolutions are computed by transforming to real space and multiplying.

Because this is a non-linear calculation, aliasing is an issue. Coupling $n$ modes with wavenumbers $k<k_\text{cutoff}$, we can generate modes up to $n\, k_\text{cutoff}$. If this wavenumber exceeds the Nyquist wavenumber, it folds back to $2 k_\text{Ny}-n\, k_\text{cutoff}$ and can lead to spurious effects.
We avoid this by using a cutoff $k_\text{cutoff}$ in $k$-space for our LPT calculation, implemented by setting to zero all the modes of the initial linear density field with $k> k_\text{cutoff}$.  If we want the modes $k \leqslant k_\text{cutoff}$ of $\phi^{(n)}$ to be unaffected by aliasing, we need to set $k_\text{cutoff} \leqslant 2/(1+n)k_\text{Ny}$.

%%%%%%%%%%%%%%%%%%%%%%%%%%%%%%%%%%%%%%%%%%%%%%%%%%%%%%%%%
\subsection{Fitting for $\alpha$}
%%%%%%%%%%%%%%%%%%%%%%%%%%%%%%%%%%%%%%%%%%%%%%%%%%%%%%%%%

\begin{figure}[t]
\centering
\includegraphics[width=0.49\textwidth]{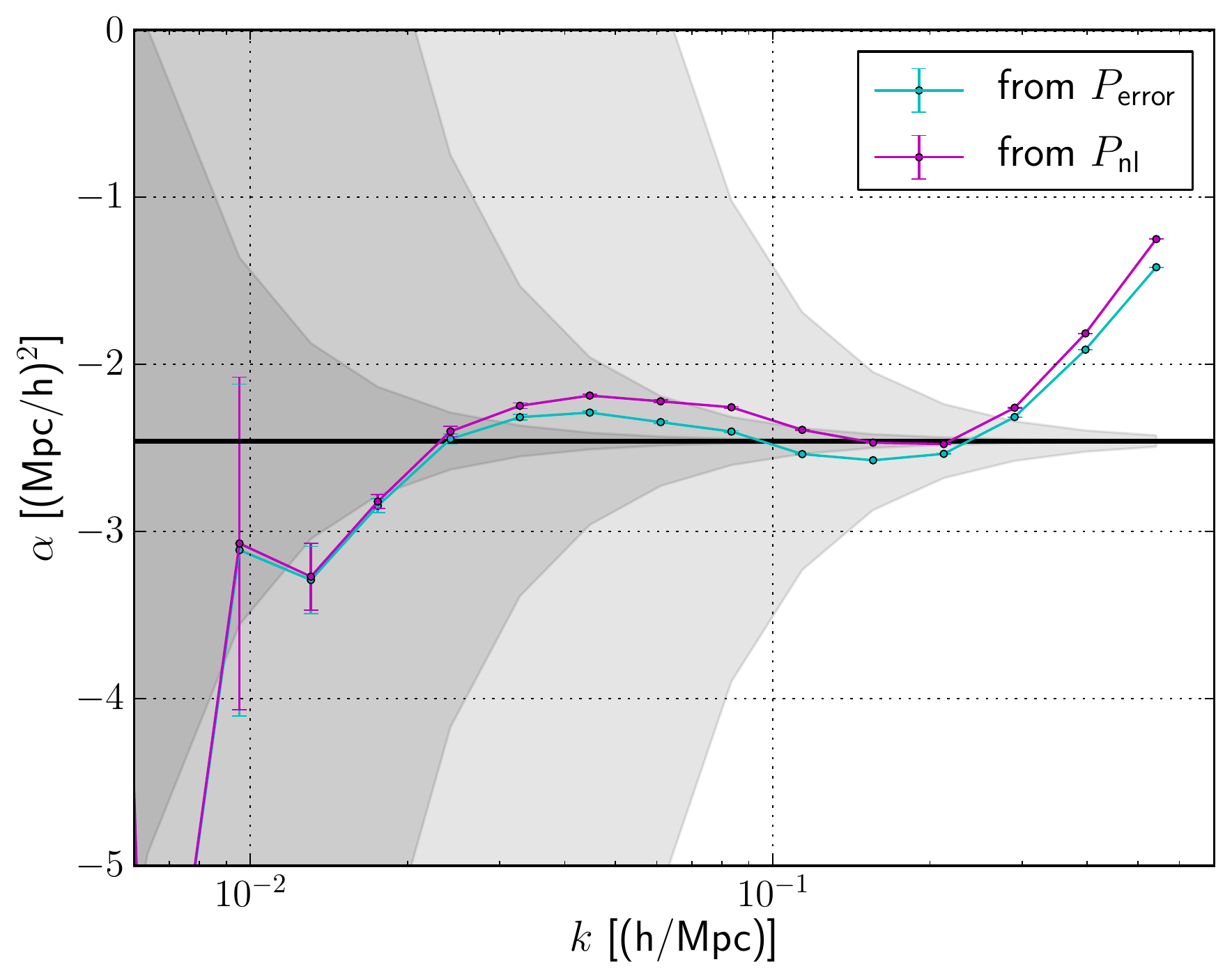}
\includegraphics[width=0.49\textwidth]{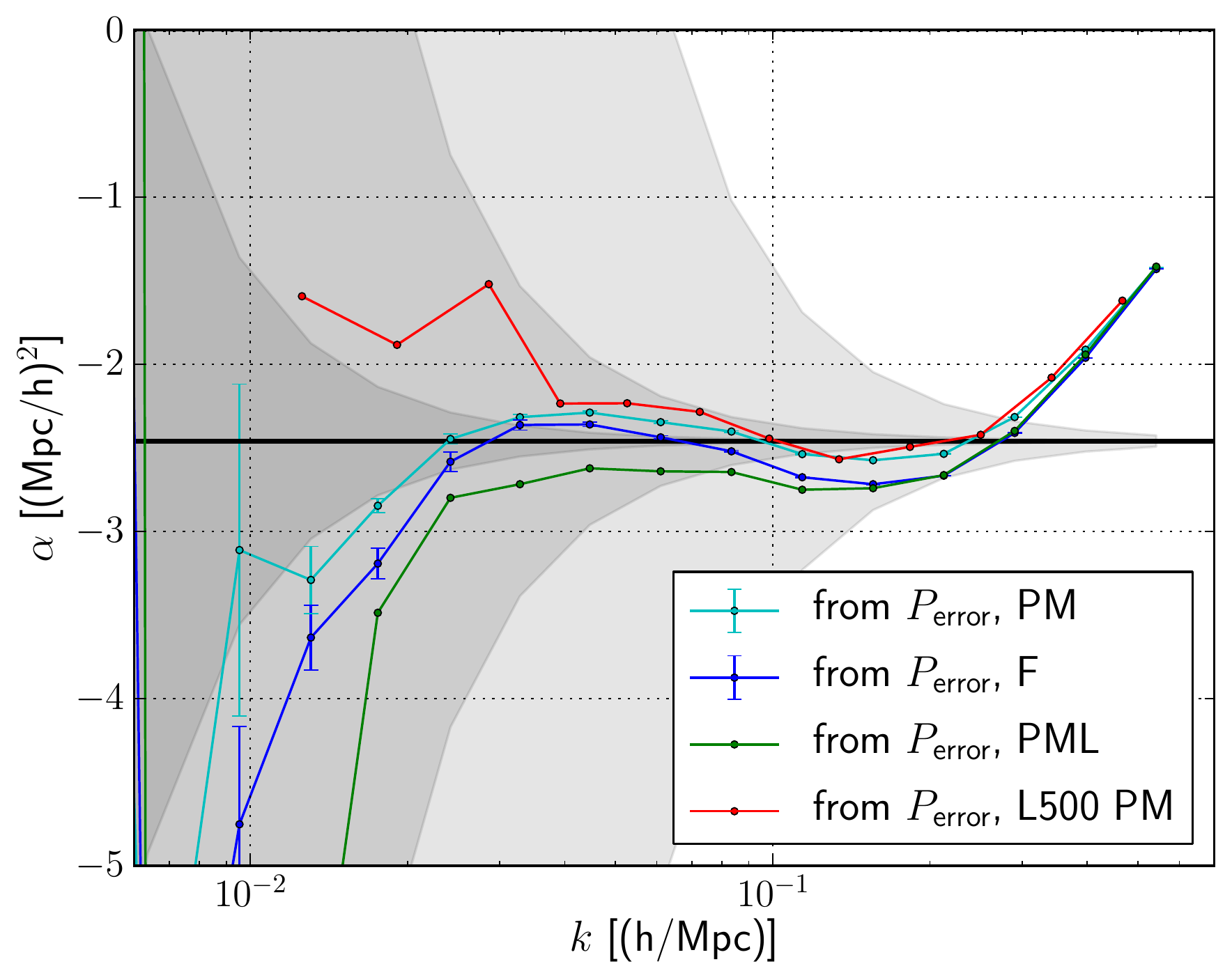}
\caption{
\emph{Left panel:} Values of $\alpha_k$ obtained from fitting to the power spectrum and from minimizing the error power spectrum for a cutoff $k_\text{cutoff}=0.61 \ihMpc$, extracted from the same realization. The two methods agree on large scales but start to differ at percent level for $k > 0.03 \ihMpc$. 
\emph{Right panel:} Values of $\alpha_k$ from minimizing the error power spectrum, for the different simulation runs and $k_\text{cutoff}=0.61 \ihMpc$. This shows that the systematic error in our measurement is sizeable for the smallest $k$.
The shaded areas indicate the domains in which changing $\alpha$ does not affect the non-linear power spectrum by more than $0.01\%$, $0.1\%$ and $1\%$ respectively.}
\label{fig:alpha_fit}
\centering
\includegraphics[width=9cm]{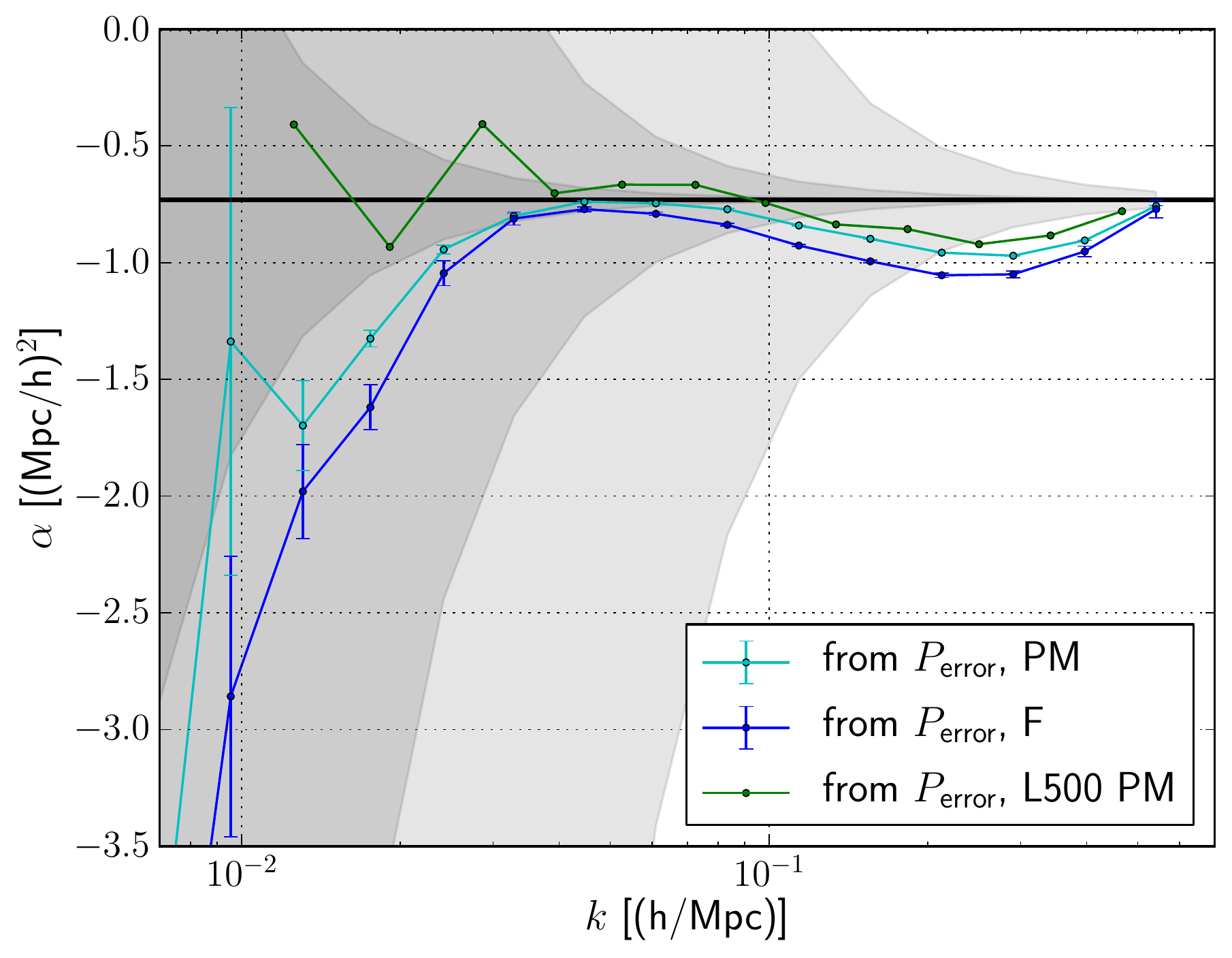}
\caption{
Values of $\alpha_k$ from minimizing the error power spectrum at redshift $z=0.5$, for the different simulation runs. The shaded areas indicate the domains in which changing $\alpha$ does not affect the non-linear power spectrum by more than $0.01\%$, $0.1\%$ and $1\%$ respectively.}
\label{fig:alpha_fit_z0_5}
\end{figure}

As a first step in our test of the Lagrangian EFT, we measure the free EFT parameter $\alpha$ that appears at 1-loop order, as defined in Eq.~\eqref{eq:def_alpha_phi}.
We fit for the coefficient $\alpha$ in two ways: by minimizing the variance between the non-linear field and the theoretical model on the one hand, encoded by the spectrum $P_\text{error}$ of $\phi_\text{error} \equiv \phi_\text{nl} - \phi_\text{model}$, and by adjusting the model power spectrum to the non-linear power spectrum $P_\text{nl}$ on the other hand. The resulting $\alpha_k$ are obtained as follows, and shown in Fig.~\ref{fig:alpha_fit}:
\beq
\left\{
\bal
&\text{minimizing  }P_\text{error} = \langle |\phi_\text{nl} - \phi_\text{3LPT} - \alpha k^2 \phi^{(1)}|^2 \rangle
&&\longrightarrow 
\alpha_\text{error} = \frac{ P_{1\times\text{nl}} - P_{1\times\text{3LPT}}}{k^2 P_{11}} ,\\
&\text{minimizing  }(P_\text{nl} - P_\text{3LPT} - 2\alpha k^2 P_{11})^2
&&\longrightarrow 
\alpha_\text{nl} = \frac{1}{2}\frac{ P_\text{nl} - P_\text{3LPT}}{k^2 P_{11}} . \\
\eal
\right.
\eeq

In both cases, we first determine the value of $\alpha_k$ for each k-bin independently, and estimate the error bars on these values by looking at the scatter across the independent realizations.
Keeping the first term $\alpha k^2$ in the EFT expansion is only supposed to be a valid approximation at low $k$, where the higher order terms are small. We would therefore expect the $\alpha_k$ curves shown in Fig.~\ref{fig:alpha_fit} to look like a constant at low $k$, and then to deviate from that constant at higher $k$.

% deviation from a constant at low k
Instead, the curves we obtain seem to deviate dramatically from a constant at the smallest $k$, for the following reason. At very low $k$, our measurement of $\alpha$ is extremely sensitive to systematic errors in the simulation, since an error of $x \%$ on the non-linear displacement translates into an error of $x/k^2 \; \%$ on $\alpha$. The grey domains on Fig.~\ref{fig:alpha_fit} represent the variations in $\alpha$ caused by variations in the non-linear displacement of $0.01\%$, $0.1\%$ and $1\%$ respectively. On large scales ($k \lesssim 0.05 \ihMpc$), the dramatic upward or downward trends seen in $\alpha$ (right panel of Fig.~\ref{fig:alpha_fit}) are the reflection of the $\sim 0.05\%$ variations in the non-linear power spectrum arising from different $N$-body settings discussed in App.~\ref{app:numerics}.

% deviation from a constant at high k
The values of $\alpha$ obtained also deviate from a constant at high $k$, which is expected due to higher order terms in the EFT expansion. 
Some of these terms -- the ones that correlate with $\phi^{(1)}$ -- affect both methods in the same way. This is the case for the $P_{15}$ term, which we will discuss in more detail below in Sec.~\ref{sec:2loop}, and the higher order order EFT counterterm of the form $k^4 \phi^{(1)}$.
A na\"{\i}ve estimate for the latter term is given by $\alpha^2 k^4 P_{11}$, which leads to a $5\%$ correction to the inferred value of $\alpha$ at $k\approx 0.14 \ihMpc$, where $\alpha$ itself leads to a $10\%$ correction to the non-linear power spectrum.

% disagreement between the two methods: reason why
However, some other higher order EFT terms affect only the value of $\alpha_k$ obtained from the non-linear power spectrum. These are the terms that are not of the form $P_{1i}$. Indeed, considering only the LPT terms, we expect
\begin{align}
\alpha_\text{error}=\frac{P_{\text{nl}\times1}-P_{11}-P_{13}}{k^2 P_{11}}, &&
\alpha_\text{nl}=\frac{P_{\text{nl}}-P_{11}-P_{22}-2P_{13}}{2k^2 P_{11}},
\end{align}
and therefore
\beq
\alpha_\text{nl}=\alpha_\text{error}+\frac{P_{33}+2P_{24}}{2 k^2P_{11}}+...
\label{eq:nlerrordiff}
\eeq
These correction terms are suppressed on large scales, but sufficient to explain the percent level difference between the $\alpha_k$ obtained from the two methods, shown in Fig.~\ref{fig:alpha_fit}. Thus, as soon as the $\alpha_k$ from the two methods differ, we know that higher order corrections are important, and the estimator from fitting to the non-linear power spectrum should not be trusted anymore. At the same time, two-loop corrections to the power spectrum will become important. 
Note that the auto-power spectrum of the stochastic term will only affect the value of $\alpha_\text{nl}$ and not $\alpha_\text{error}$.

% disagreement between the two methods: overfitting and cosmic variance
From Fig.~\ref{fig:alpha_fit}, the values of $\alpha$ inferred from both methods agree for $k \lesssim 0.03 h/$Mpc, and disagree for $k \gtrsim 0.03 h/$Mpc. This shows that for our purpose, where we focus on very large scales, there is no overfitting issue: choosing $\alpha$ to get the best displacement field or the best power spectrum on these scales is equivalent. However, this means that one should not determine the value of $\alpha$ by matching the power spectrum on scales $k \gtrsim 0.03 h/$Mpc.

Picking the value of $\alpha$ around $k=0.03 \ihMpc$, we find $\alpha = -2.46 \pm 0.2 (\hMpc)^2$, where the error bar represents the deviations between simulations run with different settings, and therefore corresponds to a systematic error. Remember also that the exact value of $\alpha$ is dependent on the cutoff used for the LPT calculation. Here we used $k_\text{cutoff} = 0.61 h/$Mpc. This can be translated to $\alpha_{\Lambda=\infty} = -3.0 \pm 0.2 (\hMpc)^2$ for $k_\text{cutoff} = \infty$, which corresponds to a 1\% correction to the power spectrum at $k=0.06 \ihMpc$.\footnote{This $k^2$ coefficient is not captured by two loop corrections. In fact there is a positive $k^2 P_{11}$ contribution from $P_{15}$, which would thus require an even more negative value of $\alpha$ after the two loop power spectrum has been included. We will come back to this issue in Sec.~\ref{sec:2loop}.}
The shaded areas in Fig.~\ref{fig:alpha_fit} show that our value of $\alpha$ allows to reproduce the non-linear power spectrum with an accuracy of 1\% up to $k\simeq 0.15 h/$Mpc.

The same measurement at $z=0.5$ yields the value $\alpha = -0.73 \pm 0.07 (\hMpc)^2$ (see Fig.~\ref{fig:alpha_fit_z0_5}), corresponding to $\alpha_{\Lambda=\infty} =  -1.13 \pm 0.07 (\hMpc)^2$ for $k_\text{cutoff} = \infty$. However, at higher redshift, the absolute value of $\alpha$ drops below the systematic error in the simulations, and we are no longer able to measure it.

\begin{figure}[t]
\centering
\includegraphics[width=0.49\textwidth]{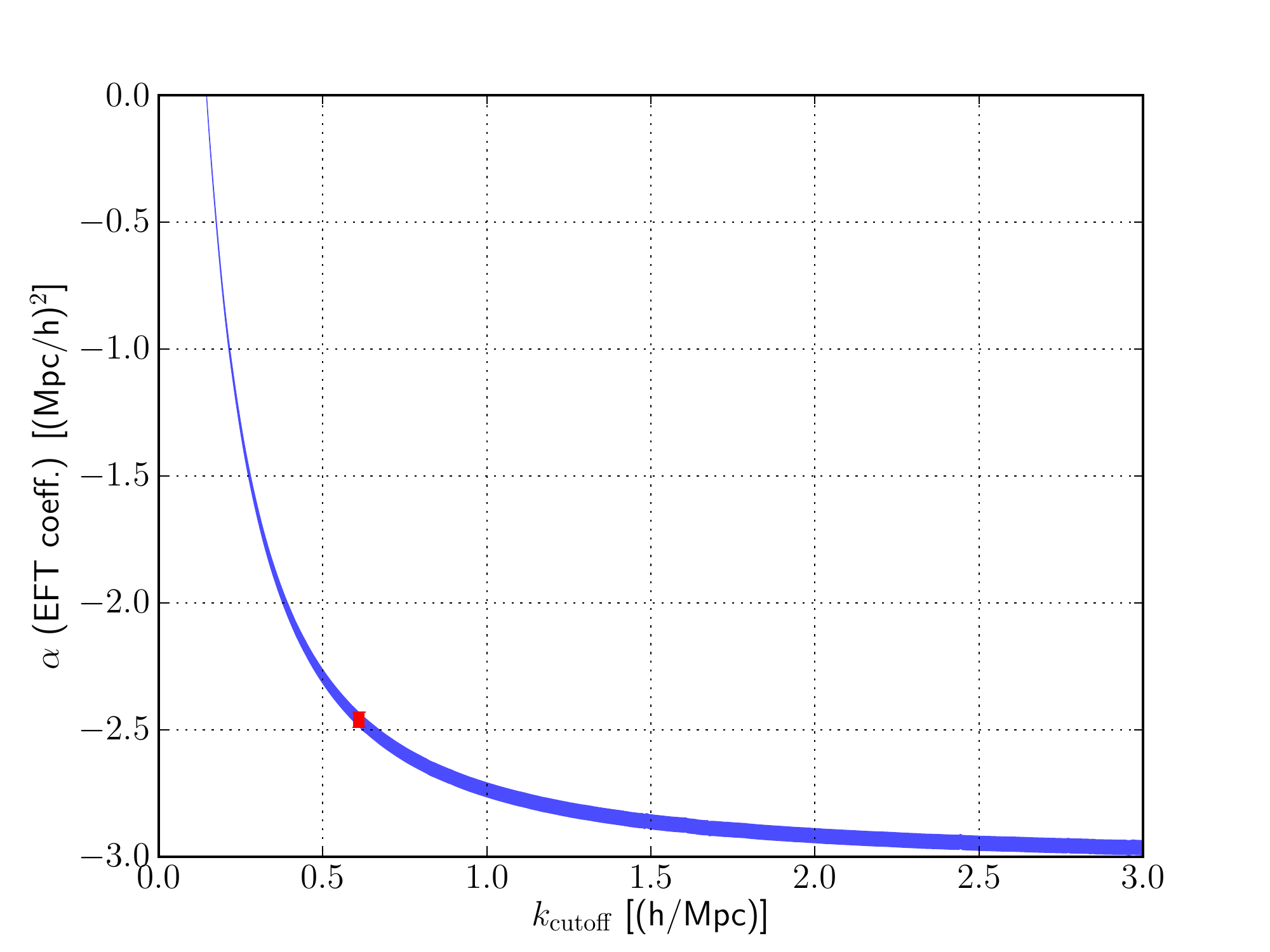}
\includegraphics[width=0.49\textwidth]{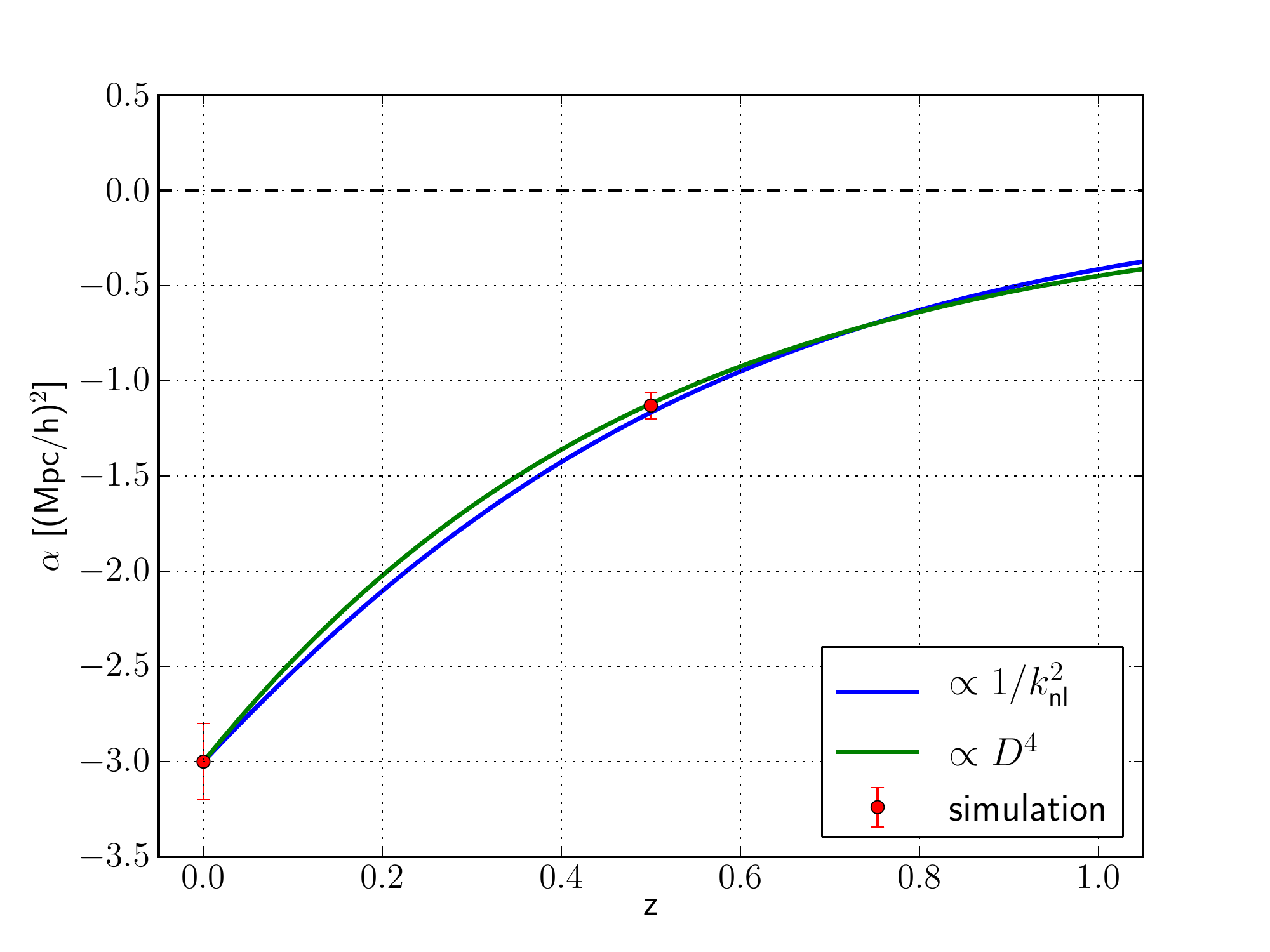}
\caption{\emph{Left panel: }Cutoff-dependence of the EFT coefficient $\alpha$. The red data point is measured from the L simulations at $z=0$. The blue band corresponds to the fitted value plus the correction from Eq.~\eqref{eq:alphascaling}. The width of the line is given by the errorbar on the data point. \emph{Right panel: }Measured value of $\alpha$, as a function of redshift. The data points have been extrapolated to $k_\text{cutoff}=\infty$.
The blue line corresponds to the na\"{\i}ve scaling $\propto 1/k_\text{nl}^2$ which we would expect for a scaling universe. Here $k_\text{nl}$ is defined by $k_\text{nl}^3P(k_\text{nl},z)/(2\pi^2)=1$.}
\label{fig:alpha_cutoff_dpdce}
\label{fig:alpha_z_dpdce}
\end{figure}

\paragraph*{Cutoff-dependence}
Loops are integrals over all momenta, and therefore introduce a mistake due to the fact that the high momenta cannot be described correctly by perturbation theory. The role of the EFT coefficient is to compensate this mistake, and thus the exact value of $\alpha$ depends on the high-$k$ cutoff used in the perturbation theory:
\beq
\alpha(\Lambda_1)=\alpha(\Lambda_2)+\frac{8}{63}\int_{\Lambda_1}^{\Lambda_2} \frac{\derd p}{2\pi^2}P_\text{lin}(p) \; .
\label{eq:alphascaling}
\eeq
This cutoff-dependence is shown in Fig.~\ref{fig:alpha_cutoff_dpdce}.
In particular, sending the cutoff to infinity in the loop integral results in a non-zero value $\alpha = -3.0 \pm 0.2\ (\text{Mpc}/h)^2$. The coefficient $\alpha$ can be set to zero by choosing $k_\text{cutoff} \simeq 0.15 h/$Mpc, but this has nothing physical, and does not imply that the EFT counterterm was not needed.

\paragraph*{Redshift-dependence}

For an EdS universe with power law initial power spectrum of index $n$, the relation between length and time scalings is $\lambda_x = \lambda_\tau ^{\frac{4}{n+3}}$ \cite{Pajer:2013jj}. We expect the EFT coefficient $\alpha$ to scale as its dimension, i.e. length$^2$, i.e. as $\tau^\frac{8}{n+3} \propto a^\frac{4}{n+3}$. 
Measurements of $\alpha_{\Lambda = \infty}$ at different redshifts are shown in Fig.~\ref{fig:alpha_z_dpdce}. The time-dependence of $\alpha_{\Lambda = \infty}$ appears compatible with that of $\text{length}^2 \propto 1/k_\text{nl}^2(a)$, and with $D^{\frac{4}{n+3}}$ with $n=-2$.
%

%%%%%%%%%%%%%%%%%%%%%%%%%%%%%%%%%%%%%%%%%%%%%%%%%%%%%%%%%
\subsection{One-loop power spectrum}

\begin{figure}[t]
\centering
\includegraphics[width=0.49\textwidth]{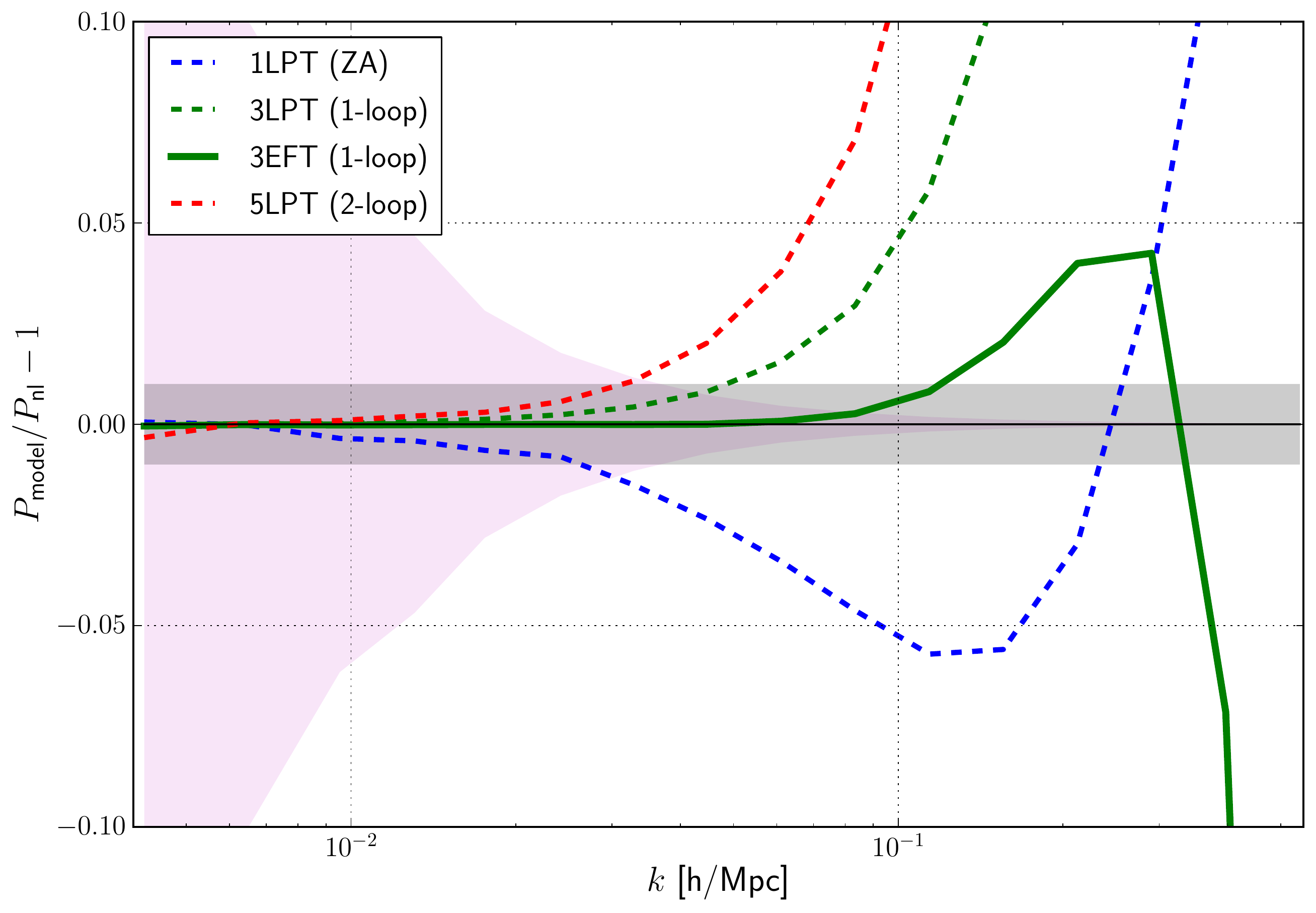}
\includegraphics[width=0.49\textwidth]{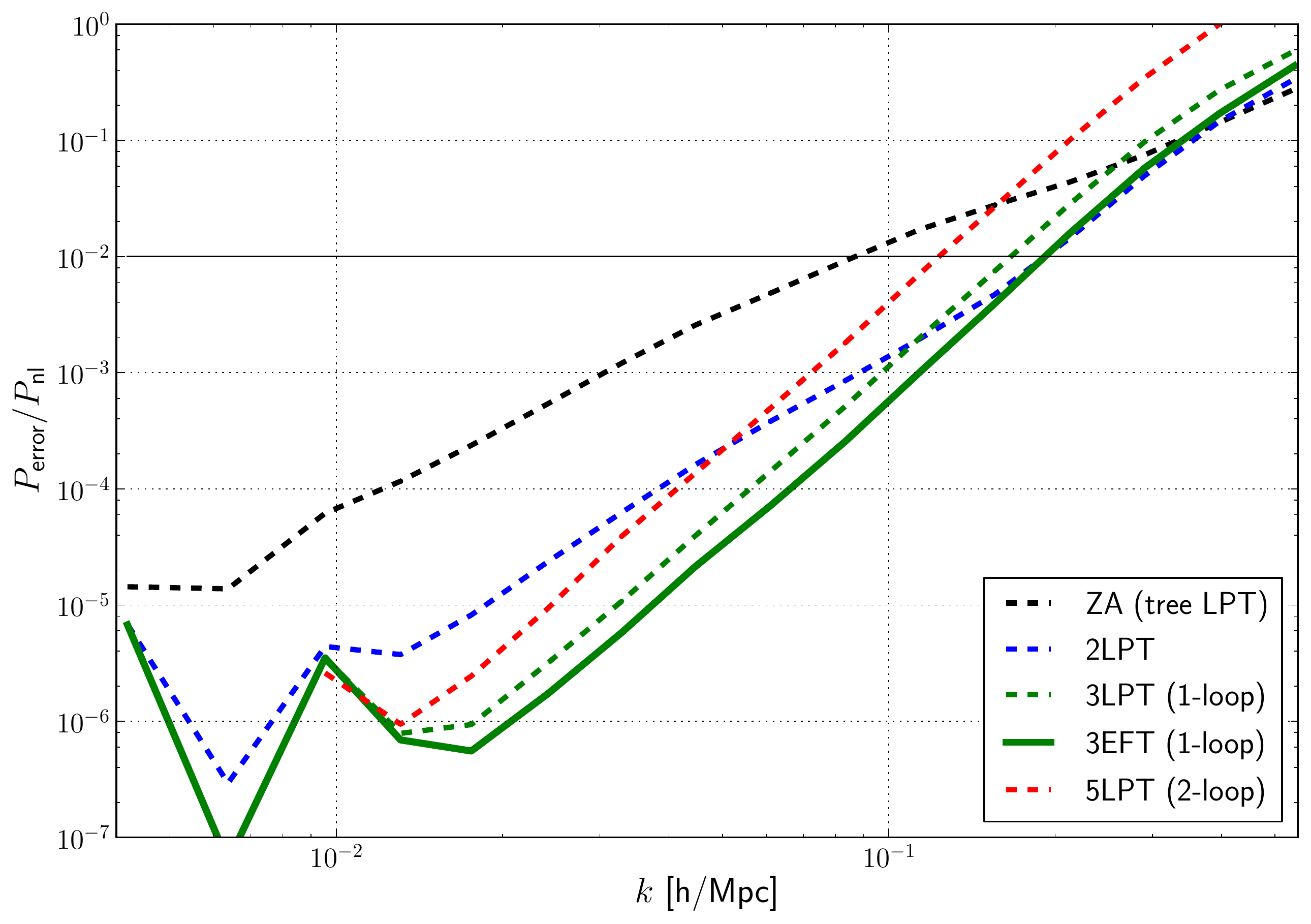}
\caption{\emph{Left panel:} Relative difference between the non-linear power spectrum from the simulation and from LPT (Zel'dovich approximation, 1 and 2-loop) and EFT (1-loop). The 1\% error domain is the shaded grey band. The maximum wave vector with accuracy of 1\% is improved by a factor of three from 1-loop LPT to 1-loop EFT, from $0.05\ h/\text{Mpc}$ to $0.15\ h/\text{Mpc}$. The 2-loop LPT worsens the agreement to simulation, compared to 1-loop LPT. The shaded magenta region indicates the scatter we would get due to cosmic variance without the LPT calculation on the simulation grid: this measurement has negligible cosmic variance.
\emph{Right panel:} Power spectrum of the error on the displacement field. Adding the second and third order to the first order displacement improves the agreement at the level of the displacement field on large scales ($k \lesssim 0.1 \ihMpc$). Including the EFT counterterm at 1-loop further improves the agreement, by correcting the UV mistake in $\phi_3$. However, going up to fifth order in LPT worsens the agreement, as expected for an asymptotic series, because of the UV mistake that is not corrected by EFT counterterms.}
\label{fig:pff_EFT_LPT_cv}
\end{figure}

% Improvement of EFT compared to LPT
We compare the non-linear power spectrum from LPT and EFT to the simulation output. Fig.~\ref{fig:pff_EFT_LPT_cv} shows that the 1-loop EFT power spectrum matches the simulation to 1\% accuracy up to $k=0.15 \ihMpc$, compared to $k=0.05 \ihMpc$ for the 1-loop LPT power spectrum. This is a factor of three improvement in the maximum wave vector giving 1\% accuracy. 
Let us stress again that the EFT coefficient $\alpha$ was not fit at the maximum wavenumber of validity $k=0.15 \ihMpc$ but on larger scales.
Notice that the 2-loop LPT power spectrum is worse than the 1-loop LPT power spectrum, which is not surprising given that it contains UV-sensitive terms that haven't been corrected by the appropriate EFT counterterms yet.

 % Agreement beyond cosmic variance
It is striking from Fig.~\ref{fig:pff_EFT_LPT_tLPT_cv} how good the agreement is at low $k$, completely devoid of cosmic variance. This shows that the EFT model works realization by realization, instead of only reproducing the mean power spectrum.

% Caveat
Thus, the third order EFT displacement provides a very good fit to the non-linear displacement on weakly non-linear scales, where the expansion in powers of $(k/k_\text{nl})$ is valid. However, one should not evaluate it at high $k$. In particular, computing the root mean square displacement from the EFT model yields wrong values, because this calculation relies on the displacement at high $k$ for which the EFT term $\propto k^2$ diverges.

%%%%%%%%%%%%%%%%%%%%%%%%%%%%%%%%%%%%%%%%%%%%%%%%%%%%%%%%%%%%%%%
\subsection{Relative importance of the various EFT terms}
%%%%%%%%%%%%%%%%%%%%%%%%%%%%%%%%%%%%%%%%%%%%%%%%%%%%%%%%%%%%%%%

As Fig.~\ref{fig:pff_EFT_LPT_cv} shows, the EFT provides a good fit not only to the non-linear power spectrum, but also to the displacement field itself.
However, in the case of the EFT power spectrum, the contribution from $\phi^{(2)}$ (i.e. the term $P_{22}$) is negligible compared to the contribution from $\phi^{(3)}$ (i.e. the term $P_{13}$).
One might therefore wonder about the relative importance of the non-linear terms $\phi^{(2)}$, $\phi^{(3)}$, $\alpha k^2 \phi^{(1)}$ present in the EFT model: do they contribute equally? Is the second order displacement $\phi^{(2)}$ helping at all in the agreement with simulation?

\begin{figure}[t]
\centering
\includegraphics[width=0.59\textwidth]{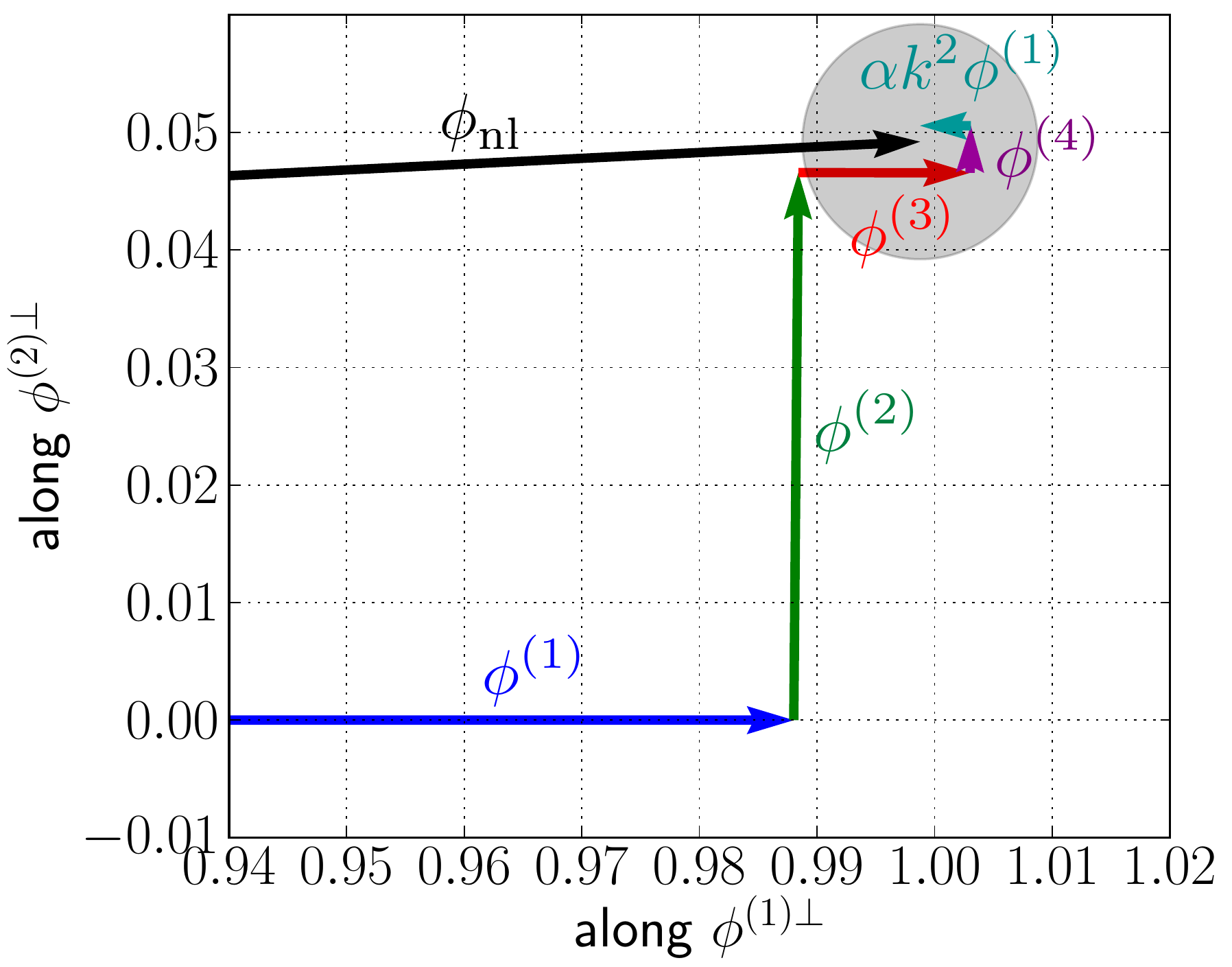}
\caption{Vector space representation of the LPT terms $\phi^{(1)}$ (blue), $\phi^{(2)}$ (green), $\phi^{(3)}$ (red), $\phi^{(4)}$ (magenta), the EFT term 
$\alpha k^2 \phi^{(1)}$ (cyan) and the non-linear displacement $\phi_\text{nl}$ (black), at $k=0.045$ h/Mpc. 
The shaded grey sphere corresponds to a displacement error such that $|| \phi_\text{error} || = 1\% || \phi_\text{nl} ||$
(i.e. $\sqrt{P_\text{error}} = 1\% \sqrt{P_\text{nl}}$). 
This representation shows that $\phi^{(2)}$ is indeed a bigger term than $\phi^{(3)}$, and contributes more than $\phi^{(3)}$ to reducing $\phi_\text{error}$. 
It also shows that the effect of $\phi^{(2)}$ on the non-linear power spectrum is smaller than that of $\phi^{(3)}$, because $\phi^{(3)}$ is more ``aligned'' with $\phi^{(1)}$ than $\phi^{(2)}$ is.
}
\label{fig:arrow_plot}
\end{figure}

The answer to these questions can be visualized as follows. The displacement fields $\phi_\text{nl}$, $\phi^{(1)}$, $\phi^{(2)}$, $\phi^{(3)}$ are functions of the wave vector $\vec k$, i.e., they are defined for each of the $N_\text{modes}$ modes in our simulation box. They can thus be understood as very high dimensional vectors $\left( \phi(\vec k_i) \right)_{i=1,\ldots,N_\text{modes}}$. We can then interpret $\langle \phi_a |\phi_b \rangle \equiv \langle \phi_a^{\star} \phi_b \rangle$ as a scalar product and $\langle |\phi_a|^2 \rangle$ as the corresponding squared norm on this vector space. Intuitively, with this scalar product, two displacement fields are aligned if they are perfectly correlated, and orthogonal if they are completely uncorrelated.
This allows a graphical representation of the displacement fields on the basis $\left(\phi^{(1)}, \phi^{(2)}, \phi^{(3)} \right)$ of the LPT terms. This basis is not orthogonal (e.g. $\langle \phi^{(1)} | \phi^{(3)} \rangle \neq 0$), so we shall instead use the orthonormal basis $\left( \phi^{(1)\perp}, \phi^{(2)\perp}, \phi^{(3)\perp} \right)$, deduced from $\left( \phi^{(1)}, \phi^{(2)}, \phi^{(3)} \right)$ through the Gram-Schmidt orthogonalization process.
Fig.~\ref{fig:arrow_plot} shows the graphical representation of the EFT terms as well as the non-linear displacement.
Fig.~\ref{fig:arrow_plot} makes it visible that the contribution of $\phi^{(2)}$ to reducing the error $\phi_\text{error}$ is more important than that of $\phi^{(3)}$, as one would expect for a well-behaved expansion. It also shows that even though $\phi^{(3)}$ is a smaller term than $\phi^{(2)}$ (i.e. $||  \phi^{(3)} || = \sqrt{P_{33}} < \sqrt{P_{22}} = || \phi^{(2)} ||$), it brings a larger contribution to the non-linear power spectrum, because it is more ``aligned'' with $\phi^{(1)}$ (i.e. $\langle \phi^{(1)} \phi^{(3)} \rangle = P_{13} > 0 = \langle\phi^{(1)} \phi^{(2)} \rangle$).
In conclusion, the second order displacement $\phi^{(2)}$ is a larger term than the third order displacement $\phi^{(3)}$, and is crucial to the agreement between EFT and simulation at the level of the displacement field. However, because $\phi^{(2)}$ is orthogonal to (i.e. not correlated with) the larger term $\phi^{(1)}$, its contribution to the displacement power spectrum is negligible.

%%%%%%%%%%%%%%%%%%%%%%%%%%%%%%%%%%%%%%%%%%%%%%%%%%%%%%%%%%%%%%%
\subsection{On the overfitting issue}
%%%%%%%%%%%%%%%%%%%%%%%%%%%%%%%%%%%%%%%%%%%%%%%%%%%%%%%%%%%%%%%

In this section, we fitted for the EFT coefficient $\alpha$, and compared various models of the displacement field to simulation, at the level of the field itself, and not only its power spectrum. This methods allows us to address the issue of overfitting.

\begin{figure}[t]
\centering
\includegraphics[width=0.59\textwidth]{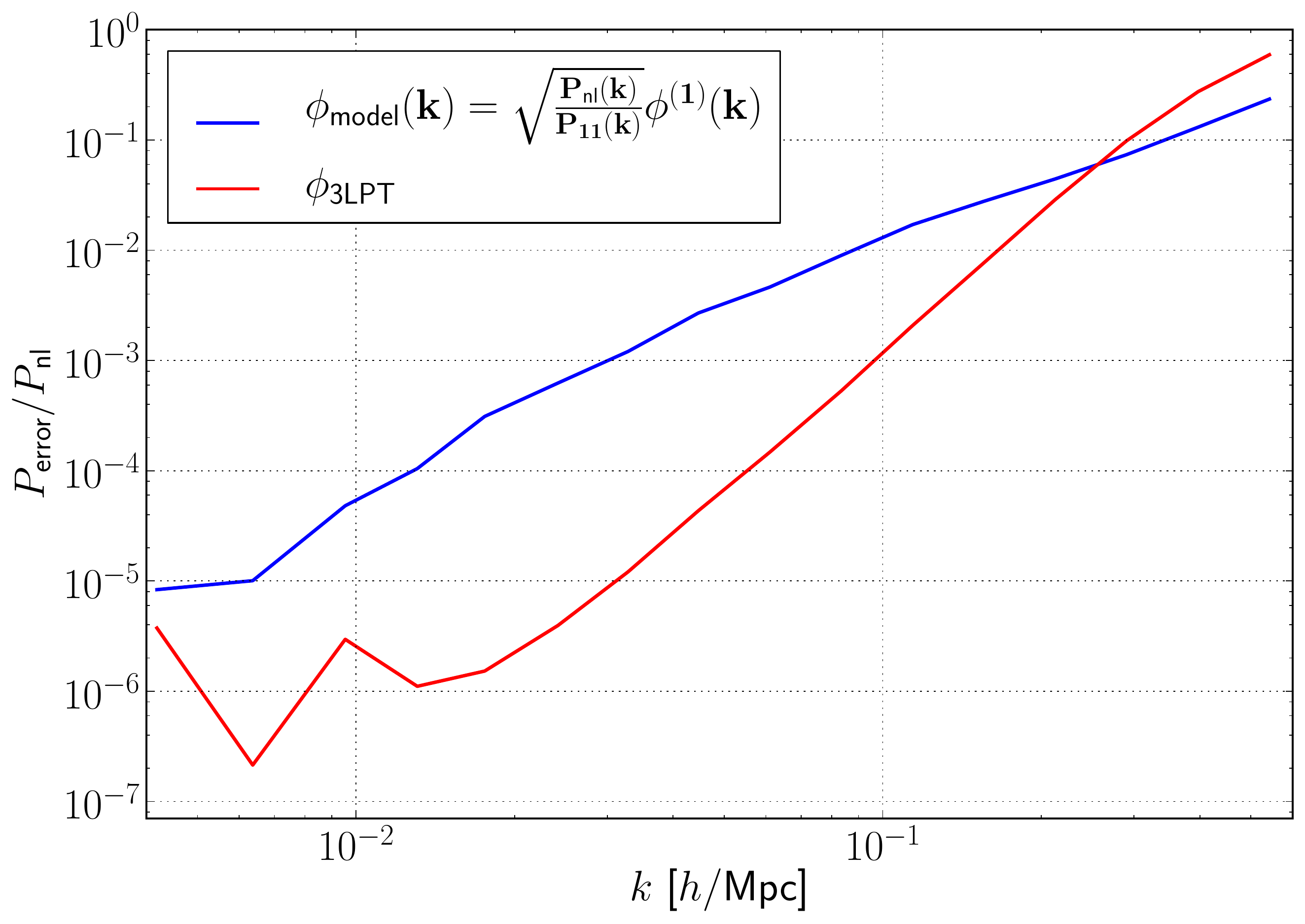}
\caption{Comparison between the error power spectra for the artificial model $\phi_\text{model}(\vec k) = \sqrt{\frac{P_\text{nl}(k)}{P_{11}(k)}}\phi^{(1)}(\vk)$ and $\phi_\text{3LPT}$. Even though the former gives exactly the right non-linear power spectrum, its error power spectrum is much larger, meaning that it is a wrong model for the displacement.}
\label{fig:overfit}
\end{figure}

Indeed, one might \textit{a priori} be concerned that the  EFT expansion for the displacement field might be incorrect, but still give the right displacement power spectrum due to the adjustable EFT parameters.
For example, the (artificial) model $\phi_\text{model}(\vec k) = \sqrt{\frac{P_\text{nl}(k)}{P_{11}(k)}}\phi^{(1)}(\vk)$ clearly predicts the correct power spectrum $P_\text{nl}$ for all $k$, but as shown in Fig~\ref{fig:overfit}, it does not correspond to the true non-linear displacement $\phi_\text{nl}$. 
One might worry that the EFT model might be similar, in that it would give an accurate non-linear power spectrum thanks to its free parameters, but be a poor description of the displacement field.
However we did not choose the EFT coefficient $\alpha$ by requiring $P_\text{model}$ to be close to $P_\text{nl}$, but instead by minimizing the error power spectrum $P_\text{error}$, which is the most stringent requirement.
Indeed, a small $P_\text{error}(k)$ means that $| \phi_\text{error} (\vk)|^2 = | \phi_\text{nl}(\vk) - \phi_\text{model}(\vk) |^2$ is small, implying that the Fourier component of the scalar displacement is correctly described by the model.
In contrast, having $P_\text{nl}$ close to $P_\text{model}$ simply means that $ | \phi_\text{nl}(\vk) |^2 - | \phi_\text{model}(\vk) |^2$ is small, i.e. the Fourier components have the same amplitude, but might have different phases.
Therefore, by minimizing $P_\text{error}$, we are guaranteed to choose the value of $\alpha$ that provides the best displacement field possible, which prevents the risk of overfitting.

In practice though, because our measurement of $\alpha$ is largely limited by systematic errors in our $N$-body simulation suite, the difference between the value of $\alpha$ obtained from minimizing the error power spectrum and from fitting to the non-linear power spectrum is small compared to the final uncertainty on $\alpha$. This was not \textit{a priori} obvious, and shows how sensitive this measurement is to systematic errors on the largest scales of the simulation.

%%%%%%%%%%%%%%%%%%%%%%%%%%%%%%%%%%%%%%%%%%%%%%%%%%%%%%%%%%%%%%%
\section{Transfer functions and `tLPT'}
\label{sec:tf}
%%%%%%%%%%%%%%%%%%%%%%%%%%%%%%%%%%%%%%%%%%%%%%%%%%%%%%%%%%%%%%%

%%%%%%%%%%%%%%%%%%%%%%%%%%%%%%%%%%%%%%%%%%%%%%%%%%%%%%%%%
\subsection{Optimal linear model from LPT -- including higher-loop terms}
%%%%%%%%%%%%%%%%%%%%%%%%%%%%%%%%%%%%%%%%%%%%%%%%%%%%%%%%%

In the previous section, we measured the EFT coefficient $\alpha$, and compared the EFT model to the various nLPT models. We found that the EFT significantly increases the range of validity of perturbation theory.
But we also wish to understand how close to optimal the EFT model is. To do so, we compare it to the ``ntLPT'' models (for ``LPT with transfer functions''), i.e. the LPT models for which the LPT displacements are multiplied by free functions of the modulus $k \equiv | \vec k |$ of the wavenumber (the ``transfer functions'', see also \cite{Tassev:2012cq}):
\beq
\phi_{n\text{tLPT}} (\vk) = a_1(k)\phi^{(1)}(\vk) + ... + a_n(k)\phi^{(n)}(\vk).
\eeq
For each of these models (LPT, EFT, tLPT), we compute the displacement error $\phi_\text{error}\equiv \phi_\text{nl} - \phi_\text{model}$. 
The transfer functions $a_i(k)$ of the tLPT models are chosen so as to minimize the error power spectrum $P_\text{error}$ for each $k$-bin. 
Notice that the transfer functions are not chosen so as to match the non-linear power spectrum. The tLPT models correspond to a lower bound on $P_\text{error}$, and thus allow to assess how close to optimal the LPT and EFT models are. 

Besides, the transfer functions allow to effectively include certain higher order LPT and EFT terms without having to compute them explicitly. For example, for the 1tLPT model $\phi_{1\text{tLPT}} = a_1 \phi^{(1)}$, the result of minimizing $P_\text{error}$ yields $a_1 = \frac{P_{1\times\text{nl}}}{P_{11}}$, so that $P_{1\text{tLPT}} = \frac{P_{1\times\text{nl}}^2}{P_{11}} \simeq P_{11} + P_{13} + \alpha k^2 P_{11} + ...$ . 
Here, the term $P_{13}$ is implicitly included in the 1tLPT power spectrum, even though evaluating the 1tLPT displacement did not require computing $\phi^{(3)}$.

\begin{figure}[t]
\centering
\includegraphics[width=0.59\textwidth]{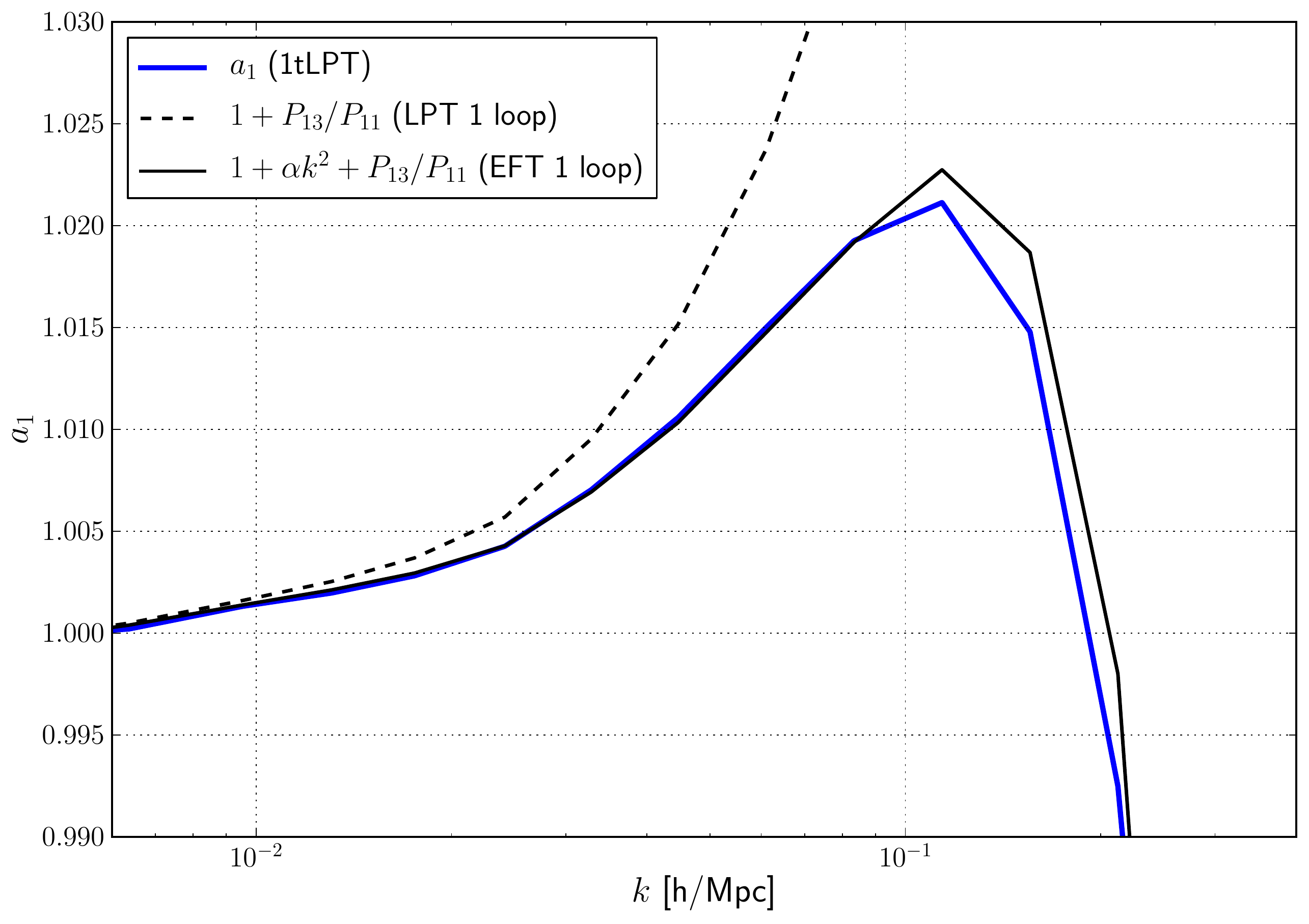}
\caption{Transfer function $a_1$ of the 1tLPT model. It corresponds to $a_1 = \frac{P_{1\times\text{nl}}}{P_{11}} \simeq 1 + \alpha k^2 +  \frac{P_{13}}{P_{11}}$, which indicates that the 1tLPT power spectrum effectively includes the $P_{13}$ term, even though $\phi^{(3)}$ is not explicitly included in the 1tLPT model.}
\label{fig:a1_perp}
\end{figure}

In turn, the LPT and EFT can predict what these transfer functions should be, as shown in Fig.~\ref{fig:a1_perp}. We see that the 3LPT model for $a_1$ overpredicts its amplitude due to the UV sensitivity of $P_{13}$, whereas the 3EFT prediction, taking into account the counterterm associated with $\alpha$, is accurate to $1\%$ even beyond $k=0.1 \ihMpc$. 

This also generalizes to higher order: as soon as the term $a_i \phi^{(i)}$ is included, the LPT contributions $P_{ij}$ to the non-linear power spectrum for all $j$ are automatically included. This is illustrated in Tab.~\ref{tab:tf_orders_loop}.
\begin{table}[H]
\centering
\begin{tabular}{c|c|c|c|c|c}
 &$a_1$&$a_2$&$a_3$&$a_4$&$a_5$\\
\hline
tree &11&&&&\\
1-loop &13&22&&&\\
2-loop &15&24&33&&\\
3-loop &17&26&35&44&\\
4-loop &19&28&37&46&55\\
\end{tabular}
\caption{List of contributions to the LPT power spectrum at tree order and various loop orders. The columns collect the terms that are effectively included by the transfer function ($a_1$, $a_2$, $a_3$, $a_4$ and $a_5$ respectively). For example, this shows that 3tLPT contains all of the 2-loop terms, including $P_{15}$ and $P_{24}$, even though $\phi_4$ and $\phi_5$ are not computed explicitly in $\phi_{3\text{tLPT}}$.}
\label{tab:tf_orders_loop}
\end{table}

In practice, we fit for the free coefficients $a_i(k)$ for each $k$-bin independently, by minimizing $P_\text{error}(k) = \langle |\phi_\text{error}|^2 \rangle = \langle \left| \phi_\text{nl} - \sum_i a_i \phi_i \right|^2 \rangle$.
If we interpret $\langle \phi_a |\phi_b \rangle \equiv \langle \phi_a^{\star} \phi_b \rangle$ as a scalar product and $\langle |\phi_a|^2 \rangle$ as a squared norm as before, we see that minimizing the norm of the displacement error $\phi_\text{error}$ (i.e. minimizing $P_\text{error}$) amounts to making the displacement error orthogonal to the LPT components $\phi_i$ (i.e. $\langle \phi_\text{error}| \phi^{(i)} \rangle = 0$):
\begin{align}
\chi^2=\langle \big| \phi_\text{nl} - \sum_i a_i \phi^{(i)} \bigr|^2\rangle 
\text{minimal}
\Leftrightarrow \frac{\partial \chi^2}{\partial a_i}=
-2\langle \phi^{(i)} | \phi_\text{nl} - \sum_i a_i \phi^{(i)} \rangle=-2\langle \phi^{(i)} | \phi_\text{error}\rangle=0.
\end{align}

Since the various $\phi^{(i)}$ do not form an orthogonal basis (i.e. $\langle \phi^{(i)} |\phi^{(j)} \rangle \neq 0$, i.e. $\phi^{(i)}$ and $\phi^{(j)}$ can be correlated), the value of the transfer function depends on the maximum order included in tLPT: for instance, the value of $a_1$ in 1tLPT and 3tLPT differs by $P_{13} / P_{11}$.
This can be avoided by defining an orthogonal basis $\phi^{(i)\perp}$ through the Gram-Schmidt orthogonalization process, applied to the LPT displacements $\phi^{(i)}$:
\beq
\left\{
\bal
&\phi^{(1)\perp} = \phi^{(1)} \\
&\phi^{(i+1)\perp} = \phi^{(i+1)} 
- \sum_{j \leq i} \frac{\langle \phi^{(i+1)}| \phi^{(j)\perp} \rangle}{\langle \phi^{(j)\perp} | \phi^{(j)\perp} \rangle} 
\phi^{(j)\perp} \\
\eal
\right.
\eeq
This way, we can define ``orthogonal'' transfer functions $a_i^\perp$, which are independent of the order in tLPT:
\beq
\phi_{n\text{tLPT}} = a_1^\perp \phi^{(1)\perp} +... + a_n^\perp \phi^{(n)\perp}
\eeq
Note that this does not affect the value of the tLPT displacement: it is a decomposition of the same tLPT model in the basis of the $\left( \phi^{(i)\perp} \right)$ instead of the $\left( \phi^{(i)} \right)$.
However, this expression will be useful in estimating the residuals $\phi_\text{nl} - \phi_{n\text{tLPT}}$, as we shall see later.

\begin{figure}[t]
\centering
\includegraphics[width=0.49\textwidth]{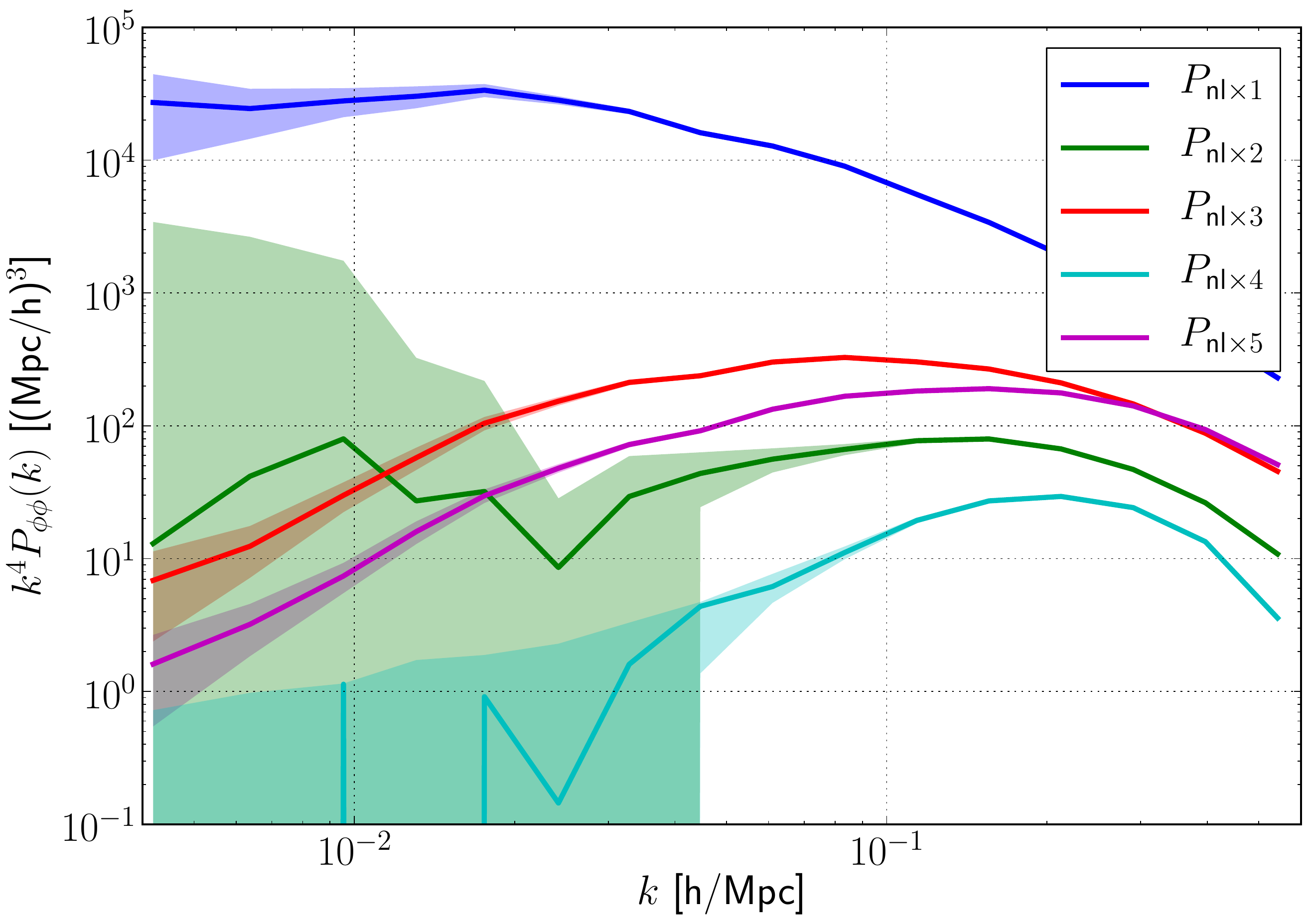}
\includegraphics[width=0.48\textwidth]{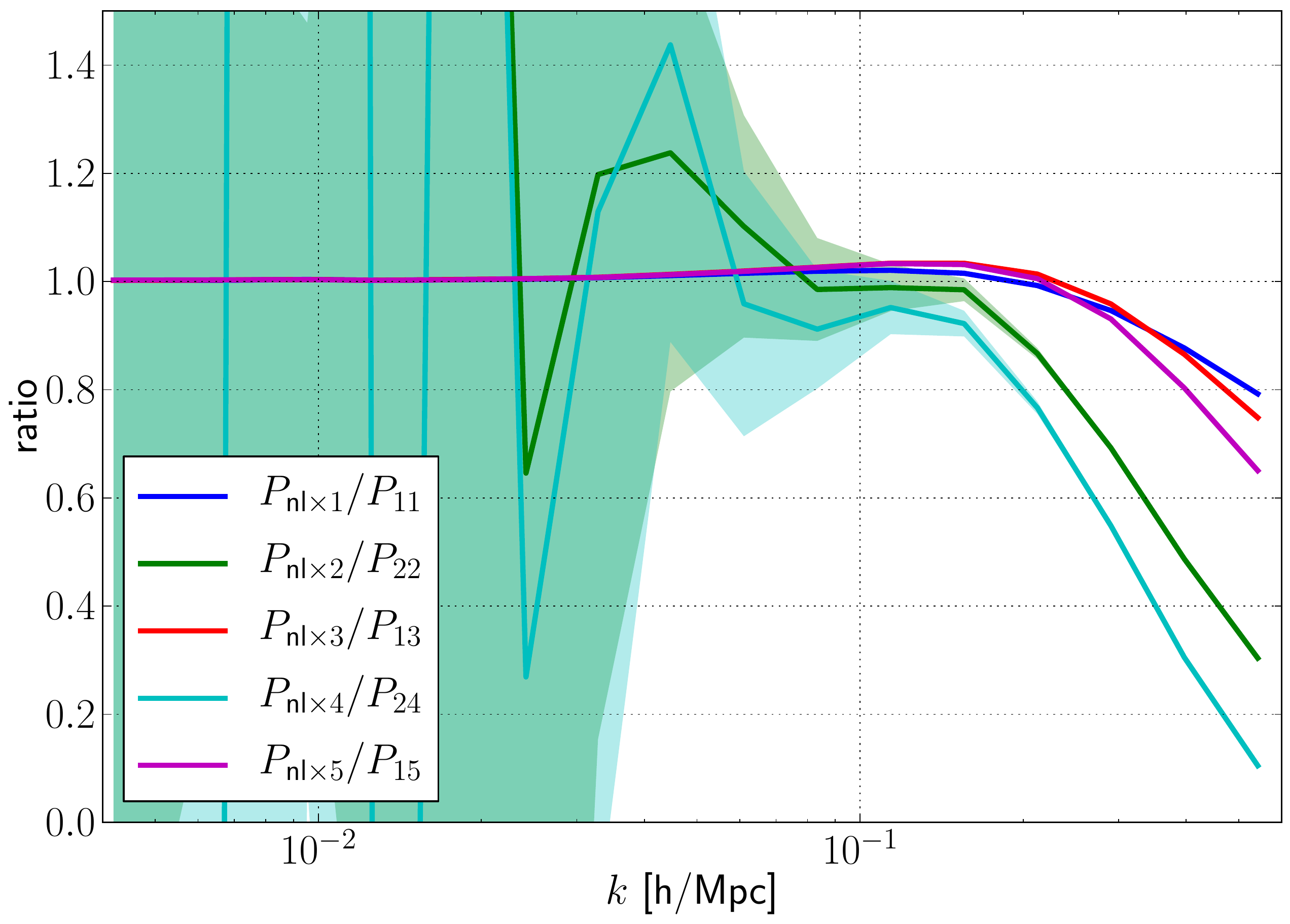}
\caption{\emph{Left panel:} cross-spectra between the LPT terms and the non-linear displacement needed to make comparisons at the field level, and to fit for the transfer functions (defined later). These were computed for one particular initial condition. The shaded area indicate the uncertainty due to cosmic variance. The cosmic variance is visibly much higher for $P_{\text{nl}\times i}$ with $i$ even, as explained in App.~\ref{sec:appendix_cosmic_variance}.
\emph{Right panel:} Ratios of the cross spectra from the left panel and the lowest order LPT prediction. The cosmic variance on the ratios is significantly reduced.}
\label{fig:simu_cross_pt}
\end{figure}

As we have seen, fitting for the transfer functions requires knowing the cross-spectra $\langle \phi^{(i)} |\phi^{(j)} \rangle$ between the various LPT terms, but also the cross-spectra $\langle \phi^{(i)} |\phi_\text{nl} \rangle$ between the LPT terms and the true non-linear displacement. For reference, we show the latter in Fig.~\ref{fig:simu_cross_pt}.

%>>>>>>>>>>>>>>>>>>>>>>>>>>>>>>>>>>>>>>>>>>>>>>>>>>>>>>>>>>>>>>>>>>>>>>>>>>>>>>>>>>>>>
%>>>>>>>>>>>>>>>>>>>>>>>>>>>>>>>>>>>>>>>>>>>>>>>>>>>>>>>>>>>>>>>>>>>>>>>>>>>>>>>>>>>>>

\begin{figure}[t]
\centering
\includegraphics[width=0.6\textwidth]{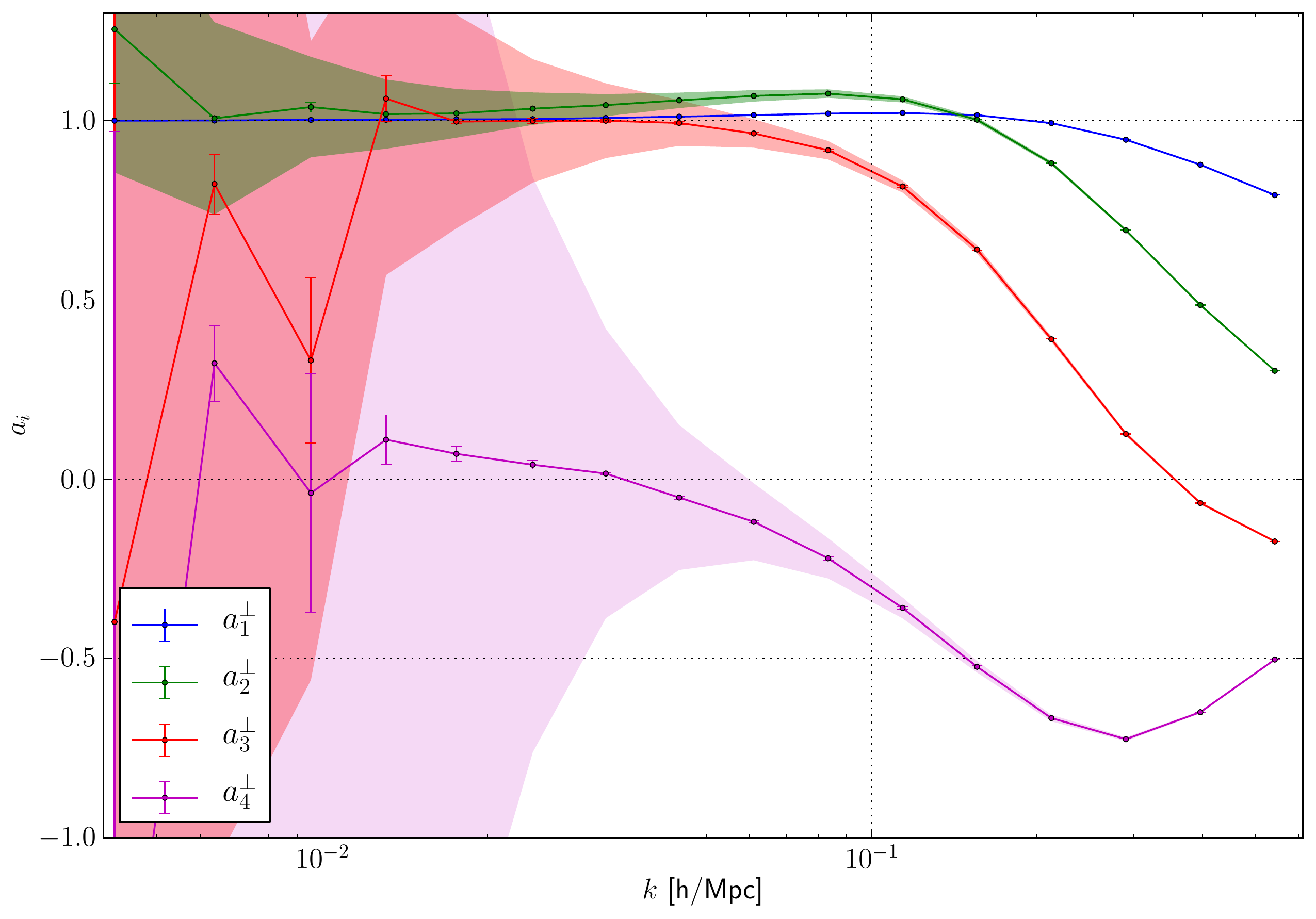}
\caption{Orthogonal transfer functions at redshift 0. The shaded area corresponds to the domains where varying $a_i$ changes the contribution to $\phi_\text{4tLPT}$ by less than $0.1\%$.}
\label{fig:transfer_functions_ortho}
\vspace{0.3cm}
\includegraphics[width=0.49\textwidth]{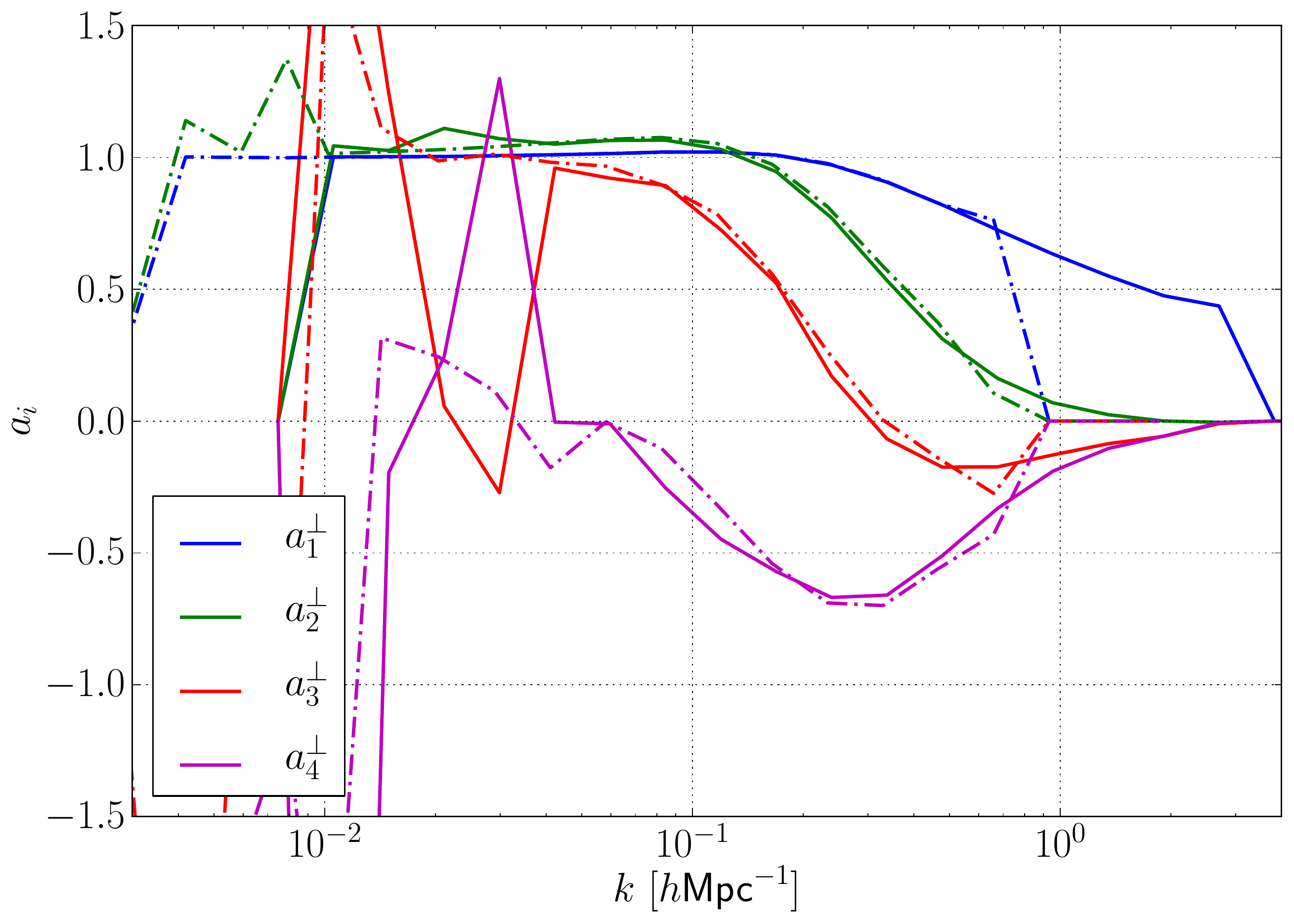}
\includegraphics[width=0.49\textwidth]{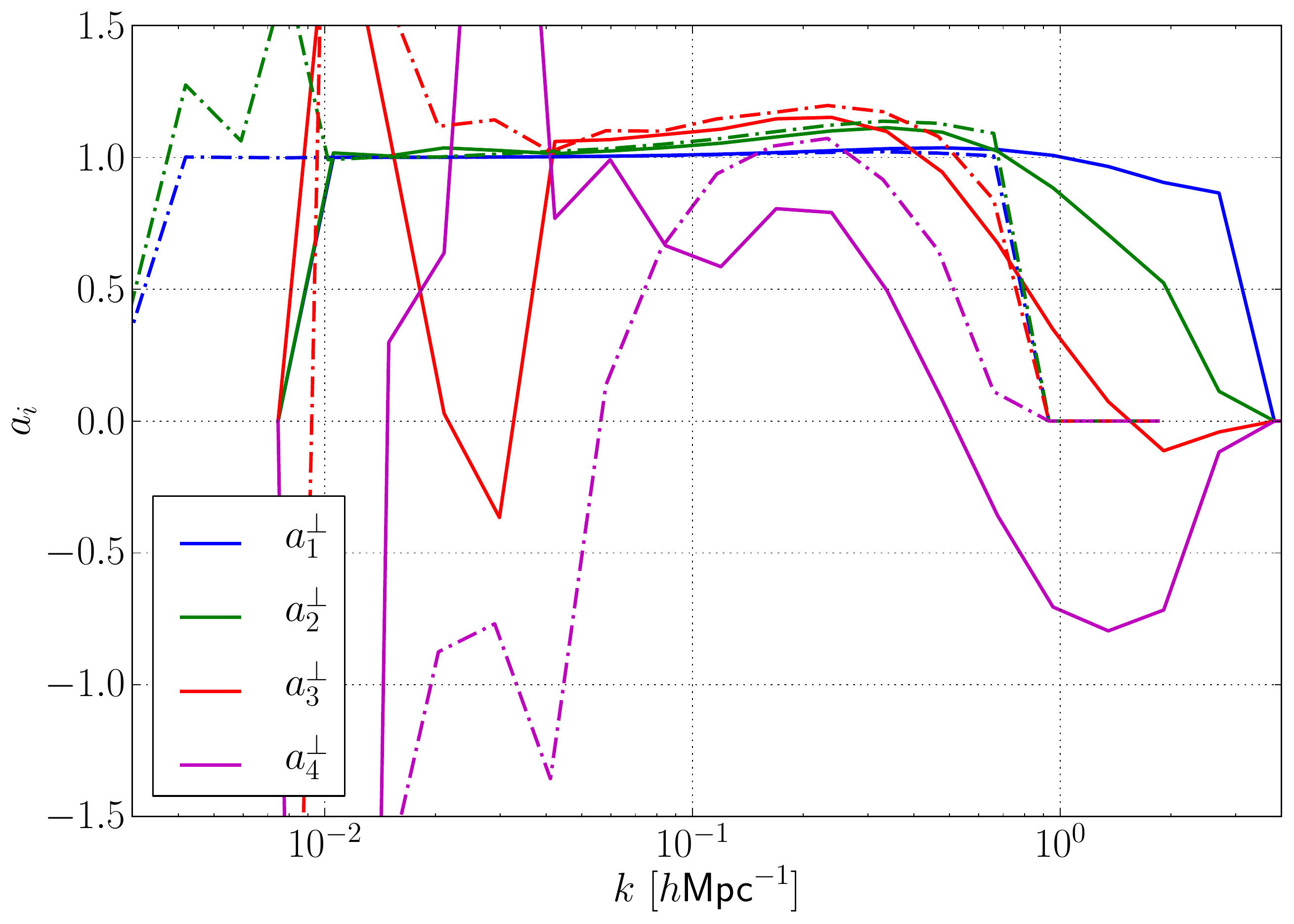}
\caption{Non-orthogonal transfer functions at redshift $z=0$ (left panel) and $z=2$ (right panel). 
The dashed lines are from the L simulation and the solid lines from the M simulation.
At redshift $z=0$ the transfer function $a_4$ of the fourth order displacement on large scales is close to zero, but it approaches unity in the M simulation at $z=2$. Such a behavior can be expected if the $k^0$ component of $P_{46}/P_{44}$ is of order unity. Note also that the transfer function $a_1$ on the Zel'dovich displacement remains close to unity up to non-linear wavenumbers.}
\label{fig:transfer_functions_4tlpt}
\end{figure}

The measured orthogonal transfer functions $a_1^\perp$, $a_2^\perp$, $a_3^\perp$ and $a_4^\perp$ at redshift $z=0$ are shown in Fig.~\ref{fig:transfer_functions_ortho}. 
The non-orthogonal transfer functions of 4tLPT at redshift $z=0$ and $z=2$ are shown in Fig.~\ref{fig:transfer_functions_4tlpt}. 
The measured transfer function $a_n$ or $a_n^\perp$ becomes increasingly sensitive to potential systematic uncertainties in the simulation as $n$ increases.

%>>>>>>>>>>>>>>>>>>>>>>>>>>>>>>>>>>>>>>>>>>>>>>>>>>>>>>>>>>>>>>>>>>>>>>>>>>>>>>>>>>>>>
%>>>>>>>>>>>>>>>>>>>>>>>>>>>>>>>>>>>>>>>>>>>>>>>>>>>>>>>>>>>>>>>>>>>>>>>>>>>>>>>>>>>>>

%%%%%%%%%%%%%%%%%%%%%%%%%%%%%%%%%%%%%%%%%%%%%%%%%%%%%%%%%
\subsection{Transfer functions in the context of the EFT}\label{sec:2loop}
%%%%%%%%%%%%%%%%%%%%%%%%%%%%%%%%%%%%%%%%%%%%%%%%%%%%%%%%%

We mentioned above that 3tLPT includes all the terms present in a consistent two-loop EFT calculation, including the next-to-leading order counterterms.
Here we make the connection between EFT and tLPT explicit, by showing that the transfer functions are correctly described by LPT power spectra and their EFT counterterms on large scales. For clarity, we focus on the orthogonal transfer functions $a_n^\perp$, which are independent on the order used in tLPT.

\begin{figure}[t]
\centering
\includegraphics[width=0.49\textwidth]{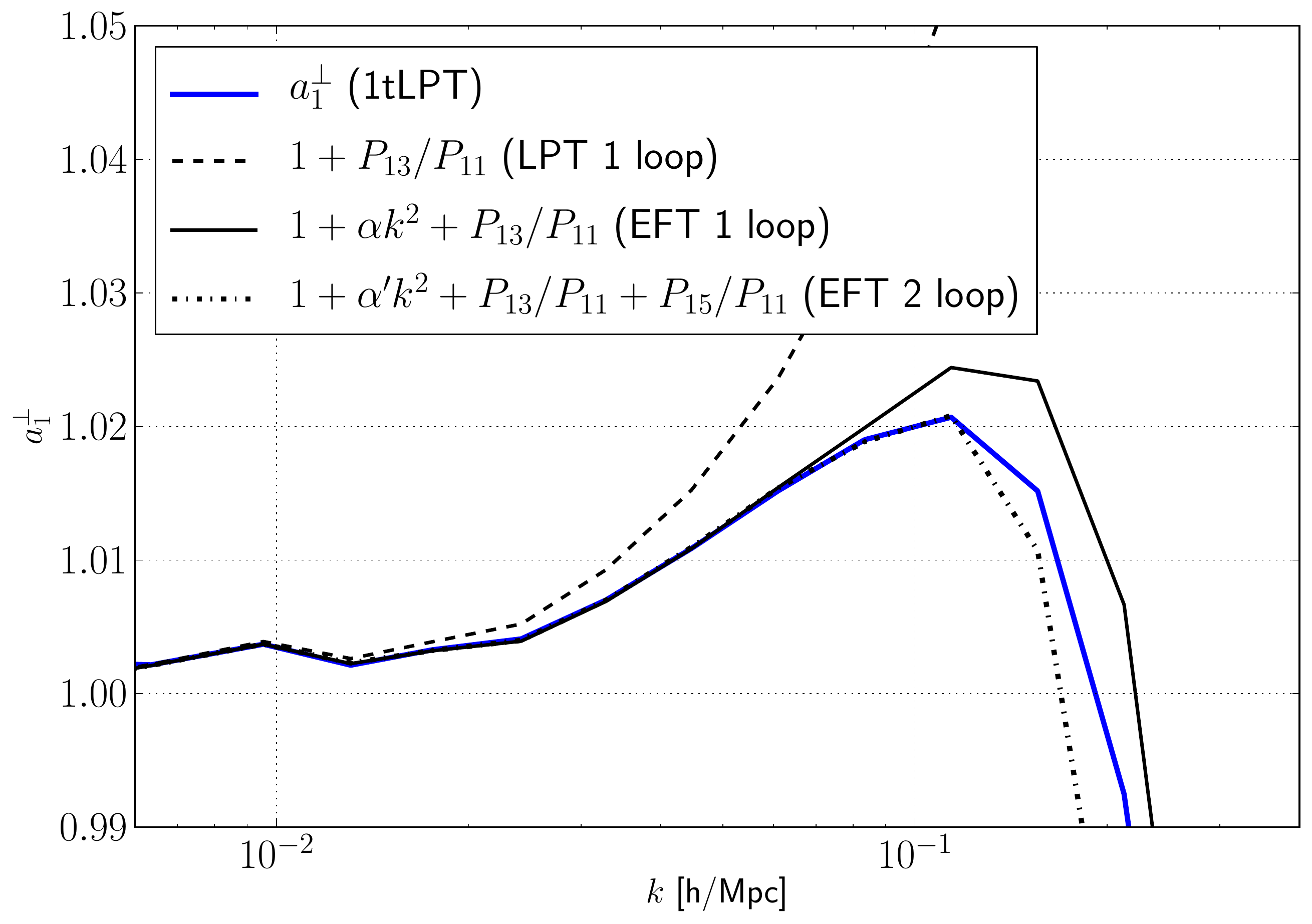}
\includegraphics[width=0.48\textwidth]{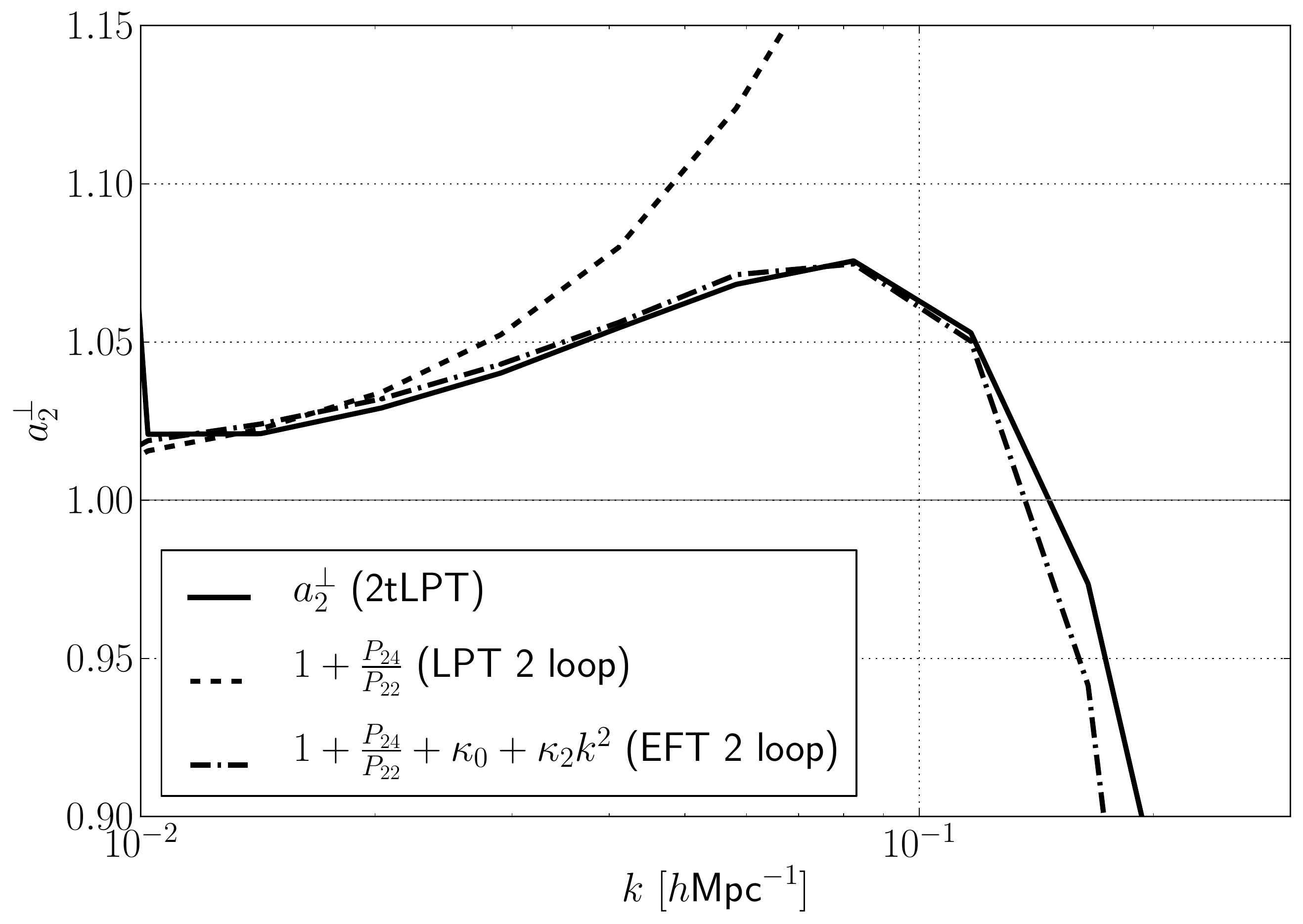}
\caption{\emph{Left panel: }Transfer function $a_1^\perp$ of the 1tLPT model. Adding $P_{15}$ modifies the value of $\alpha$, as expected, and improves the agreement. \emph{Right panel: }Transfer function $a_2^\perp$ of the 1tLPT model. We clearly see a percent level deviation on the largest scales, that is accounted for by adding $P_{24}/P_{22}$ to the model for this term. The latter however overpredicts the enhancement in the mildly non-linear regime, which is in turn fixed by the EFT counterterms $E_{2,i}$. As we pointed out before, they lead to $k^0$ and $k^2$ corrections through $P_{\tilde 2 2}/P_{22}$.}
\label{fig:a1_perp_2loop}
\end{figure}

The EFT predicts the first order orthogonal transfer function as follows:
\beq
\bal
a_1^\perp
\equiv  \frac{P_{1\times\text{nl}}}{P_{11}}
&\simeq 1+ \frac{P_{13}}{P_{11}}+\alpha k^2 \\
&\simeq 1+ \frac{P_{13}}{P_{11}}+  \frac{P_{15}}{P_{11}}+\alpha ' k^2 . \\
\eal
\eeq
Here, the first line corresponds to the 1-loop EFT prediction for $a_1^\perp$, where $\alpha$ is the EFT coefficient measured above. The second line corresponds to the 2-loop prediction (see Fig.~\ref{fig:a1_perp_2loop}). 
Since the LPT terms such as $P_{13}$ and $P_{15}$ are cutoff-dependent and have potentially wrong UV contributions, they are associated with a counterterm.
Thus $\alpha$ and $\alpha '$ differ by the $k^2$ coefficient of $ P_{15}/P_{11}$ which is approximately $3\, h^{-2}\text{Mpc}^2$ for a cutoff of $\Lambda=0.6 \ihMpc$. 
Beyond that, $P_{13}/P_{11}$ and $ P_{15}/P_{11}$ have $k^4$ contributions which are cutoff dependent, and should be corrected by a counterterm of the form $\beta k^4$. However, we found that including such a counterterm does not improve the prediction for $a_1^\perp$, and we therefore do not include it in the following discussion.
%
%>>>>>>>>>>>>>>>>>>>>>>>>>>>>>>>>>>>>>>>>>>>>>>>>>>>>>>>>>>>>>>>>>>>>>>>>>>>>>>>>>>>>>
%>>>>>>>>>>>>>>>>>>>>>>>>>>>>>>>>>>>>>>>>>>>>>>>>>>>>>>>>>>>>>>>>>>>>>>>>>>>>>>>>>>>>>

Similarly, the deviations of the transfer function for the second order displacement can be modeled by a combination of the next-to-leading order LPT term and the corresponding EFT counterterms
\beq
a_2^\perp=1 +  \frac{P_{24}}{P_{22}} +  \frac{P_{\tilde22}}{P_{22}}.
\eeq
Note that $P_{24}/P_{22}$ starts as $k^0$ (actually predicting deviations of 1\% on large scales at $z=0$). This means that the transfer function on $\phi_{2}$ can deviate from unity even on very large scales.
The cutoff dependence of $P_{24}$ needs to be captured by the corresponding counterterms -- 
$P_{\tilde22}/P_{22}$ is the sum of three next-to-leading order counterterms with their free coefficients. As discussed in Sec.~\ref{sec:nloctr},  they scale as $k^0$ and $k^2$ for small wavenumbers.  
We show the EFT description of the second order transfer function in Fig.~\ref{fig:a1_perp_2loop}. $P_{24}$ captures the large scale offset quite well, but overpredicts the scale dependence of the transfer function. This mistake can be corrected by adding the quadratic counterterms in the form of a $k^0$ term and a $k^2$ term with free coefficients. 

More generally, the transfer functions beyond the one on $\phi^{(1)}$ do not necessarily go to unity on large scales: they can be renormalized at the $k^0$ level by higher order LPT and EFT corrections. The reason for the distinction between $a_1$ and $a_{>1}$ is that the LPT kernels already ensure mass and momentum conservation in the coupling between initial modes and can thus be modified at the $k^0$ level.
For the linear field $\phi^{(1)}$, corrections need to explicitly conserve mass and momentum and thus need to start as $k^2 \phi^{(1)}$ \cite{Peebles:LSS,Mercolli:2013bsa}.

%%%%%%%%%%%%%%%%%%%%%%%%%%%%%%%%%%%%%%%%%%%%%%%%%%%%%%%%%
\subsection{Optimality of the EFT at 1-loop}
%%%%%%%%%%%%%%%%%%%%%%%%%%%%%%%%%%%%%%%%%%%%%%%%%%%%%%%%%

As explained earlier, we compare the EFT model to the tLPT models to assess how close to optimal the EFT model is.
In terms of the error power spectrum, which quantifies the agreement with simulation at the level of the displacement field, the 1-loop EFT and tLPT both have $P_\text{error} \leqslant 1\% P_\text{nl}$ up to $k \simeq 0.2 \ihMpc$. This shows that the EFT displacement is very close to the 3tLPT displacement, showing that the EFT model is close to optimal at 1-loop order: no expansion of the form $a_1(k) \phi^{(1)}(\vk) + a_2(k) \phi^{(2)}(\vk) + a_3(k) \phi^{(3)}(\vk)$ can significantly outperform it.
In terms of the non-linear power spectrum, the 1-loop EFT provides a 1\% fit to the power spectrum up to $k \gtrsim 0.1 \ihMpc$, compared to $k \simeq 0.2 \ihMpc$ for the 3tLPT model. One might be able to achieve this factor of two extension in the range over which the power spectrum can be described at the $1\%$ level using 2-loop EFT. 
\begin{figure}[t]
\centering
\includegraphics[width=0.49\textwidth]{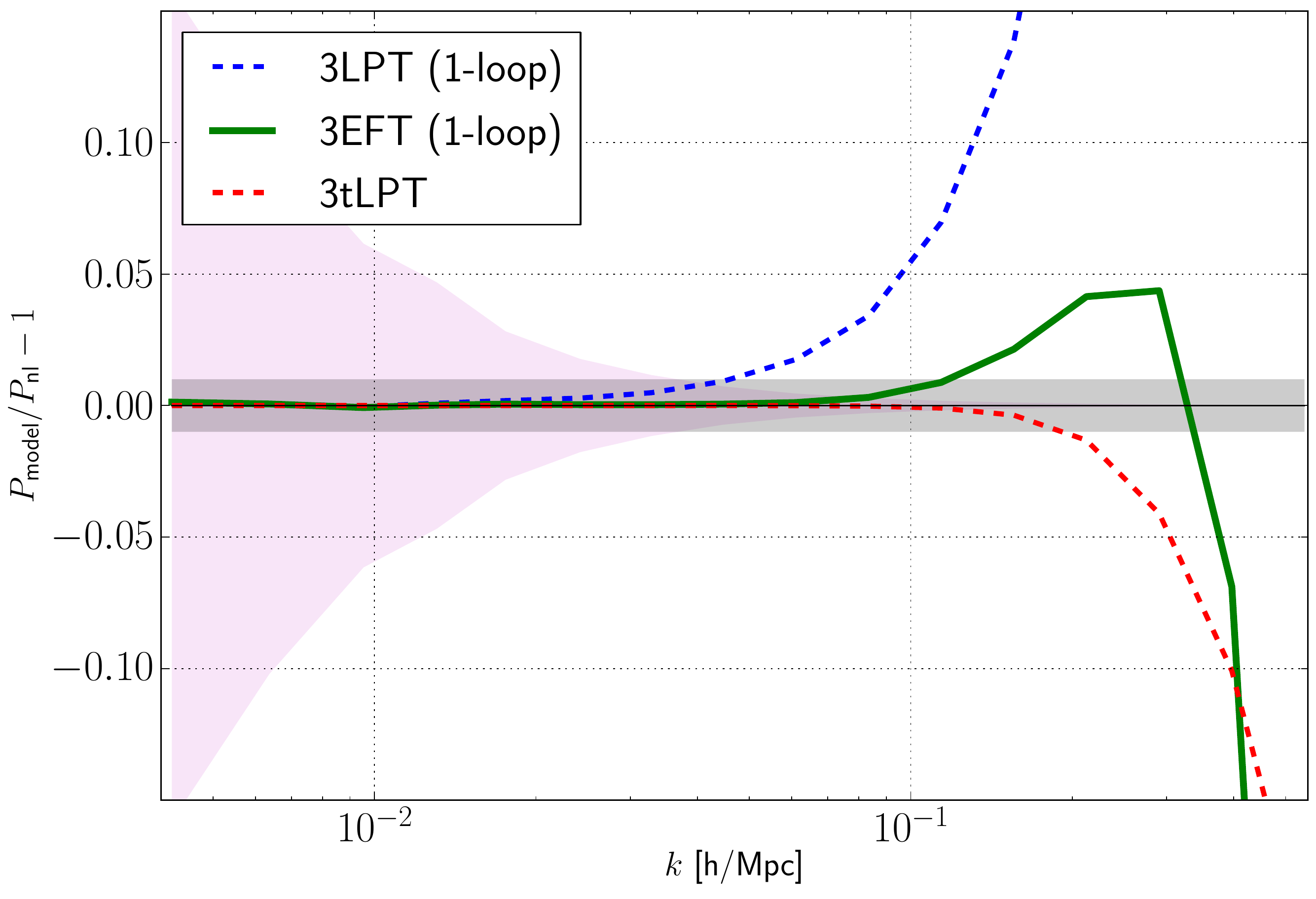}
\includegraphics[width=0.48\textwidth]{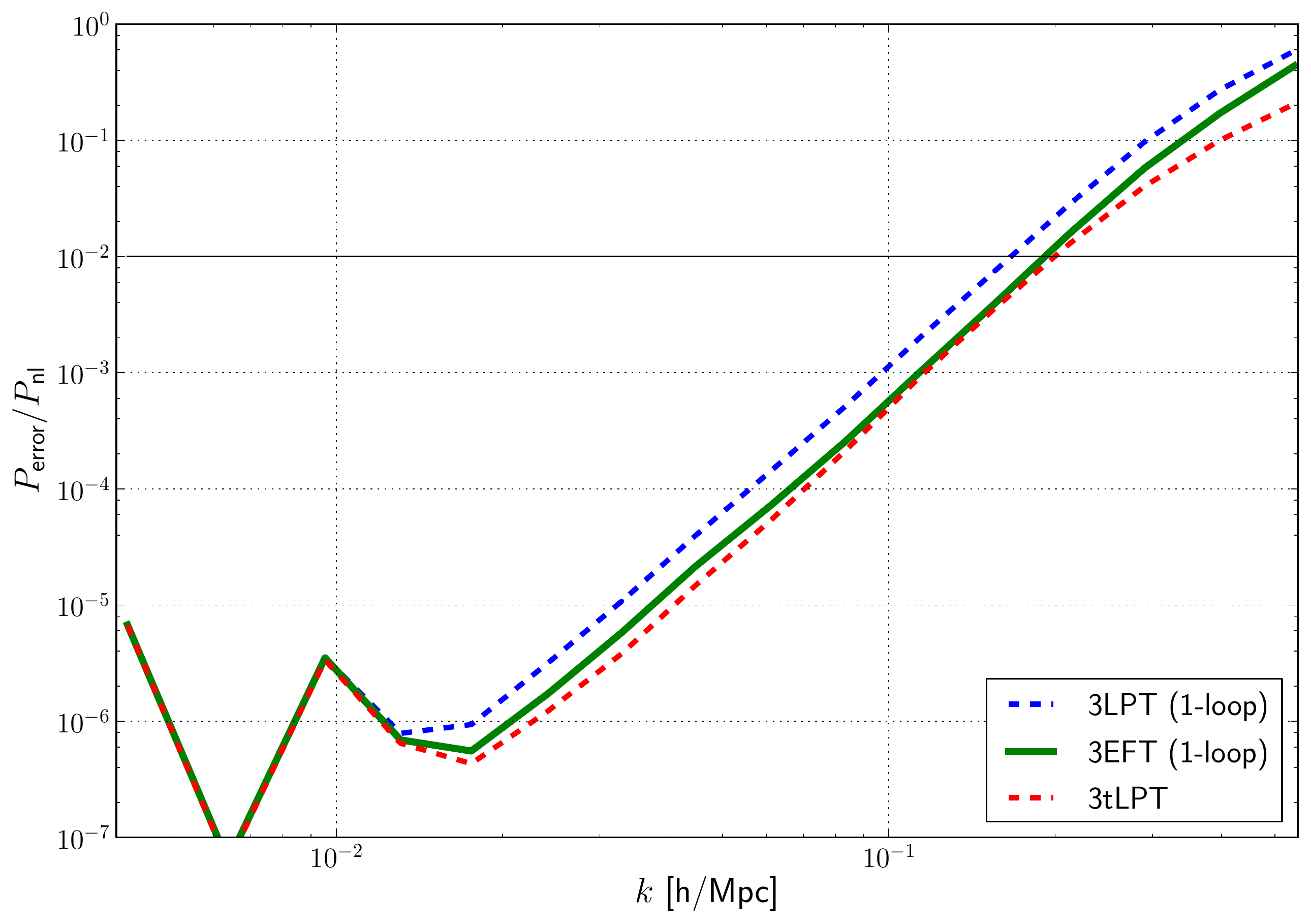}
\caption{Comparison between 1-loop LPT, EFT and tLPT, in terms of their agreement with simulation for the power spectrum (left panel), and the displacement field itself (right panel). The EFT improves on the LPT, and is close to tLPT, with the latter performing slightly better on the power spectrum.
This shows that the 1-loop EFT cannot be outperformed by any expansion of the form $a_1(k) \phi^{(1)}(\vk) + a_2(k) \phi^{(2)}(\vk) + a_3(k) \phi^{(3)}(\vk)$ by more than a factor of two in the wavenumber up to which we can trust the theory at percent level.}
\label{fig:pff_EFT_LPT_tLPT_cv}
\end{figure}

%%%%%%%%%%%%%%%%%%%%%%%%%%%%%%%%%%%%%%%%%%%%%%%%%%%%%%%%%
\section{The stochastic term}
\label{sec:stoch}
%%%%%%%%%%%%%%%%%%%%%%%%%%%%%%%%%%%%%%%%%%%%%%%%%%%%%%%%%

%%%%%%%%%%%%%%%%%%%%%%%%%%%%%%%%%%%%%%%%%%%%%%%%%%%%%%%%%
\subsection{A floor in the error power spectrum}
%%%%%%%%%%%%%%%%%%%%%%%%%%%%%%%%%%%%%%%%%%%%%%%%%%%%%%%%%

\paragraph*{Detection of the stochastic term}

We wish to understand the error power spectra $P_\text{error}^\text{tLPT}$ for the various tLPT models.
We make use of the orthogonal basis $ \phi^{(i)\perp}$ defined above. In this basis, the transfer functions $a_i^\perp(k)$ are independent of the tLPT order used (i.e. $a_1^\perp(k)$ is the same for 1tLPT, 2tLPT, etc).
Since $\phi_\text{4tLPT}$ is our best model for the true $\phi_\text{nl}$, we will write:
\beq
\phi_\text{nl} \simeq a_1^\perp \phi^{(1)\perp} + a_2^\perp \phi^{(2)\perp} + a_3^\perp \phi^{(3)\perp} + a_4^\perp \phi^{(4)\perp} + \phi_\text{stoch}
\label{eq:4tLPT}
\eeq
Thus we can estimate the displacement errors for 1tLPT, 2tLPT and 3tLPT as follows:
\beq
\bal
&\phi_\text{error}^\text{1tLPT} \simeq a_2^\perp \phi^{(2)\perp} + a_3^\perp \phi^{(3)\perp} + a_4^\perp \phi^{(4)\perp} + \phi_\text{stoch} \\
&\phi_\text{error}^\text{2tLPT} \simeq a_3^\perp \phi^{(3)\perp} + a_4^\perp \phi^{(4)\perp} + \phi_\text{stoch} \\
&\phi_\text{error}^\text{3tLPT} \simeq a_4^\perp \phi^{(4)\perp} + \phi_\text{stoch} \\
\eal
\eeq

\begin{figure}[t]
\centering
\includegraphics[width=10cm]{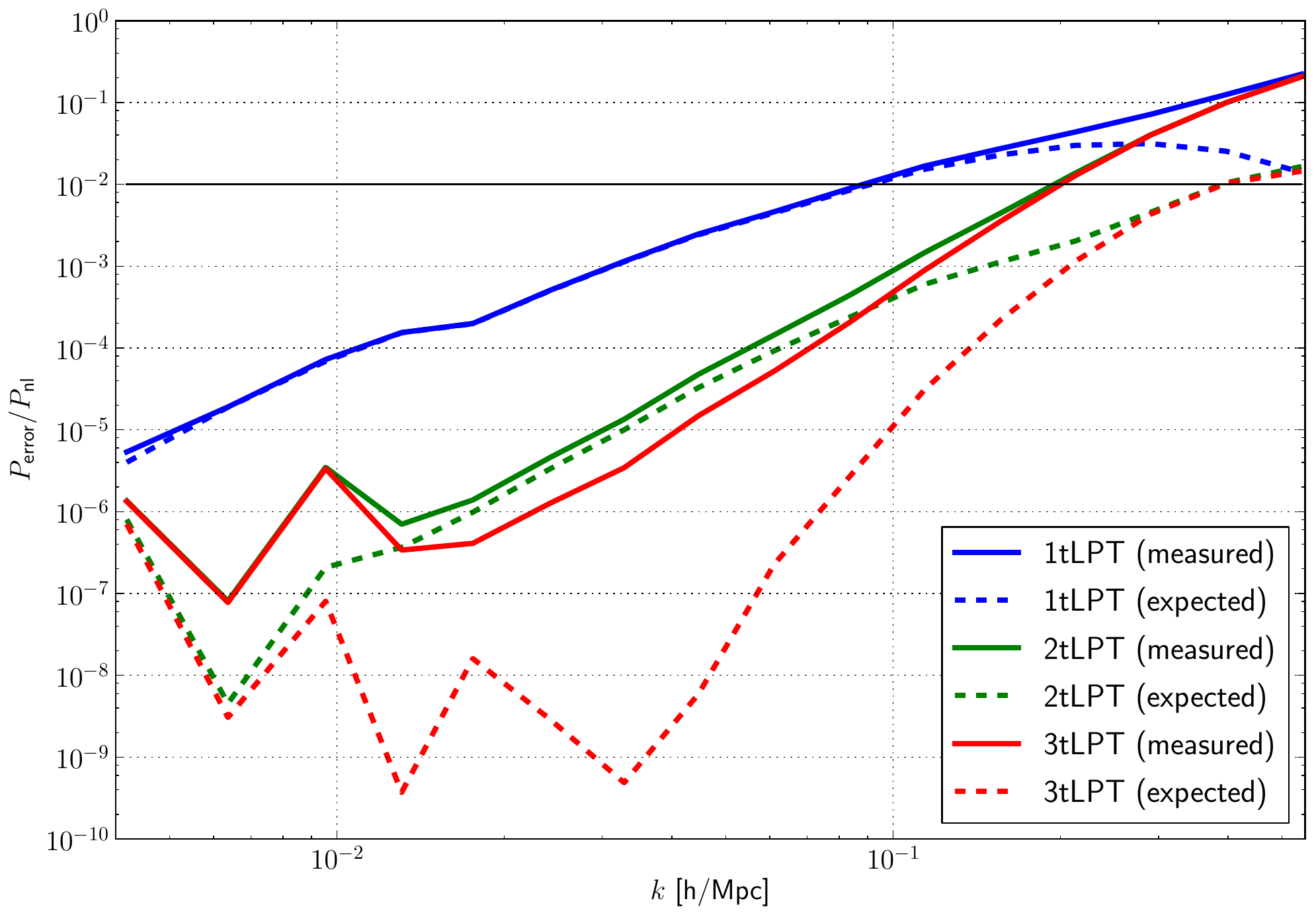}
\caption{Error power spectrum for the various tLPT models (solid lines), compared to the na\"\i ve expectation of equations~\eqref{eq:4tLPT} and~\eqref{eq:4tLPT_nostoch} (dashed lines), relative to the non-linear power spectrum. The na\"\i ve expectation underestimates the measurement, which indicates the presence and the exact size of the stochastic term $P_\text{stoch}$.}
\label{fig:floor_perror}
\end{figure}

\begin{figure}[t]
\centering
\includegraphics[width=10cm]{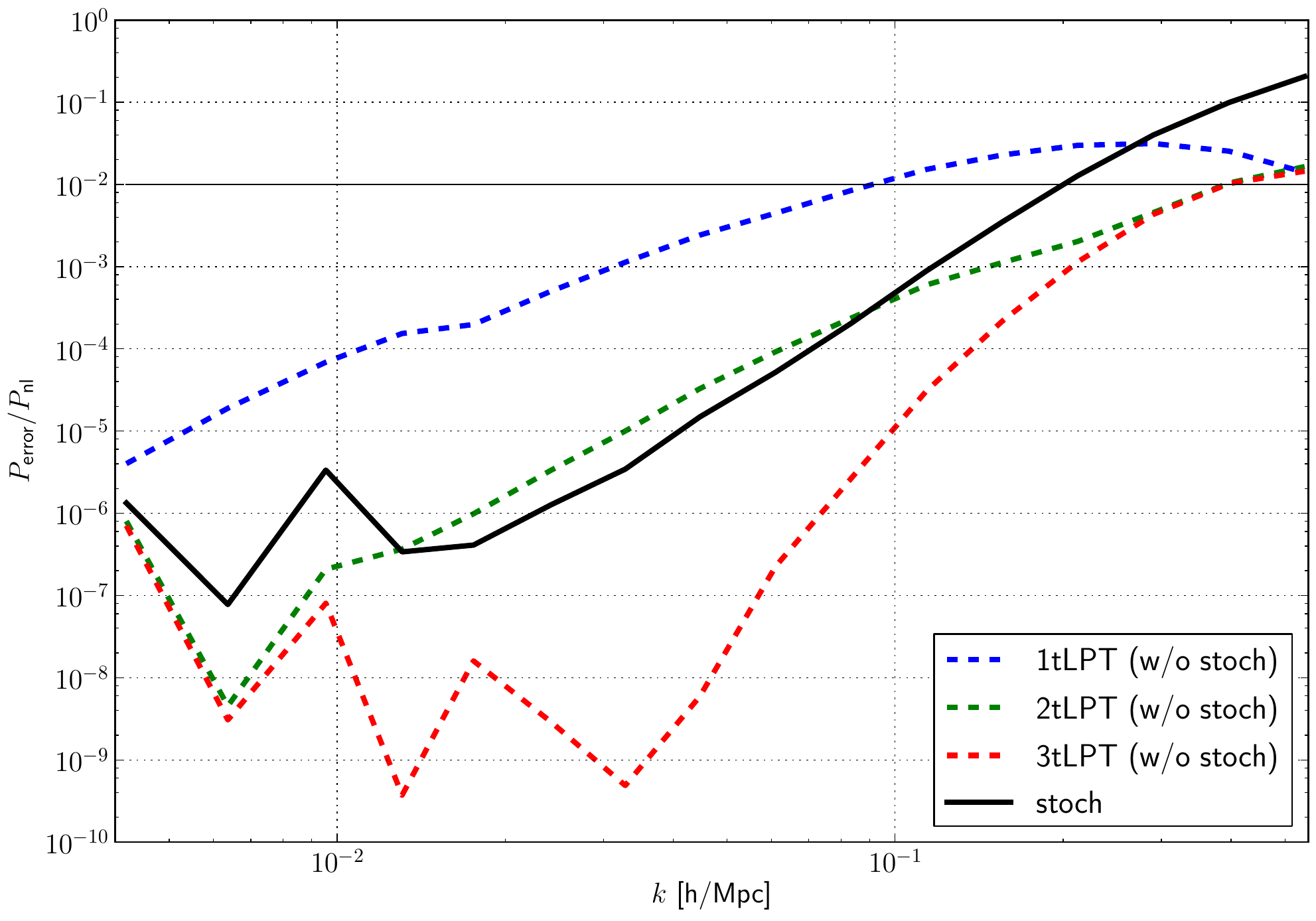}
\caption{Ratio of the error power spectrum for 1tLPT, 2tLPT and 3tLPT and the non-linear power spectrum after the stochastic term has been removed (dashed lines, same as Fig.~\ref{fig:floor_perror}). The stochastic term $P_\text{stoch}$ (black solid line) corresponds to $P_\text{error} = 1\% \, P_\text{nl}$ at $k\simeq 0.2$ h/Mpc. Beyond 3tLPT, the new LPT terms have contributions much smaller than the stochastic term, and therefore do not bring any significant improvement on the error power spectrum.}
\label{fig:limit_lpt}
\end{figure}

For a well-behaved perturbative series, one would assume that the various terms on the r.h.s. of \eqref{eq:4tLPT} are ranked in decreasing order. Keeping only the dominant term and neglecting the stochastic term then leads to the following estimate for $P_\text{error}$ for the tLPT models: 
\begin{equation}
\bal
& P_\text{error}^\text{1tLPT} \simeq \bigl[a_2^{\perp}\bigr]^2 P_{2\perp 2\perp} \\
& P_\text{error}^\text{2tLPT} \simeq \bigl[a_3^{\perp}\bigr]^2 P_{3\perp 3\perp} \\
& P_\text{error}^\text{3tLPT} \simeq \bigl[a_4^{\perp}\bigr]^2 P_{4\perp 4\perp} \\
\eal
\label{eq:4tLPT_nostoch}
\end{equation}
These estimates are shown in Fig.~\ref{fig:floor_perror} (dashed lines). The fact that they respect $a_2^2 P_{22}^\text{ortho} > a_3^2 P_{33}^\text{ortho} > a_4^2 P_{44}^\text{ortho}$ shows that this expansion is well defined: higher order terms are indeed smaller.
However, Fig.~\ref{fig:floor_perror} also shows that these estimates for the $P_\text{error}^{n\text{tLPT}}$ are much lower than the measured error power spectra (solid lines). There is clearly a floor in the measured error power spectra, which we associate with the stochastic displacement $\phi_\text{stock}$.
Fig.~\ref{fig:floor_perror} shows that the correct ranking of the terms in the 4tLPT model on large scales  ($k<0.1 \ihMpc$) is not that of \eqref{eq:4tLPT}, but instead
\beq
\phi_\text{nl} = a_1^\perp \phi^{(1)\perp} + a_2^\perp \phi^{(2)\perp} + a_3^\perp \phi^{(3)\perp} + \phi_\text{stoch} + a_4^\perp \phi^{(4)\perp},
\eeq
i.e., the stochastic term is not negligible and even exceeds the amplitude of the fourth order displacement.
These findings indicate that the error power spectra should scale according to:
\beq
\bal
& P_\text{error}^\text{1tLPT} \simeq \bigl[a_2^{\perp}\bigr]^2 P_{2\perp 2\perp} +\bigl[a_3^{\perp}\bigr]^2 P_{3\perp 3\perp}  + P_\text{stoch} \\
& P_\text{error}^\text{2tLPT} \simeq \bigl[a_3^{\perp}\bigr]^2 P_{3\perp 3\perp} + P_\text{stoch} \\
& P_\text{error}^\text{3tLPT} \simeq P_\text{stoch} \\
\eal
\eeq
Indeed we can define a unique function of $k$, the stochastic power spectrum, for which all three of the above equations are satisfied. We show the stochastic power spectrum from the L simulation as the solid black line in Fig.~\ref{fig:limit_lpt}.

\begin{figure}[H]
\centering
\includegraphics[width=0.59\textwidth]{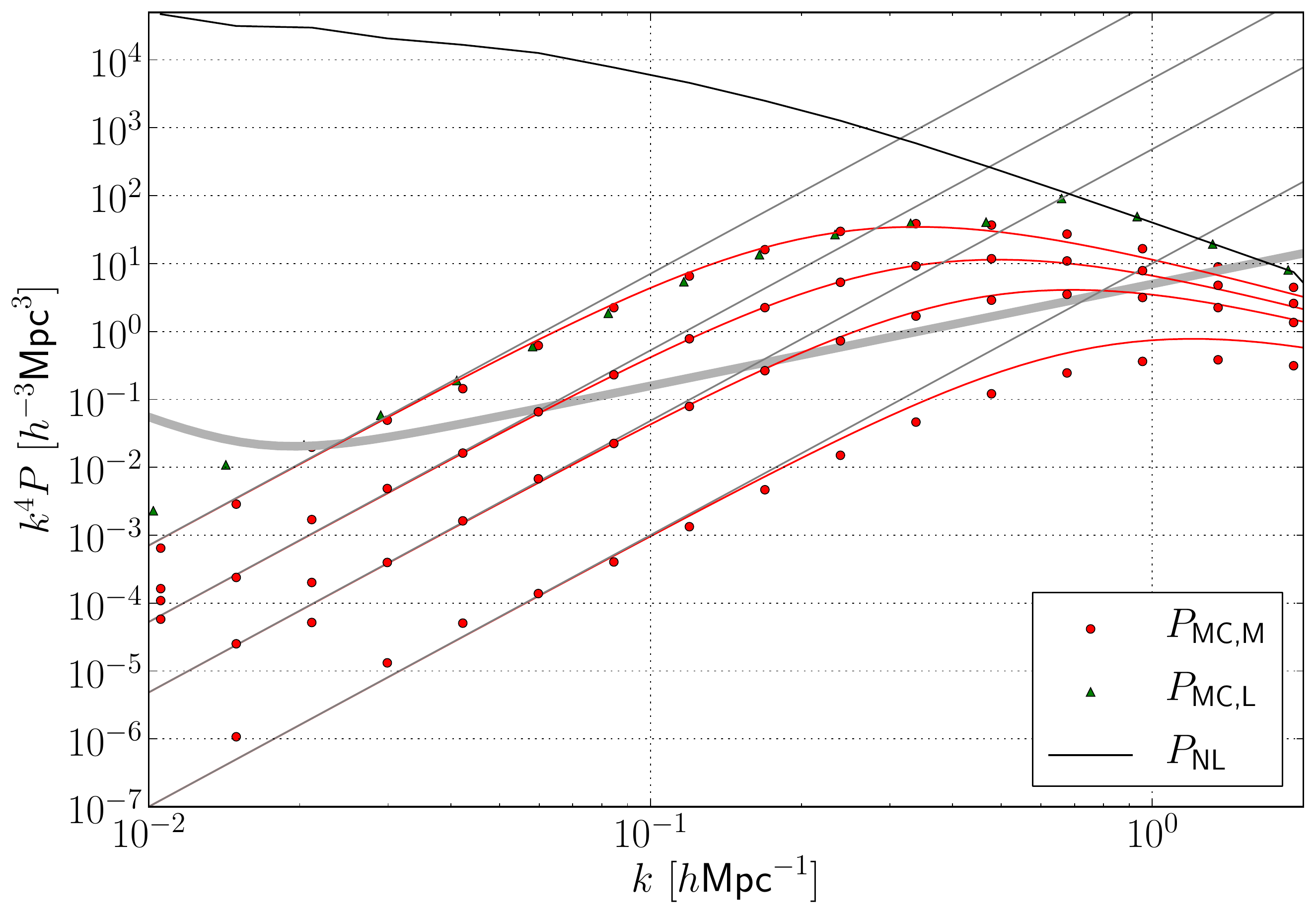}
\caption{Power spectrum of the stochastic term. The red points show the stochastic term in the M simulation at redshifts $z=0,0.5,1,2$ from top to bottom. The stochastic term of the L simulation at $z=0$ is shown by green triangles and agrees perfectly for $k<0.6 \ihMpc$ except for a small upturn on large scales.
The thick shaded line shows the systematic error on the L simulation at $z=0$, which exceeds the stochastic term on the largest scales, where the two simulations disagree. The red lines show the phenomenological model for the stochastic term from Eq.~\eqref{eq:stochmodel}.}
\label{fig:stochterm}
\end{figure}

The term $\phi_\text{stoch}$ is by definition not correlated with the LPT terms, and acts as a noise that limits the accuracy achievable with tLPT (see Fig.~\ref{fig:limit_lpt}): because of this term, the improvement in $P_\text{error}$ from 2tLPT to 3tLPT is limited, and there is no improvement at all beyond 3tLPT.
The ragged features of $P_\text{stoch}$ visible in Fig.~\ref{fig:limit_lpt} are likely a consequence of systematic errors in our $N$-body simulation. They correspond to very small errors on large scales: $P_\text{error} \sim 10^{-6} P_\text{nl}$ at $k\sim 0.01 \ihMpc$.
In Fig.~\ref{fig:stochterm}, we show the stochastic power spectrum for the L and M simulations, and its time dependence in the M simulation. We clearly see that the stochastic power spectra agree between the two simulations.

\begin{figure}[t]
\centering
\includegraphics[width=0.49\textwidth]{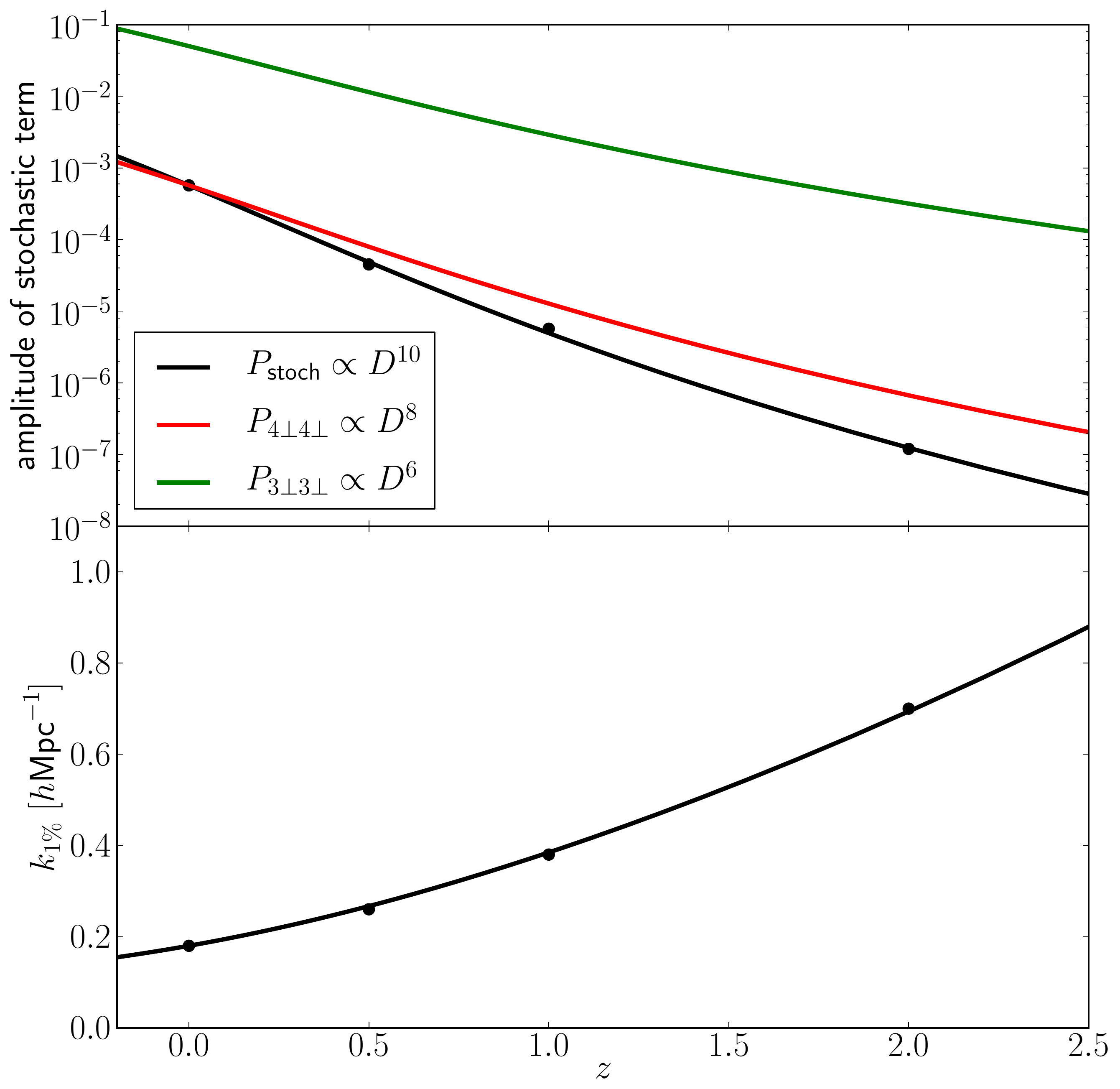}
\caption{\emph{Upper panel:} Time/redshift dependence of the large scale amplitude of the stochastic term.
\emph{Lower panel:} Time/redshift dependence of the scale, where the stochastic term amounts to 1\% of the total power.}
\label{fig:stochasticsmall}
\end{figure}

\paragraph*{Scale-dependence:}
On large scales, we can fit the stochastic term in Fig.~\ref{fig:stochterm} by
\beq
k^4 P_\text{stoch}\approx 5.7\times10^4 \left(\hMpc\right)^3 \ D^{10} \left(\frac{k}{1\  \ihMpc}\right)^4 \; .
\label{eq:fit_Pstoch}
\eeq
On the largest scales, we thus find that $P_\text{stoch}$ is independent of $k$. This is what we expect from mass and momentum conservation: for a mass and momentum-conserving perturbation, the lowest order correction to the density field is expected to scale as $k^2$ \cite{Mercolli:2013bsa,Peebles:LSS}, which corresponds to a $k$-independent correction to $\phi$. The stochastic term predicted by the EFT comes from a small-scale reshuffling of the matter, which is not describable in terms of the large-scale displacements, but conserves mass and momentum on large scales. 
The scaling observed in \eqref{eq:fit_Pstoch} is thus consistent with it being the stochastic term expected in the EFT framework. More specifically, in a scaling Universe we expect
\beq
\Delta^2_\text{stoch}=\frac{k^3 (k^4 P_\text{stoch})}{2\pi^2}=\mathcal{O}(1)\left(\frac{k}{k_\text{nl}}\right)^7
\eeq
and self-similarity dictates $k_\text{nl}\propto a^{-2/(n+3)}$, where $n$ is the slope of the initial power spectrum. The fitted time dependence $D^{10}$ is reproduced by a slope of $n=-1.6$, which corresponds to the slope of our input power spectrum at $k=0.1 \ihMpc$. Furthermore, from $\Delta^2_\text{stoch}= (k/k_\text{nl})^7$ we can deduce $k_\text{nl}\approx 0.32 \ihMpc$ at $z=0$. Due to the steep scaling ($k^7$), an order one prefactor does not change $k_\text{nl}$ significantly. 
We show the time dependence of the power law part of the stochastic term and the time dependence of the scale where it amounts to a $1\%$ change in the power spectrum in Fig.~\ref{fig:stochasticsmall}.

\paragraph*{Amplitude of the stochastic displacement}

To get an idea of the size of the stochastic term, we use the measured power spectra to infer the corresponding root mean square displacement
\beq
\sigma^2_\text{stoch}=\frac{1}{3}\int \frac{\derd^3 k}{(2\pi)^3}\left\langle \psi_i |\psi_i \right\rangle=\frac{1}{3}\int \frac{\derd^3 k}{(2\pi)^3}k^2 P_\text{stoch}
\label{eq:stochdisp}
.
\eeq
and find a rms stochastic displacement $\sigma_\text{stoch} \simeq 0.8$ Mpc/h (compared to $\sigma_\text{nl} \simeq 6.0$ Mpc/h for the full non-linear rms displacement). The integrand in Eq.~\eqref{eq:stochdisp} is peaked at $k\approx 0.6\ \ihMpc$, i.e. far beyond the range of scales where the stochastic term follows the $k^4$ scaling.\\
We do not expect perturbation theory to be able to capture the motion of particles within halos. These motions will thus contribute a (probably significant) fraction of the stochastic displacement. A crude estimate for this source of stochastic displacement can be obtained by assuming that particles within a halo of mass $m$ have a root-mean-square displacement of roughly two virial radii $R_\text{vir}$, and to average over all halos using the halo mass function:
\beq
\sigma^2_\text{stoch.,h} = \int \derd m \frac{\derd n}{\derd m} m \biggl(2 R_\text{vir}(m)\biggr)^2/\int \derd m \frac{\derd n}{\derd m} m
\eeq
Integrating over all masses, this estimate yields $\sigma_\text{stoch. motion in halo} \simeq 1.1 \hMpc$, similar to $\sigma_\text{stoch}$. The integral peaks at $M\approx 2\times 10^{14}M_\odot$, i.e., for haloes of Lagrangian radius $r\approx 9 \hMpc$. In Fourier space this corresponds to a wavenumber $k\approx \pi/r= 0.35\ihMpc$.
If we assume that this stochastic term induces a Gaussian smoothing in the resulting Eulerian space density field, we conclude that it produces a $1 \%$ change in the density power spectrum at $k\approx 0.2\ihMpc$ and will thus contribute to the Eulerian EFT sound speed $c_\text{s}$.
The time dependence can be introduced into the halo model by scaling the variance as $\sigma(M)\to D(a)\sigma(M)$. The resulting $\sigma^2_\text{stoch.,h}$ then scales as $D^{1.6}$.
For this scaling we employed a Sheth-Tormen mass function \cite{Sheth:1999mn}.

\paragraph*{Fitting function}
The stochastic term can be accounted for by a simple fitting function that is constructed from the following requirements:
\begin{itemize}
\item on large scales it scales as $k^4$;
\item on small scales it scales as $k^{m}$ with $m<-1$;
\item it integrates to the stochastic displacement dispersion $\sigma_\text{stoch}^2$.
\end{itemize}
One function that satisfies these constraints and smoothly transitions between the low- and high-$k$ regimes is given by
\beq
k^4 P=\left(\frac{k}{k_\text{s}}\right)^4 \sigma_\text{stoch}^2 \frac{1}{(1+(k/k_\text{s})^2)^n}\frac{2^4\pi^{3/2}}{k_\text{s}}\frac{\Gamma(n)}{\Gamma(n-5/2)},
\label{eq:stochmodel}
\eeq
where $n>2.5$.
As we show in Fig.~\ref{fig:stochterm} this functional form provides a very good description of the measured stochastic term at all redshifts once the scalings $\sigma_\text{stoch}\propto D^{1.5}$ and $k_\text{s}\propto D^{-1.5}$ are accounted for. For the scale we employed $k_\text{s}\approx k_\text{nl}=0.26\ihMpc$.

%%%%%%%%%%%%%%%%%%%%%%%%%%%%%%%%%%%%%%%%%%%%%%%%%%%%%%%%%
\subsection{Origin of the stochastic term}
%%%%%%%%%%%%%%%%%%%%%%%%%%%%%%%%%%%%%%%%%%%%%%%%%%%%%%%%%

What is the origin of this observed stochastic term $\phi_\text{stoch}$? Such a term is expected in the framework of the EFT, due to the the small-scale fluctuations which are not amenable to perturbation theory and are treated as a random noise. This term could also arise from higher order LPT terms which we haven't computed. Finally, it could also be a mere artefact of numerical errors in the $N$-body simulation.

In order to exclude the possibility that this stochastic term is simply due to numerical errors, we have run simulations with two different box sizes. The stochastic terms $P_\text{stoch}$ identified at redshift $z=0$ in these two simulations agree, as can be seen in Fig.~\ref{fig:stochterm}. In this context we should mention that the change in the simulation box size also changes the wavenumber at which we have to cut our initial power spectrum in order to avoid aliasing. Thus, while the L simulation has $k_\text{max}=0.6\ \hMpc$ the M simulation has $k_\text{max}=2.4\ \hMpc$. This difference in the cutoff wavenumber translates into slightly different LPT contributions and slightly different transfer functions. Yet, at the scales where we trust the simulations, the power spectrum of the stochastic term does \emph{not} change. We interpret this observation as a strong indication of this term being truly related to highly non-linear motions that are uncorrelated with the perturbation theory prediction.
Virialized motions are an example of such motions uncorrelated with the LPT terms, since LPT is not able to capture shell crossing.

We also considered transients in the $N$-body simulation \cite{Crocce:2006ve} as a possible source of stochastic term. Indeed, the $N$-body simulations are initialized with 2LPT (at redshift $z_\text{init} = 99$), so we expect the simulations to contain transient errors in the third and higher order displacement fields. This error can be estimated in LPT in terms of an error on the growth factor of each LPT term (see App.~\ref{sec:transients}): $\phi_1$ and $\phi_2$ are not affected, $\phi_3$ is underestimated by $2 \times10^{-4}$, the $\mathcal{L}^{2,2}$ and $\mathcal{M}^{1,1,2}$ contributions to $\phi_4$ are underestimated by $10^{-6}$ while the $\mathcal{L}^{3,1}$ contribution to $\phi_4$ is underestimated by $4 \times 10^{-4}$. These errors are too small to account for the measured stochastic term, and they are clearly correlated with the LPT terms, so they should be automatically corrected for by the transfer functions in tLPT. In conclusion, transients cannot account for the stochastic term.

\subsection{Link with the curl part of the displacement}
\begin{figure}[t]
\centering
\includegraphics[width=0.59\textwidth]{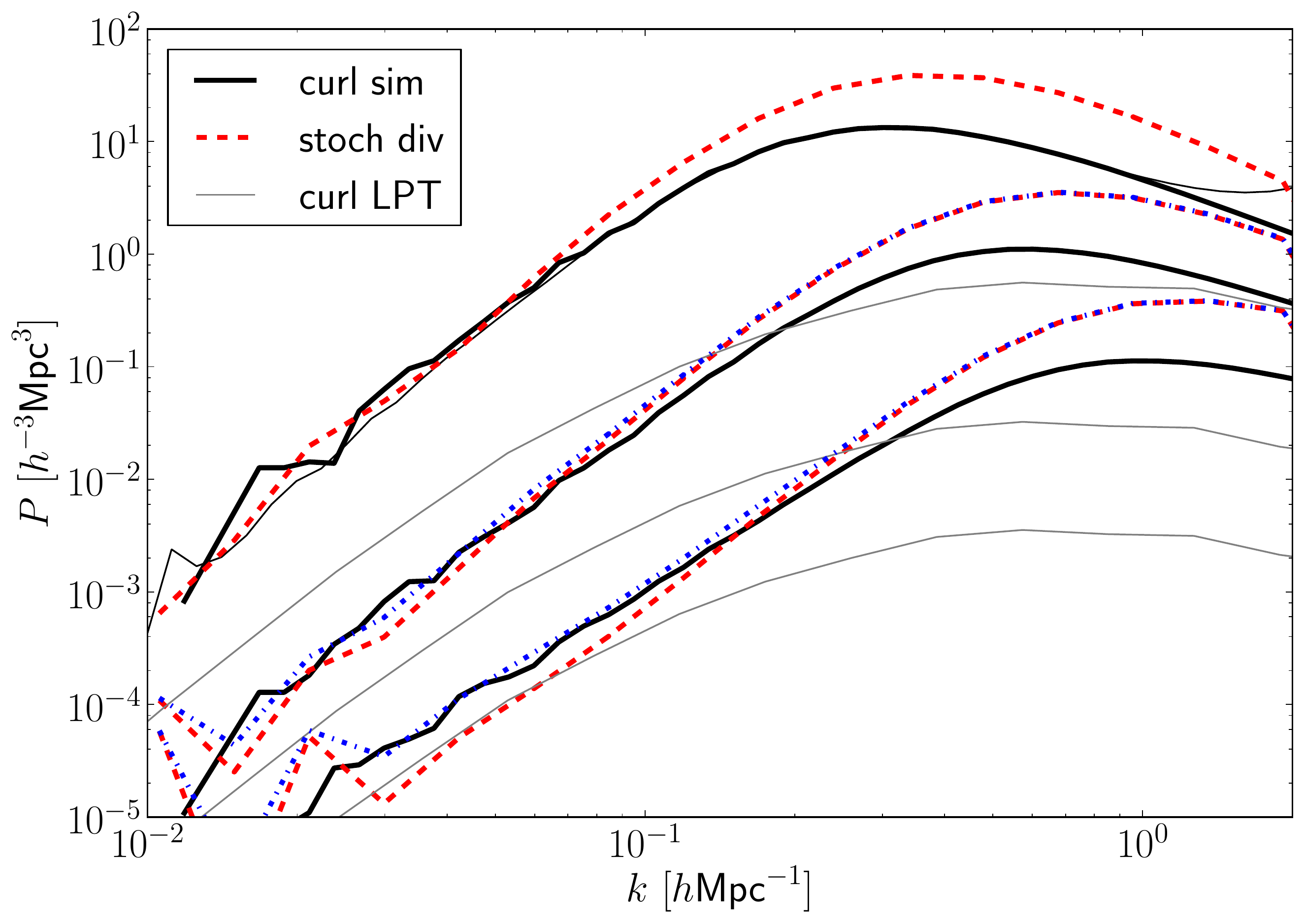}
\caption{Power spectra of the stochastic displacement divergence $\vec \nabla \cdot \vec \psi_\text{stoch}$ (red dashed), the full curl $\vec \nabla \times \vec \psi$ (black solid) and the LPT prediction for the curl (gray). From top to bottom we show measurements and predictions for $z=0$, $z=1$ and $z=2$. At redshift $z=0$ the measurement of the curl exceeds the LPT prediction by roughly one order of magnitude but agrees with the divergence of the stochastic displacement on large scales. The dot-dashed lines at $z=1$ and $z=2$ show the sum of the LPT curl and the stochastic part of the displacement divergence.}
\label{fig:curl}
\end{figure}

Besides the divergence, we can also consider the curl of the displacement field. For the correlators we have
\beq
\left \langle (\vec \nabla \times \vec \psi)_i |(\vec \nabla \times \vec \psi)_i\right\rangle=\left\langle\Delta \omega_i|\Delta \omega_i\right\rangle\; ,
\eeq
where we used Coulomb gauge. Thus, we have in Fourier space
\beq
P_{\vec \nabla \times \vec \psi}=\left \langle (\vec k \times \vec \psi)_i |(\vec k \times \vec \psi)_i\right\rangle=k^4\left\langle \omega_i| \omega_i\right\rangle\; .
\eeq
Correlating the curl components of the field itself, we have
\beq
\left\langle\psi_{c,i}|\psi_{c,i}\right\rangle=\left\langle(\vec \nabla \times\vec  \omega)_i|(\vec \nabla \times \vec \omega)_i\right\rangle
\eeq
or in Fourier space
\begin{align}
P_{\psi_c}=\left\langle(\vec k \times \vec \omega)_i|(\vec k \times \vec \omega)_i\right\rangle=&\left\langle \epsilon_{ijl}k_j \omega_l | \epsilon_{imn}k_m \omega_n\right\rangle=(\delta_{jm}\delta_{ln}-\delta_{jn}\delta_{lm})k_jk_m \left\langle\omega_l | \omega_n\right\rangle\nonumber\\
=&k^2\left\langle\omega_i|\omega_i\right\rangle\; ,
\end{align}
where we again used the Coulomb gauge condition $k_i \omega_i=0$.
The first quantity is easier for measurements in simulations while the latter is more closely related to perturbation theory. As we saw above, their power spectra are related by multiplicative factors of $k$.\\
We measure the curl power spectrum $P_{\vec \nabla\times \vec \Psi}$ in both the M and L simulations and find very good agreement where the spectra overlap. In agreement with \cite{Chan:2013vao} we find that the curl power spectra scale as $D^{10}$. As shown in Fig.~\ref{fig:curl},  at redshift $z=0$ the measured curl power spectrum exceeds the LPT prediction from $P_{3c3c}$ by about a factor of 10 on large scales. This indicates that the curl component of the displacement is completely dominated by small scale,  non-perturbative virialized structures. Since the perturbative prediction scales as $D^6$, we expect the perturbative and non-linear curl power spectra to be of the same order only at $z=2$. Indeed we see a slight upturn of the curl power spectrum at low $k$ at $z=2$, where the measured curve approaches the perturbative prediction.\\
Since the leading curl contribution is fully non-perturbative, it is interesting to compare its power spectrum to the stochastic term of the displacement divergence, which we argued to be sourced by virialized motions. Amazingly, both the large scale amplitude and the redshift scaling of the curl are in very good agreement with the scalar stochastic term. This further solidifies our confidence in having identified the true stochastic term arising from virialized, non-perturbative motions.

%%%%%%%%%%%%%%%%%%%%%%%%%%%%%%%%%%%%%%%%%%%%%%%%%%%%%%%%%
%%%%%%%%%%%%%%%%%%%%%%%%%%%%%%%%%%%%%%%%%%%%%%%%%%%%%%%%%
\section{Conclusions}
\label{sec:conclusions}
%%%%%%%%%%%%%%%%%%%%%%%%%%%%%%%%%%%%%%%%%%%%%%%%%%%%%%%%%
%%%%%%%%%%%%%%%%%%%%%%%%%%%%%%%%%%%%%%%%%%%%%%%%%%%%%%%%%

In this paper we studied Lagrangian Perturbation theory and its regularization in the effective field theory approach.
We numerically evaluated the LPT displacement fields up to fifth order on a regular grid that has the same phases as a corresponding suite of $N$-body simulations.
This allowed us to test the 1-loop LPT and EFT at the level of the displacement field itself, which is a much more stringent test of the perturbative approach than matching the power spectrum only and also reduces the cosmic variance significantly. As an added benefit, we were able to evaluate LPT terms (such as $P_{55}$, which is formally a 4-loop term) that would be barely tractable with a Fourier space loop calculation.

We verified that the LPT expansion is well-behaved up to third order, in that the error on the displacement field decreases from 1LPT to 2LPT, to 3LPT, on scales $k \lesssim 0.1 \ihMpc$.
In doing so, we highlighted the importance of the second order displacement $\phi_2$, which is crucial in reducing the error power spectrum (even more so than $\phi_3$),
despite its small contribution to the non-linear power spectrum ($P_{22} \ll P_{13}$).
We found that 1-loop LPT provides a 1\%-accurate description of the non-linear power spectrum up to $k=0.05\ihMpc$.

We carefully tested the 1-loop EFT. 
We validated the simple scalings of the various LPT terms and the EFT counterterm up to 1-loop, estimated for an EdS Universe. 
We reliably detected the leading EFT counterterm $\alpha k^2 \phi_1$ with a non-zero coefficient $\alpha$ on large scales ($k<0.03\ihMpc$). 
The EFT provides a significant improvement over LPT, with a better convergence up to 1-loop, and a threefold increase in the maximum wave vector where the non-linear power spectrum is reproduced to 1\%: the 1-loop EFT power spectrum is accurate to 1\% up to $k=0.15\ihMpc$. The EFT also reduces the error power spectrum by up to a factor of two around $k=0.1 \ihMpc$. 
The measured time-dependence of the EFT coefficient $\alpha$ matches reasonably well the one expected from simple scalings in an EdS Universe.
We showed that the EFT is close to optimal, in the sense that it achieves similarly small error power spectrum $P_\text{error}$ and similarly accurate non-linear power spectrum as the 3tLPT and higher order tLPT models do.

By looking at the displacement field and not only its power spectrum, we were able to detect a stochastic term $\phi_\text{stoch}$ that is uncorrelated with the LPT terms and seems to constitute a limit to the reach of LPT and EFT. 
Its power spectrum is slightly smaller than that of $\phi^{(3)\perp}$ but larger than that of $\phi^{(4)\perp}$, and the corresponding root-mean-square displacement in real space is consistent with the typical random motion within halos.
Its scale-dependence on large scales matches what is expected of a momentum-conserving displacement, consistent with the stochastic term expected in the EFT framework. However, the size of this effect seems larger than the na\"{\i}ve expectation based on power law Universe scalings. 
To the best of our understanding this term is not due to numerical errors or transients in our simulations. For example, its amplitude matches in simulations that have very different resolutions.
If not modeled explicitly, this term would imply a limit to the accuracy to which the displacement field can be described by LPT or EFT, corresponding to a 1\% error on the non-linear power spectrum at $k=0.2\ihMpc$  at $z=0$.  In this case, there is no benefit in going to higher order than third order in the displacement with the corresponding one-loop counterterm. In particular, no tLPT-like model fit to the displacement can predict the non-linear power spectrum to better than $1\%$ beyond $k=0.2$ h/Mpc, unless it overfits the power spectrum. This conclusion extends the result from Fig.~\ref{fig:pff_EFT_LPT_tLPT_cv} to all orders in tLPT, and not only at one-loop.
However, one might be able to model the power spectrum of this stochastic term with a few free parameters or a fitting function and thereby extend the range of validity of the EFT displacement power spectrum.

For practical calculations of the density field, Eulerian or Standard Perturbation Theory and the corresponding Effective Field Theory are very useful. Performing a similar test at the level of the fields rather than power spectra would be desirable, but is complicated by the decorrelation due to long wavelength motions. Thus the long wavelength displacements need to be resummed \cite{Senatore:2014via}. We will address this issue in a forthcoming paper.

Besides testing the EFT approach, we also pushed the simulations to their limits. The EFT, or more generally its underlying symmetry arguments, predict how corrections to the linear power spectrum should behave on large scales. 
These constraints are not fully satisfied by the simulation measurements at the sub-percent level. In this context, we pointed out that the measurement of the $\alpha$ coefficient on large scales is very difficult, because it is highly sensitive to systematic errors in the simulations.

%%%%%%%%%%%%%%%%%%%%%%%%%%%%%%%%%%%%%%%%%%%%%%%%%%%%%%%%%%%%%%%
%%%%%%%%%%%%%%%%%%%%%%%%%%%%%%%%%%%%%%%%%%%%%%%%%%%%%%%%%%%%%%%
\section*{Acknowledgements}
%%%%%%%%%%%%%%%%%%%%%%%%%%%%%%%%%%%%%%%%%%%%%%%%%%%%%%%%%%%%%%%
%%%%%%%%%%%%%%%%%%%%%%%%%%%%%%%%%%%%%%%%%%%%%%%%%%%%%%%%%%%%%%%

The authors would like to thank Francis Bernardeau, Simone Ferraro, Renee Hlozek, Lorenzo Mercolli, Uro\v{s} Seljak, David Spergel, Svetlin Tassev, Zvonimir Vlah and Martin White for fruitful discussions.
T.B. is supported by the Institute for Advanced Study through a Corning Glass Works foundation fellowship.
M.Z. is supported in part by the NSF grants  PHY-1213563 and AST-1409709.
E.S. is supported by the NSF grant AST1311756 and the NASA grant NNX12AG72G.

%%%%%%%%%%%%%%%%%%%%%%%%%%%%%%%%%%%%%%%%%%%%%%%%%%%%%%%%%%%%%%%
%%%%%%%%%%%%%%%%%%%%%%%%%%%%%%%%%%%%%%%%%%%%%%%%%%%%%%%%%%%%%%%
\bibliographystyle{JHEP}
\bibliography{lpt}

\providecommand{\href}[2]{#2}\begingroup\raggedright\begin{thebibliography}{10}

\bibitem{Planck:2015params}
{Planck Collaboration}, P.~A.~R. {Ade}, N.~{Aghanim}, M.~{Arnaud},
  M.~{Ashdown}, J.~{Aumont}, C.~{Baccigalupi}, A.~J. {Banday}, R.~B.
  {Barreiro}, J.~G. {Bartlett}, and et~al., {\it {Planck 2015 results. XIII.
  Cosmological parameters}},  {\em ArXiv e-prints} (Feb., 2015)
  [\href{http://arxiv.org/abs/1502.01589}{{\tt arXiv:1502.01589}}].

\bibitem{Planck:2015nonGauss}
{Planck Collaboration}, P.~A.~R. {Ade}, N.~{Aghanim}, M.~{Arnaud}, F.~{Arroja},
  M.~{Ashdown}, J.~{Aumont}, C.~{Baccigalupi}, M.~{Ballardini}, A.~J. {Banday},
  and et~al., {\it {Planck 2015 results. XVII. Constraints on primordial
  non-Gaussianity}},  {\em ArXiv e-prints} (Feb., 2015)
  [\href{http://arxiv.org/abs/1502.01592}{{\tt arXiv:1502.01592}}].

\bibitem{Alvarez:2014vva}
M.~Alvarez, T.~Baldauf, J.~R. Bond, N.~Dalal, R.~de~Putter, et~al., {\it
  {Testing Inflation with Large Scale Structure: Connecting Hopes with
  Reality}},  \href{http://arxiv.org/abs/1412.4671}{{\tt arXiv:1412.4671}}.

\bibitem{Dore:2014cca}
O.~Dore, J.~Bock, P.~Capak, R.~de~Putter, T.~Eifler, et~al., {\it {Cosmology
  with the SPHEREX All-Sky Spectral Survey}},
  \href{http://arxiv.org/abs/1412.4872}{{\tt arXiv:1412.4872}}.

\bibitem{Zeldovich:1969sb}
Y.~Zeldovich, {\it {Gravitational instability: An Approximate theory for large
  density perturbations}},  {\em Astron.Astrophys.} {\bf 5} (1970) 84--89.

\bibitem{Peebles:LSS}
P.~J.~E. {Peebles}, {\em The large-Scale Structure of the Universe}.
\newblock Princeton University Press, Princeton, NJ, 1980.

\bibitem{Bernardeau:2001qr}
F.~Bernardeau, S.~Colombi, E.~Gaztanaga, and R.~Scoccimarro, {\it {Large-scale
  structure of the universe and cosmological perturbation theory}},  {\em Phys.
  Rept.} {\bf 367} (2002) 1--248,
  [\href{http://arxiv.org/abs/astro-ph/0112551}{{\tt astro-ph/0112551}}].

\bibitem{Baumann:2010tm}
D.~Baumann, A.~Nicolis, L.~Senatore, and M.~Zaldarriaga, {\it {Cosmological
  Non-Linearities as an Effective Fluid}},  {\em JCAP} {\bf 1207} (2012) 051,
  [\href{http://arxiv.org/abs/1004.2488}{{\tt arXiv:1004.2488}}].

\bibitem{Carrasco:2012cv}
J.~J.~M. Carrasco, M.~P. Hertzberg, and L.~Senatore, {\it {The Effective Field
  Theory of Cosmological Large Scale Structures}},  {\em JHEP} {\bf 1209}
  (2012) 082, [\href{http://arxiv.org/abs/1206.2926}{{\tt arXiv:1206.2926}}].

\bibitem{Porto:2013qua}
R.~A. Porto, L.~Senatore, and M.~Zaldarriaga, {\it {The Lagrangian-space
  Effective Field Theory of Large Scale Structures}},  {\em JCAP} {\bf 1405}
  (2014) 022, [\href{http://arxiv.org/abs/1311.2168}{{\tt arXiv:1311.2168}}].

\bibitem{Carrasco:2013mua}
J.~J.~M. Carrasco, S.~Foreman, D.~Green, and L.~Senatore, {\it {The Effective
  Field Theory of Large Scale Structures at Two Loops}},  {\em JCAP} {\bf 1407}
  (2014) 057, [\href{http://arxiv.org/abs/1310.0464}{{\tt arXiv:1310.0464}}].

\bibitem{Pajer:2013jj}
E.~Pajer and M.~Zaldarriaga, {\it {On the Renormalization of the Effective
  Field Theory of Large Scale Structures}},  {\em JCAP} {\bf 1308} (2013) 037,
  [\href{http://arxiv.org/abs/1301.7182}{{\tt arXiv:1301.7182}}].

\bibitem{Carrasco:2013sva}
J.~J.~M. Carrasco, S.~Foreman, D.~Green, and L.~Senatore, {\it {The 2-loop
  matter power spectrum and the IR-safe integrand}},  {\em JCAP} {\bf 1407}
  (2014) 056, [\href{http://arxiv.org/abs/1304.4946}{{\tt arXiv:1304.4946}}].

\bibitem{Baldauf:2014qfa}
T.~Baldauf, L.~Mercolli, M.~Mirbabayi, and E.~Pajer, {\it {The Bispectrum in
  the Effective Field Theory of Large Scale Structure}},
  \href{http://arxiv.org/abs/1406.4135}{{\tt arXiv:1406.4135}}.

\bibitem{Angulo:2014tfa}
R.~E. Angulo, S.~Foreman, M.~Schmittfull, and L.~Senatore, {\it {The One-Loop
  Matter Bispectrum in the Effective Field Theory of Large Scale Structures}},
  \href{http://arxiv.org/abs/1406.4143}{{\tt arXiv:1406.4143}}.

\bibitem{Tassev:2012cq}
S.~Tassev and M.~Zaldarriaga, {\it {Estimating CDM Particle Trajectories in the
  Mildly Non-Linear Regime of Structure Formation. Implications for the Density
  Field in Real and Redshift Space}},  {\em JCAP} {\bf 1212} (2012) 011,
  [\href{http://arxiv.org/abs/1203.5785}{{\tt arXiv:1203.5785}}].

\bibitem{Bouchet:1995}
F.~R. {Bouchet}, S.~{Colombi}, E.~{Hivon}, and R.~{Juszkiewicz}, {\it
  {Perturbative Lagrangian approach to gravitational instability.}},  {\em AAP}
  {\bf 296} (Apr., 1995) 575,
  [\href{http://arxiv.org/abs/astro-ph/9406013}{{\tt astro-ph/9406013}}].

\bibitem{Catelan:1995}
P.~{Catelan}, {\it {Lagrangian dynamics in non-flat universes and non-linear
  gravitational evolution}},  {\em MNRAS} {\bf 276} (Sept., 1995) 115--124,
  [\href{http://arxiv.org/abs/astro-ph/9406016}{{\tt astro-ph/9406016}}].

\bibitem{Matsubara:2008re}
T.~{Matsubara}, {\it {Resumming cosmological perturbations via the Lagrangian
  picture: One-loop results in real space and in redshift space}},  {\em
  Phys.Rev.} {\bf D77} (Mar., 2008) 063530,
  [\href{http://arxiv.org/abs/0711.2521}{{\tt arXiv:0711.2521}}].

\bibitem{Rampf:2012xa}
C.~Rampf and T.~Buchert, {\it {Lagrangian perturbations and the matter
  bispectrum I: fourth-order model for non-linear clustering}},  {\em JCAP}
  {\bf 1206} (2012) 021, [\href{http://arxiv.org/abs/1203.4260}{{\tt
  arXiv:1203.4260}}].

\bibitem{Zheligovsky:2013eca}
V.~Zheligovsky and U.~Frisch, {\it {Time-analyticity of Lagrangian particle
  trajectories in ideal fluid flow}},  {\em J.Fluid Mech.} {\bf 749} (2014)
  404, [\href{http://arxiv.org/abs/1312.6320}{{\tt arXiv:1312.6320}}].

\bibitem{Eisenstein:1997ik}
D.~J. Eisenstein and W.~Hu, {\it {Baryonic features in the matter transfer
  function}},  {\em Astrophys.J.} {\bf 496} (1998) 605,
  [\href{http://arxiv.org/abs/astro-ph/9709112}{{\tt astro-ph/9709112}}].

\bibitem{Crocce:2006ve}
M.~Crocce, S.~Pueblas, and R.~Scoccimarro, {\it {Transients from Initial
  Conditions in Cosmological Simulations}},  {\em Mon.Not.Roy.Astron.Soc.} {\bf
  373} (2006) 369--381, [\href{http://arxiv.org/abs/astro-ph/0606505}{{\tt
  astro-ph/0606505}}].

\bibitem{Springel2005}
V.~{Springel}, {\it {The cosmological simulation code GADGET-2}},  {\em
  Mon.Not.Roy.Astron.Soc.} {\bf 364} (2005) 1105--1134.

\bibitem{WMAP7}
{\bf WMAP Collaboration} Collaboration, {Komatsu, E. et al.}, {\it {Seven-Year
  Wilkinson Microwave Anisotropy Probe (WMAP) Observations: Cosmological
  Interpretation}},  {\em Astrophys.J.Suppl.} {\bf 192} (2011) 18,
  [\href{http://arxiv.org/abs/1001.4538}{{\tt arXiv:1001.4538}}].

\bibitem{Mercolli:2013bsa}
L.~Mercolli and E.~Pajer, {\it {On the velocity in the Effective Field Theory
  of Large Scale Structures}},  {\em JCAP} {\bf 1403} (2014) 006,
  [\href{http://arxiv.org/abs/1307.3220}{{\tt arXiv:1307.3220}}].

\bibitem{Sheth:1999mn}
R.~K. Sheth and G.~Tormen, {\it {Large scale bias and the peak background
  split}},  {\em Mon.Not.Roy.Astron.Soc.} {\bf 308} (1999) 119,
  [\href{http://arxiv.org/abs/astro-ph/9901122}{{\tt astro-ph/9901122}}].

\bibitem{Chan:2013vao}
K.~C. Chan, {\it {Helmholtz Decomposition of the Lagrangian Displacement}},
  {\em Phys.Rev.} {\bf D89} (2014), no.~8 083515,
  [\href{http://arxiv.org/abs/1309.2243}{{\tt arXiv:1309.2243}}].

\bibitem{Senatore:2014via}
L.~Senatore and M.~Zaldarriaga, {\it {The IR-resummed Effective Field Theory of
  Large Scale Structures}},  {\em JCAP} {\bf 1502} (2015), no.~02 013,
  [\href{http://arxiv.org/abs/1404.5954}{{\tt arXiv:1404.5954}}].

\bibitem{Smith:2012uz}
R.~E. Smith, D.~S. Reed, D.~Potter, L.~Marian, M.~Crocce, et~al., {\it
  {Precision cosmology in muddy waters: Cosmological constraints and N-body
  codes}},  \href{http://arxiv.org/abs/1211.6434}{{\tt arXiv:1211.6434}}.

\bibitem{Matsubara:2015re}
T.~{Matsubara}, {\it {Recursive Solutions of Lagrangian Perturbation Theory}},
  {\em ArXiv e-prints} (May, 2015) [\href{http://arxiv.org/abs/1505.01481}{{\tt
  arXiv:1505.01481}}].

\bibitem{Scoccimarro:1998}
R.~{Scoccimarro}, {\it {Transients from initial conditions: a perturbative
  analysis}},  {\em Mon.Not.Roy.Astron.Soc.} {\bf 299} (Oct., 1998) 1097--1118,
  [\href{http://arxiv.org/abs/astro-ph/9711187}{{\tt astro-ph/9711187}}].

\end{thebibliography}\endgroup
\appendix
%%%%%%%%%%%%%%%%%%%%%%%%%%%%%%%%%%%%%%%%%%%%%%%%%%%%%%%%%%%%%%%
%%%%%%%%%%%%%%%%%%%%%%%%%%%%%%%%%%%%%%%%%%%%%%%%%%%%%%%%%%%%%%%

%%%%%%%%%%%%%%%%%%%%%%%%%%%%%%%%%%%%%%%%%%%%%%%%%%%%%%%%%%%%%%%
%%%%%%%%%%%%%%%%%%%%%%%%%%%%%%%%%%%%%%%%%%%%%%%%%%%%%%%%%%%%%%%
\section{Numerical tests}
\label{app:numerics}
%%%%%%%%%%%%%%%%%%%%%%%%%%%%%%%%%%%%%%%%%%%%%%%%%%%%%%%%%%%%%%%
%%%%%%%%%%%%%%%%%%%%%%%%%%%%%%%%%%%%%%%%%%%%%%%%%%%%%%%%%%%%%%%

\begin{figure}[t]
\centering
\includegraphics[width=0.49\textwidth]{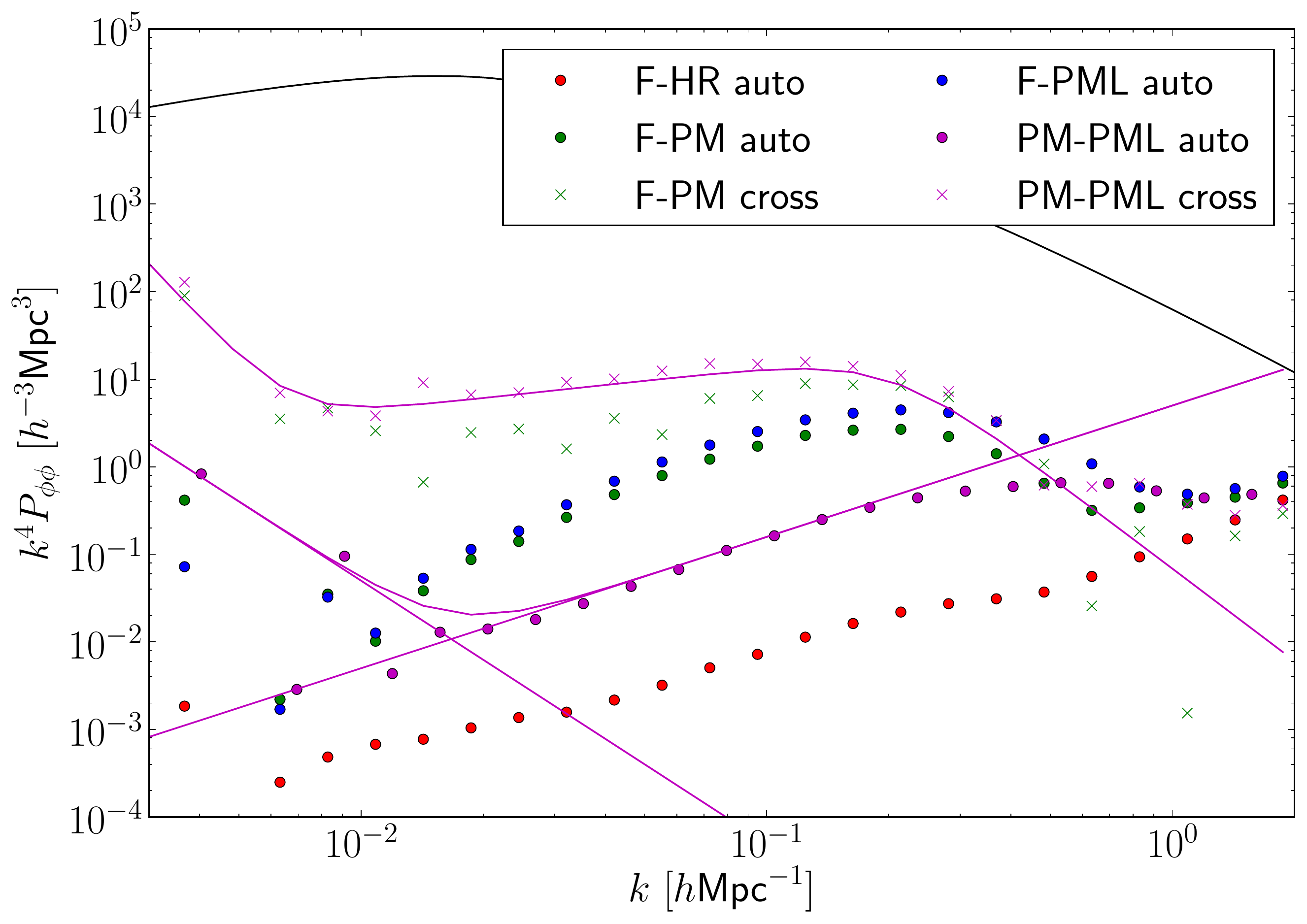}
\includegraphics[width=0.49\textwidth]{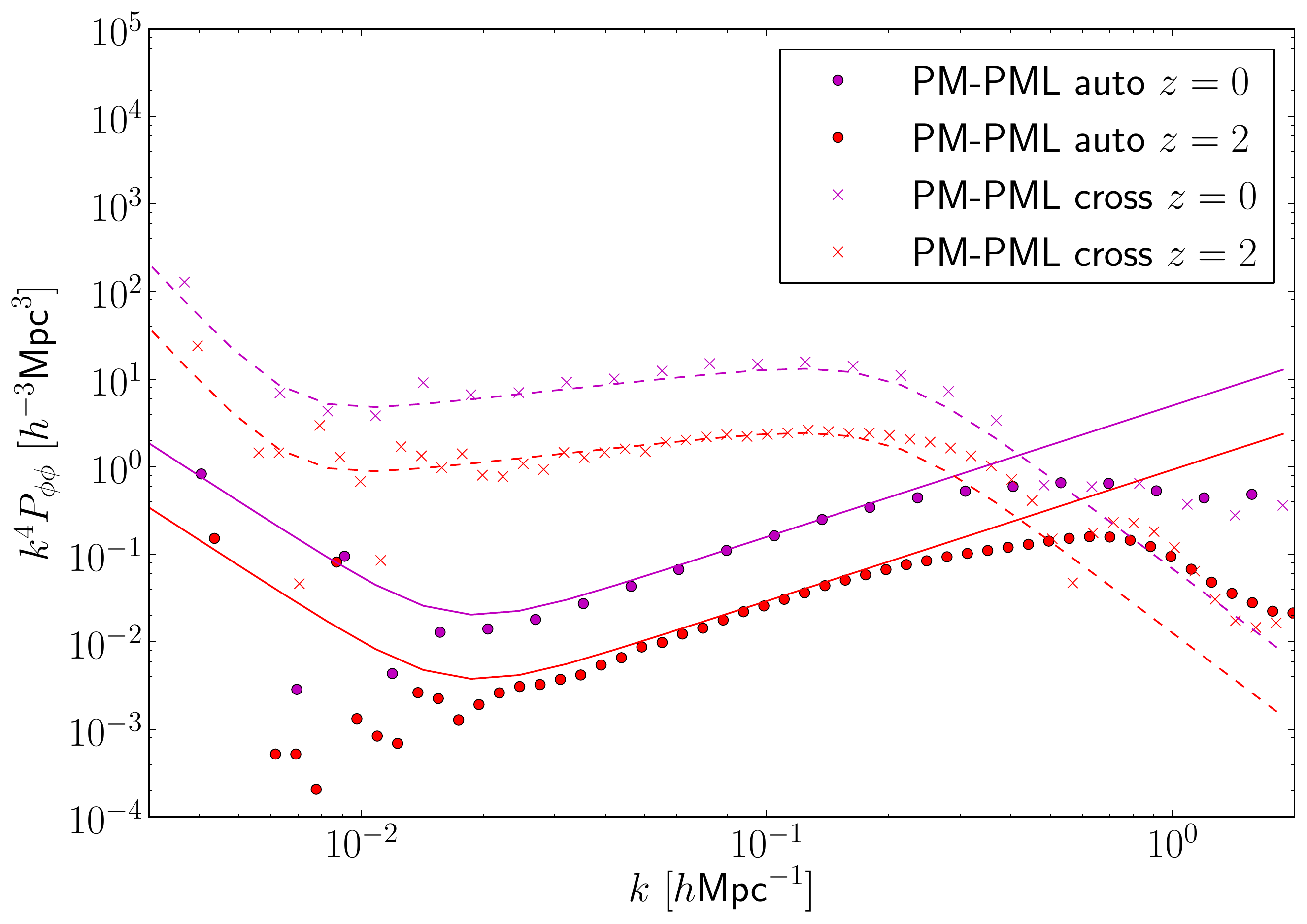}
\caption{Power spectrum of the difference displacement divergence between simulations with variant parameters. The solid points show the auto power of the error, while the crosses show the cross correlation between the error and one of the displacement divergencies.
\emph{Right panel:} Time dependence of the error from measurements at $z=0$ and $z=2$. Both errors on the auto and the cross power scale as $D^2$.}
\label{fig:errors}
\end{figure}

We estimate the systematic error inherent to our simulations by varying several parameters of the Gadget $N$-body simulation and study the deviations between these simulations and our fiducial case. The corresponding systematic error on the power spectrum is shown in Fig.~\ref{fig:errors}. In particular we estimate the cross power and auto power of the differences between two simulations run with parameter choices A and B
\begin{align}
P_{A-B \times B}=\left \langle (\phi_A-\phi_B)| \phi_B\right\rangle\; , &&
P_{A-B}=\left \langle (\phi_A-\phi_B)^2\right\rangle\; .
\end{align}
It turns out that the results are very sensitive to the number of grid cells that Gadget uses to calculate the long range potential. This parameter is called PMGRID and has to be set in the Gadget makefile. We had initially set $N_\text{PM,F}=N_\text{c}$, which we shall refer to as the fiducial case. We also changed this parameter to $N_\text{PM,PM}=2N_\text{c}$ or $N_\text{PM,PML}=3/4N_\text{c}$. As was noted previously in \cite{Smith:2012uz}, the PMGRID parameter has a big and non-monotonic effect on the matter density power spectrum. As shown in Fig.~\ref{fig:errors}, we confirm a similar behavior for the displacement divergence. In particular we find that the auto power of the difference between the PM and PML cases is smaller than the respective errors between the F and PM/PML cases. We also considered a simulation with an improved time stepping and error tolerance (denoted HR in the Figure) and find a considerably smaller impact on the displacement power spectrum.
We also consider the time dependence of the error power spectra and find that both the auto and cross power spectra scale roughly as $D^2$.
We find fitting functions that roughly reproduce the shape and amplitude of the errors. These fitting functions are used in the main text to assess the systematic error on the measurements of the EFT coefficients and the stochastic term. In particular, the cross power of the error and the field $P_{A-B \times B}$ will affect the transfer functions and the EFT coefficients, while the auto power $P_{A-B \times A-B}$ affects the stochastic term. 
%

%%%%%%%%%%%%%%%%%%%%%%%%%%%%%%%%%%%%%%%%%%%%%%%%%%%%%%%%%%%%%%%
%%%%%%%%%%%%%%%%%%%%%%%%%%%%%%%%%%%%%%%%%%%%%%%%%%%%%%%%%%%%%%%
\section{Perturbative Displacement Fields}
\label{sec:appendix_lpt}
%%%%%%%%%%%%%%%%%%%%%%%%%%%%%%%%%%%%%%%%%%%%%%%%%%%%%%%%%%%%%%%
%%%%%%%%%%%%%%%%%%%%%%%%%%%%%%%%%%%%%%%%%%%%%%%%%%%%%%%%%%%%%%%
In this Appendix we will rederive the equations for the scalar and vector components of the displacement, as they will be needed to calculate the higher order solutions in presence of EFT counterterms in App.~\ref{app:ctr}.

\paragraph*{Scalar component of the displacement:}
For the scalar equation we take the Eulerian derivative of the EoM, yielding
\beq
\frac{\partial}{\partial x_i}\left(\ddot \psi_i+\confh \dot \psi_i\right)=-\Delta \varphi=-\frac{3}{2}\confh^2 \Omega_\text{m}\delta
\eeq
From the mapping between Lagrangian and Eulerian space $\vec x=\vec q+\vec \psi$, we have that $\delta=1/J-1$, where $J$ is the determinant of the Jacobian matrix
\beq
A_{ij}\equiv\frac{\partial x_i}{\partial q_j}=\delta^\text{(K)}_{ij}+\psi_{i,j}
\eeq
If not explicitly mentioned otherwise, partial derivatives are with respect to the Lagrangian coordinate $q$. 
The determinant can then be related to the invariants of the displacement field
\begin{align}
J=\text{Det}[A]=&\frac{1}{3!}\epsilon_{ijl}\epsilon_{stu}A_{is}A_{jt}A_{lu}\nonumber\\
=&1+\psi_{i,i}+\frac12 \left(\psi_{i,i}\psi_{j,j}-\psi_{i,j}\psi_{j,i}\right)+\frac{1}{3!}\epsilon_{ijl}\epsilon_{stu}\psi_{i,s}\psi_{j,t}\psi_{l,u}\label{eq:determinant}\\
=&1+\mathcal{K}+\mathcal{L}+\mathcal{M}\nonumber
\end{align}
We will need the mapping from Eulerian to Lagrangian derivatives
\beq
\frac{\partial}{\partial x_j}=A_{ij}^{-1}\frac{\partial}{\partial q_i}
\eeq
with
\begin{align}
A_{ij}^{-1}=\frac{C_A^\text{T}}{\text{Det}[A]}=&\frac{1}{2!\; \text{Det}[A]}\epsilon_{jmn}\epsilon_{ist}A_{ms}A_{nt}\\
=&\frac{1}{J}\biggl(\delta^\text{(K)}_{ij}+\frac{1}{2}(\delta^\text{(K)}_{ij}\delta^\text{(K)}_{nt}-\delta^\text{(K)}_{jt}\delta^\text{(K)}_{in})\psi_{n,t}+\frac{1}{2}\epsilon_{jmn}\epsilon_{ist}\psi_{m,s}\psi_{n,t}\biggr)
\end{align}
For the EoM we now have
\beq
J A_{ji}^{-1}\left(\ddot \psi_{i,j}+\confh \dot \psi_{i,j}\right)=\frac{3}{2}\Omega_\text{m}\mathcal{H}^2\left(\mathcal{K}+\mathcal{L}+\mathcal{M}\right)
\eeq
Let us define the time derivative operator
\beq
\mathcal{D}=\frac{\derd^2}{\derd (\ln a)^2}+\frac 12 \frac{\derd }{\derd \ln a}\; .
\eeq
We can further simplify this equation by going to an EdS Universe and keeping only the $n$-th order displacement    field on the left hand side
\beq
\begin{split}
\mathcal{D} \psi_{i,i}^{(n)}=&\frac{3}{2}\left(\mathcal{K}^{(n)}+\mathcal{L}^{(n)}+\mathcal{M}^{(n)}\right)-2\frac{1}{2}\sum_{m=1}^{n-1}\Bigl[\psi_{i,i}^{(n-m)}\mathcal{D}\psi_{j,j}^{(m)}-\psi_{i,j}^{(n-m)}\mathcal{D}\psi_{j,i}^{(m)}\Bigr]\\
&-3\sum_{m=1}^{n-2}\sum_{k+l=m-n\atop{k>0,l>0}}\frac{1}{3!}\epsilon_{ijl} \epsilon_{stu} \psi_{i,s}^{(k)} \psi_{j,t}^{(l)}\mathcal{D} \psi_{l,u}^{(m)}
\label{eq:eomderivs}
\end{split}
\eeq
Using that $\mathcal{D} a^m=(m^2+m/2)a^m$ we have
\beq
\begin{split}
\left(n^2+\frac{n}{2}\right) \psi_{i,i}^{(n)}=&\frac{3}{2}\left(\mathcal{K}^{(n)}+\mathcal{L}^{(n)}+\mathcal{M}^{(n)}\right)-2\sum_{m=1}^{n-1}\left(m^2+\frac{m}{2}\right)\mathcal{L}^{n-m,m}\\
&-3\sum_{m=1}^{n-2}\sum_{k+l=m-n\atop{k>0,l>0}}\left(m^2+\frac{m}{2}\right)\mathcal{M}^{k,l,m}\; .
\end{split}
\eeq
As a last step we can symmetrize the right hand side and collect the invariants $\mathcal{L}$ and $\mathcal{M}$
\beq
\begin{split}
\psi_{i,i}^{(n)}=&
-\sum_{n_1+n_2=n\atop{n_1,n_2>0}}\left(\frac{2(n_1^2+n_2^2)+n-3}{(n-1) (2 n+3)}\right)\mathcal{L}^{n_1,n_2}\\&-\sum_{n_1+n_2+n_3=n\atop{n_1,n_2,n_3>0}}\left(\frac{2 (n_1^2+n_2^2+n_3^2)+n-3}{(n-1) (2 n+3)}\right)\mathcal{M}^{n_1,n_2,n_3}\\
=&-\sum_{n_1+n_2=n\atop{n_1,n_2>0}}\left(\frac{2(n_1^2+n_2^2)+n-3}{(n-1) (2 n+3)}\right)\frac{1}{2}\epsilon_{mis}\epsilon_{mjt}\psi_{i,j}^{(n_1)}\psi_{s,t}^{(n_2)}\\&-\sum_{n_1+n_2+n_3=n\atop{n_1,n_2,n_3>0}}\left(\frac{2 (n_1^2+n_2^2+n_3^2)+n-3}{(n-1) (2 n+3)}\right)\frac{1}{3!}\epsilon_{i_1i_2i_3}\epsilon_{j_1j_2j_3}\psi_{i_1,j_1}^{(n_1)} \psi_{i_2,j_2}^{(n_2)} \psi_{i_3,j_3}^{(n_3)}
\; .
\label{eq:rec_lpt_scalar_real}
\end{split}
\eeq
In the second part of the above equation, we have restored the full index structure, which will become useful for writing the recursion relations in Fourier space later on.
%

%%%%%%%%%%%%%%%%%%%%%%%%%%%%%%%%%%%%%%%%%%%%%%%%%%%%%%%%%%%%%%%%%%%%%%%%%%

\paragraph*{Vector component of the displacement:}
For the curl part of the displacement field, we start again from the equation of motion:
\beq
\ddot{x}_i + \mathcal{H} \dot{x}_i = -\frac{\partial \Phi}{\partial x_i} .
\eeq
The Eulerian gradient on the r.h.s. can be converted to a Lagrangian gradient by multiplying the equation by the Jacobian matrix $A_{ij} = \frac{\partial x_i}{\partial q_j}$:
\beq
\frac{\partial{x_i}}{\partial q_j}
\left( \ddot{x}_i + \mathcal{H} \dot{x}_i \right)
= -\frac{\partial \Phi}{\partial q_j} .
\eeq
With integration by parts with respect to time and space, this equation can be reexpressed as:
\beq
\left( \frac{\partial}{\partial \tau} +\mathcal{H} \right) 
\frac{\partial{x_i}}{\partial q_j} \dot{x}_i
=
\frac{\partial}{\partial q_j} 
\left( \frac{| \dot{\vec x} |^2}{2} - \Phi \right) .
\eeq
The point of this transformation is that the r.h.s is now a pure Lagrangian gradient, and therefore its Lagrangian curl vanishes:
\beq
\epsilon_{lmj} \frac{\partial }{\partial q_m}
\left( \frac{\partial}{\partial \tau} +\mathcal{H} \right) 
\frac{\partial{x_i}}{\partial q_j} \dot{x}_i
=
0 ,
\eeq
i.e.:
\beq
\left( \frac{\partial}{\partial \tau} +\mathcal{H} \right) 
\epsilon_{lmj}
\frac{\partial{x_i}}{\partial q_j} 
\frac{\partial \dot{x}_i }{\partial q_m}
=
0 .
\eeq
This differential equation in time is readily integrated, since the initial condition vanishes:
\beq
\epsilon_{lmj}
\frac{\partial{x_i}}{\partial q_j} 
\frac{\partial \dot{x}_i }{\partial q_m}
=
0,
\eeq
which can be rewritten in vector notation as:
\beq
\vec{\nabla}_q \dot{x}_i
\times
\vec{\nabla}_q x_i
=
\vec{0}.
\eeq
Substituting $\vec{x} = \vec{q} + \vec{\psi}$ then leads to:
\beq
\vec{\nabla}_q \times \dot{\vec{\psi}}
=
\vec{\nabla}_q  \psi_i 
\times
\vec{\nabla}_q  \dot{\psi}_i .
\eeq
Assuming that the time-dependence of the displacement follows that for an EdS universe, this yields the configuration space recursion relation for the LPT curl part:
\beq
\vec{\nabla}_q \times \vec \psi^{(n)}
=
\sum_{m=1}^{n-1} \frac{n-2m}{2n} 
\vec{\nabla}_q \psi^{(m)}_i \times \vec{\nabla}_q \psi^{(n-m)}_i \, .
\label{eq:rec_lpt_vector_real}
\eeq

%%%%%%%%%%%%%%%%%%%%%%%%%%%%%%%%%%%%%%%%%%%%%%%%%%%%%%%%%%%%%%%%%%%%%%%%%%
% recursion relations for the LPT kernels
\paragraph*{Recursion relations in Fourier space:}
Finally, we deduce the Fourier space recursion relations for the kernels of the scalar $l_{n}$ and vector part $\vec{t}_n$ of the displacement. Both are sourced by the total displacement $\vec{S}_{n} \equiv i\vk L_{n} + i\vk \times \vec{T}_n$, which leads to a coupling between scalar and vector modes starting from third order.\\
The scalar kernels $L_{n},\, n>1$ are readily obtained by Fourier transforming Eq.~\eqref{eq:rec_lpt_scalar_real}:
\beq
\begin{split}
L_{n}(\vp_1, \ldots, \vp_n)
=&-\sum_{n_1+n_2=n\atop{n_1,n_2>0}}\left(\frac{2(n_1^2+n_2^2)+n-3}{2(n-1) (2 n+3)}\right)
\frac{n!}{n_1! n_2!} \times \\
&
\frac{\Bigl[ \vp|_1^{n_1} \times \vp|_{n_1+1}^{n_1+n_2} \Bigr]}{\bigl[p|_1^n\bigr]^2}
\cdot
\Bigl[ \vec{S}_{n_1}(\vp_1, \ldots, \vp_{n_1}) \times \vec{S}_{n_2}(\vp_{n_1+1},\ldots,\vp_{n_1+n_2}) \Bigr] \\
&-i\sum_{n_1+n_2+n_3=n\atop{n_1,n_2,n_3>0}}\left(\frac{2 (n_1^2+n_2^2+n_3^2)+n-3}{6 (n-1) (2 n+3)}\right)
\frac{n!}{n_1! n_2! n_3!}  \frac{\text{Det} \left[ \vp|_1^{n_1}, \vp|_{n_1+1}^{n_1+n_2}, \vp|_{n_1+n_2+1}^{n_1+n_2+n_3} \right]}{\bigl[p|_1^n\bigr]^2}
\times \\
& 
\text{Det} \Bigl[ 
\vec{S}_{n_1}(\vp_1, \ldots, \vq_{n_1}),
\vec{S}_{n_2}(\vp_{n_1+1}, \ldots, \vp_{n_1+n_2}),
\vec{S}_{n_3} (\vp_{n_1+n_2+1}, \ldots , \vp_{n_1+n_2+n_3})
\Bigr]
 \;.
\label{eq:rec_lpt_scalar}
\end{split}
\eeq
with starting conditions $L_1=1/k^2$ and $T_1=T_2=0$. Here, $\text{Det}[\vec a,\vec b,\vec c]=\vec a \cdot(\vec b \times \vec c)$ stands for the determinant of the matrix, whose columns are the vectors $\vec a,\ \vec b$ and $\vec c$ and we defined $\vec p|_1^m\equiv p_1+\ldots+p_m$.
For the vector kernel $\vec{T}_n$, we get from Eq.~\eqref{eq:rec_lpt_vector_real}
\beq
 \vec{T}_{n}(\vp_1, \ldots, \vp_n)
=
-\sum_{m=1}^{n-1} \frac{n-2m}{2n} 
\frac{n!}{n_1! n_2!}
\frac{\Bigl[ \vp|_{1}^{n_1} \times \vp|_{n_1+1}^{n_2} \Bigr]}{\left[p|_1^n\right]^2}
\Bigl[ \vec{S}_{n_1}(\vp_1, \ldots, \vp_{n_1}) \cdot \vec{S}_{n_2}(\vp_{n_1+1},\ldots,\vp_{n_1+n_2}) \Bigr] \, .
\label{eq:rec_lpt_vector}
\eeq
Similar relations were recently given by \cite{Matsubara:2015re}.\footnote{These expressions correspond to Eqs.~(67--69) in \cite{Matsubara:2015re} once the following mapping has been used 
$\vec L_{n,\text{Matsubara}}=\vec S_n$, $S_{n,\text{Matsubara}}=i k^2 L_n$ and $\vec T_{n,\text{Matsubara}}=i k^2 \vec T_n$, where $\vec k=\vec p|_1^n$ is the sum of the momenta in the kernel.}
For the purely scalar part Eq.~\ref{eq:rec_lpt_scalar} simplifies to\footnote{Here we are using that
\[\text{Det}[\vp_1,\vp_2,\vp_3]^2=\text{Det}\begin{pmatrix}\vp_1 \cdot \vp_1& \vp_1 \cdot \vp_2 & \vp_1 \cdot \vp_3\\\vp_1 \cdot \vp_2& \vp_2 \cdot \vp_2 & \vp_2 \cdot \vp_3\\
\vp_1 \cdot \vp_3& \vp_2 \cdot \vp_3 & \vp_3 \cdot \vp_3\\
\end{pmatrix}=p_1^2 p_2^2 p_3^2 \kappa_3(\vp_1,\vp_2,\vp_3)\]
}
\beq
\begin{split}
L_n(\vp_1, \ldots, \vp_n)=&+\sum_{n_1+n_2=n\atop{n_1,n_2>0}}\left(\frac{2(n_1^2+n_2^2)+n-3}{2(n-1) (2 n+3)}\right)\frac{n!}{n_1! n_2!}\frac{\bigl[\vec p|_1^{n_1}\bigr]^2\bigl[\vec p|_{n_1+1}^{n}\bigr]^2}{\bigl[\vec p|_1^{n}\bigr]^2}\kappa_2(\vec p|_1^{n_1},\vec p|_{n_1+1}^{n})\\
&L_{n_1}(\vp_1,\ldots,\vp_{n_1})L_{n_2}(\vp_{n_1+1},\ldots,\vp_n)\\
-&\sum_{n_1+n_2+n_3=n\atop{n_1,n_2,n_3>0}}\left(\frac{2 (n_1^2+n_2^2+n_3^2)+n-3}{6 (n-1) (2 n+3)}\right)\frac{n!}{n_1! n_2!n_3!}\kappa_3(\vec p|_1^{n_1},\vec p|_{n_1+1}^{n_1+n_2},\vec p|_{n_1+n_2+1}^{n})\\
&\frac{\bigl[\vec p|_1^{n_1}\bigr]^2\bigl[\vec p|_{n_1+1}^{n_1+n_2}\bigr]^2\bigl[\vec p|_{n_1+n_2+1}^{n}\bigr]^2}{\bigl[\vec p|_1^{n}\bigr]^2}L_{n_1}(\vp_1,\ldots,\vp_{n_1}) L_{n_2}(\vp_{n_1+1},\ldots,\vp_{n_1+n_2})\\
& L_{n_3}(\vp_{n_1+n_2+1},\ldots,\vp_{n})
\end{split}
\eeq
This equation can be used to calculate the full scalar displacement up to third order and the dominant part of the fourth order displacement.\\

%%%%%%%%%%%%%%%%%%%%%%%%%%%%%%%%%%%%%%%%%%%%%%%%%%%%%%%%%%%%%%%%%%%%%%%%%%

In order to validate the LPT code on the simulation grid, we compare it to the analytical 1-loop and 2-loop predictions. They can be computed using the following explicit relations corresponding to the diagrams in Fig.~\ref{fig:diagrams_1loop_2loop}
\beq
\begin{split}
P_{11}(k) =& \frac{P_\text{lin}(k)}{k^4} \\
P_{22}(k) =&\frac{1}{2} \int \frac{\derd^3\vp}{(2\pi)^3}\left[L_2(\vk-\vp, \vp)\right]^2 P_\text{lin}(q) P_\text{lin}(|\vk-\vp|)\\
=& \frac{9}{98 (2\pi)^2} \frac{1}{k}  
\int_0^\infty \derd x x^2 P_\text{lin}(kx) 
\int_{-1}^1 \derd\mu P_\text{lin}(k \sqrt{1+x^2 - 2x\mu})
\left(  \frac{1-\mu^2}{1+x^2 - 2x\mu}  \right)^2  \\
P_{13}(k) =& \frac{1}{2}\frac{P_\text{lin}(k)}{k^2}
\int \frac{\derd^3\vp}{(2\pi)^3} L_3(\vp, -\vp, \vk) P_\text{lin}(q)\\
=& \frac{5}{2016 (2\pi)^2}
 \frac{P_\text{lin}(k)}{k}
\int_0^\infty \derd x  \frac{ P_\text{lin}(kx) }{x^3}
[
4x(-3 + 11x^2 + 11x^4 -3x^6)
+3 (1-x^2)^4 \ln \left[ \left(  \frac{1+x}{1-x}  \right)^2 \right]
] \\
P_{24}(k) =& \frac{1}{4}
\int \frac{\derd^3\vp_1 \derd^3\vp_2}{(2\pi)^6}
L_2(-\vp_1, \vp_1-\vk) L_4(\vk-\vp_1, \vp_1, \vp_2, -\vp_2)
P_\text{lin}(p_1) P_\text{lin}(p_2) P_\text{lin}(\vk-\vp_1) \\
P_{33-I}(k) =&\frac{1}{6} 
\int \frac{\derd^3\vp_1 \derd^3\vp_2}{(2\pi)^6}
\left[ L_3(\vp_1, \vp_2, \vk-\vp_1-\vp_2)\right]^2
P_\text{lin}(p_1) P_\text{lin}(p_2) P_\text{lin}(|\vk -\vp_1-\vp_2|) \\
P_{33-II}(k) =& \frac{\left(P_{13}(k)/2 \right)^2}{P^{11}(k)}\\
P_{15}(k) =& \frac{1}{8} \frac{P_\text{lin}(k)}{k^2}
\int \frac{\derd^3\vp_1 \derd^3\vp_2}{(2\pi)^6}
L_5(\vp_1, -\vp_1, \vp_2, -\vp_2, \vk) 
P_\text{lin}(p_1) P_\text{lin}(p_2) \\
P_{3c3c}=&\frac{1}{6} 
\int \frac{\derd^3\vp_1 \derd^3\vp_2}{(2\pi)^6}
\left[ \vec T_3(\vp_1, \vp_2, \vk-\vp_1-\vp_2)\right]^2
P_\text{lin}(p_1) P_\text{lin}(p_2) P_\text{lin}(|\vk -\vp_1-\vp_2|) 
\end{split}
\eeq

\begin{figure}[H]
\centering
\includegraphics[width=10cm]{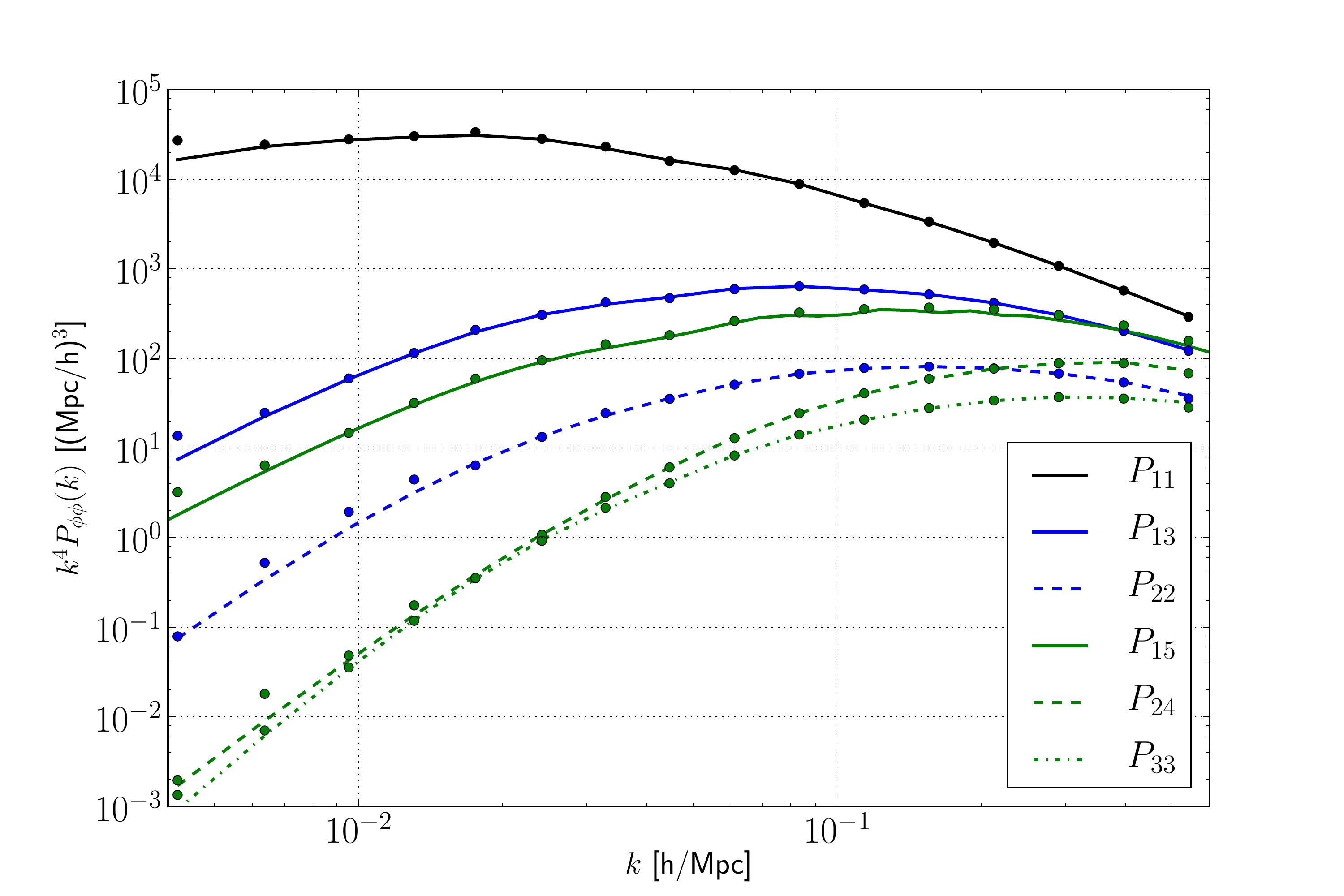}
\caption{Comparison between the LPT on the grid and the analytical one, for the linear power spectrum, 1-loop terms, and some 2-loop terms. This is a test for our expressions of $\phi^{(1)}$, $\phi^{(2)}$, $\phi^{(3)}$, $\phi^{(4)}$ and  $\phi^{(5)}$.}
\label{fig:grid_vs_theory}
\end{figure}

We find a good agreement between our code on the grid and the loop calculations, as seen in Fig.~\ref{fig:grid_vs_theory}.
For definiteness, we have for the one and two loop LPT contributions to the power spectrum
\beq
\begin{split}
P_\text{1-loop}&=2P_{13}+P_{22}\; ,\\
P_\text{2-loop}&=2P_{15}+2P_{24}+P_{33-I}+P_{33-II}\; .
\end{split}
\eeq

The calculation on the grid has effective cutoffs: the fundamental wave vector $k_\text{f} = (2\pi)/L_\text{box}$ is the minimum non-zero wave vector present, and if no explicit upper cutoff is used, the Nyquist wave vector $k_\text{Ny}$ is the highest wave vector present. In order to get agreement between the theory LPT and the calculation on the grid, one has to impose these same cutoffs to the loop integrals.
Since the LPT expansion is a non-linear calculation, aliasing is also an issue, as explained in the main text. Finally, reaching an agreement at the percent level between theory LPT and LPT on the grid requires binning the theory power spectra with the same bins as for the power spectrum estimation on the grid.

%%%%%%%%%%%%%%%%%%%%%%%%%%%%%%%%%%%%%%%%%%%%%%%%%%%%%%%%%%%%%%%
%%%%%%%%%%%%%%%%%%%%%%%%%%%%%%%%%%%%%%%%%%%%%%%%%%%%%%%%%%%%%%%
\section{Cosmic variance}
\label{sec:appendix_cosmic_variance}
%%%%%%%%%%%%%%%%%%%%%%%%%%%%%%%%%%%%%%%%%%%%%%%%%%%%%%%%%%%%%%%
%%%%%%%%%%%%%%%%%%%%%%%%%%%%%%%%%%%%%%%%%%%%%%%%%%%%%%%%%%%%%%%

In this appendix we estimate the uncertainties on our measurements due to cosmic variance, in order to show the difficulty of measuring the EFT coefficient at the level of the power spectrum (instead of the field itself), and to understand the scatter in our measurements of the cross power spectra between the displacement from simulation and from perturbation theory. 

The variance of our measurements can be estimated from their scatter across the various realizations.
On the other hand, these variances can be predicted as:
\beq
\text{Cov}\left( P_{ab}, P_{cd} \right) = \frac{1}{\Nmodes} \left[ P_{ac}P_{bd} + P_{ad}P_{bc} \right] + \frac{1}{V}T_{abcd},
\label{eq:sample_cov_pab}
\eeq
where $\Nmodes$ is the number of modes in each $k$-bin, $V$ is the volume of the box, and $T_{abcd}$ is the trispectrum. The first term is the Gaussian covariance, while the second term is the non-Gaussian term, which would vanish if the fields of interest were Gaussian. In what follows, we only evaluate the Gaussian covariance, although we will keep the trispectrum terms in the equations for completeness.

\begin{figure}[t]
\centering
\includegraphics[width=0.49\textwidth]{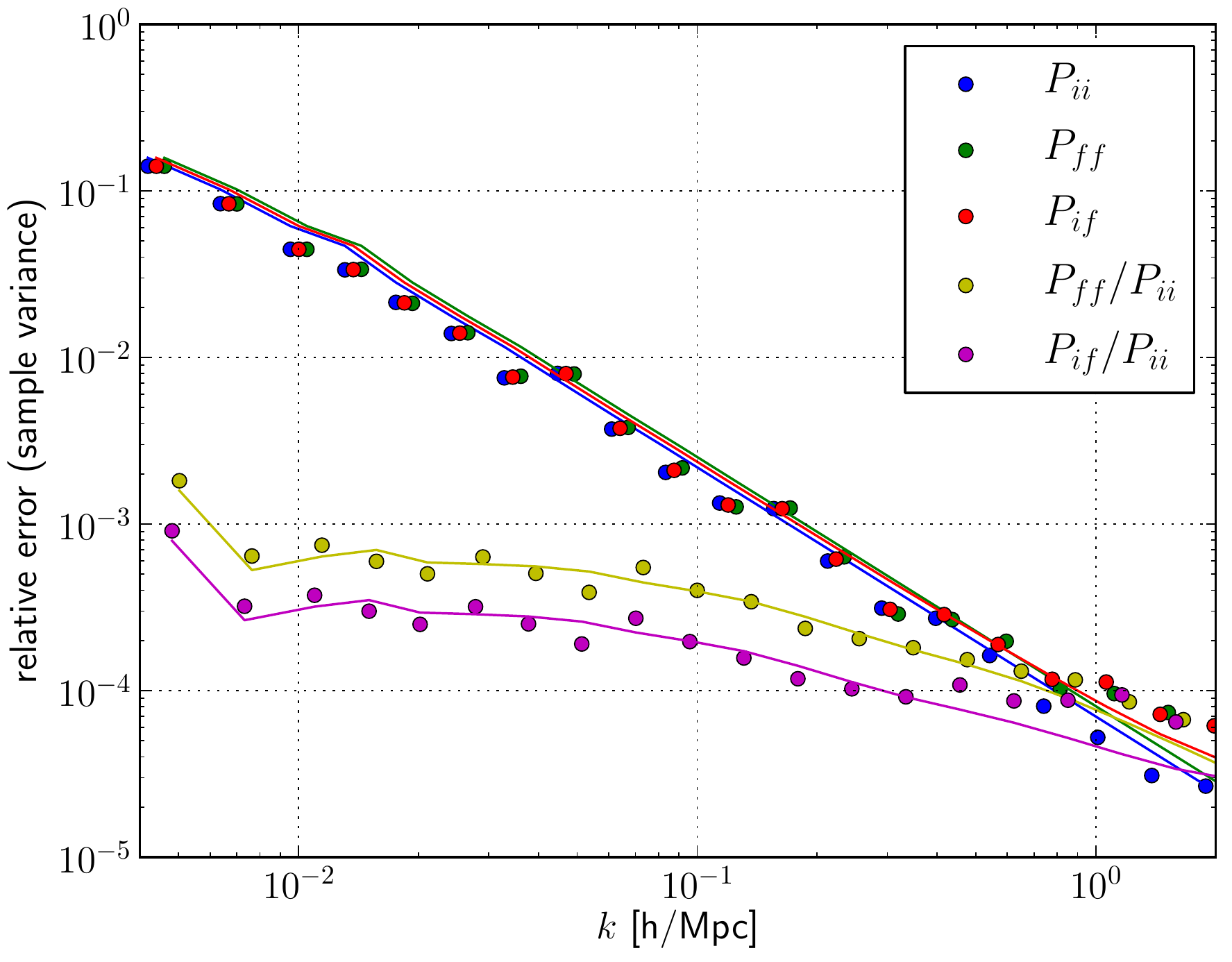}
\caption{Relative scatter of the various power spectra and ratios of power spectra, due to sample variance.  On large scales, considering ratios of power spectra cancels the sample variance by up to a factor of a hundred.}
\label{fig:errorP}
\centering
\includegraphics[width=0.49\textwidth]{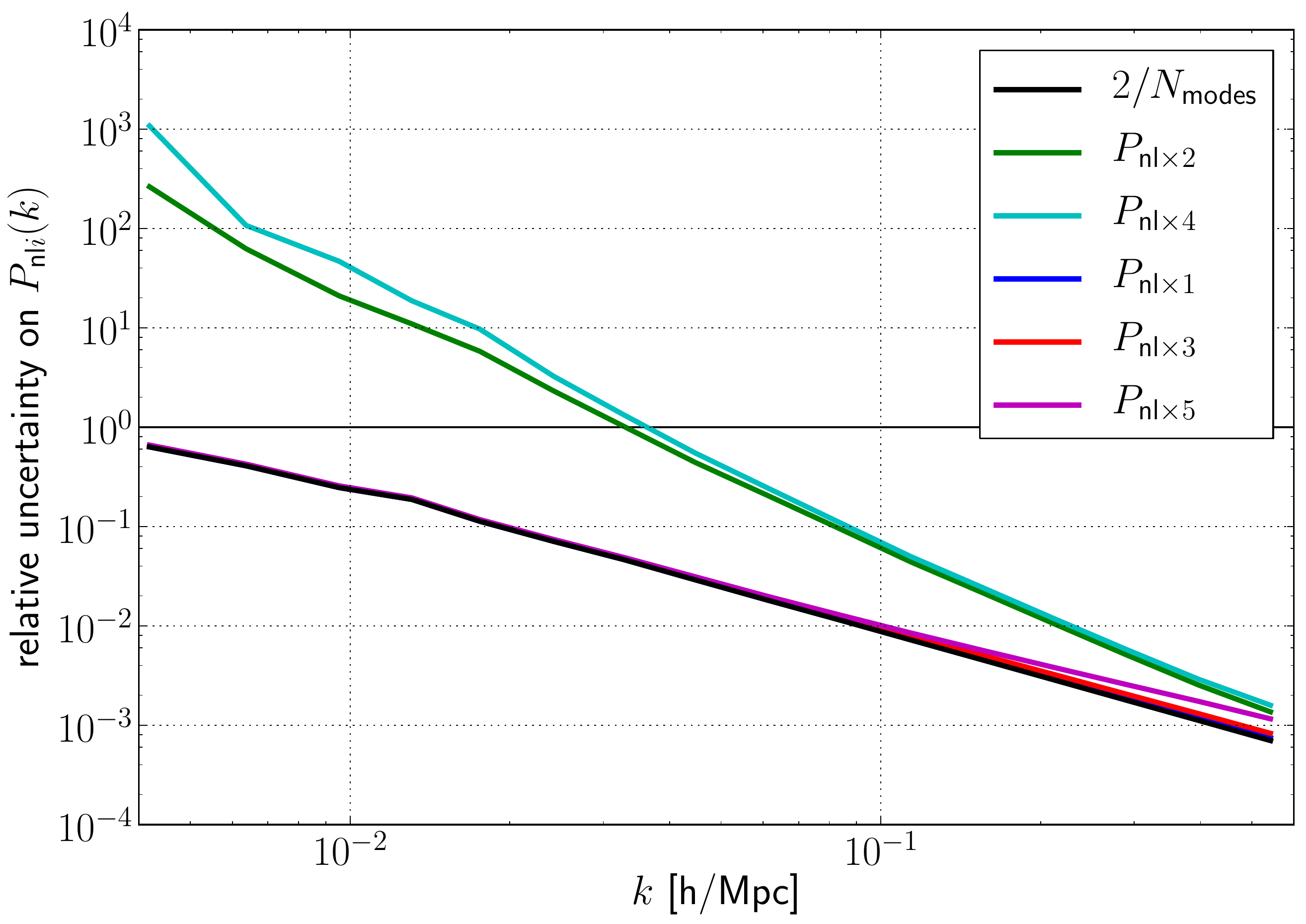}
\caption{Relative uncertainty on $P_{\text{nl}i}$ due to cosmic variance (Gaussian term only), showing that the relative uncertainty is much greater for $i$ even. This explains the scatter seen in the measurements of Fig.~\ref{fig:simu_cross_pt}.}
\label{fig:snr_simu_cross_pt}
\end{figure}

In particular:
\beq
\frac{ \sigma^2(P_{ab}) }{ P_{ab}^2 } = \frac{1}{\Nmodes} \left[ 1 + \frac{1}{r_{ab}^2} \right] + \frac{1}{V} \frac{T_{aabb}}{P_{ab}^2},
\label{eq:sample_var_pab}
\eeq
where $r_{ab}$ is the correlation coefficient $r_{ab} = \frac{P_{ab}}{\sqrt{P_{aa} P_{bb}}} \leqslant 1$. A simple consequence of Eq.~\eqref{eq:sample_var_pab} is that $\frac{ \sigma^2(P_{ab}) }{ P_{ab}^2 } \geqslant \frac{2}{\Nmodes}$. This is a lower bound on how well one can measure any power spectrum on large scales, and it corresponds to a relative error of $\sim 10\%$ on the largest scales of our simulations ($k\sim 5\times 10^{-3} \ihMpc$), as shown in Fig.~\ref{fig:errorP}. Such a high uncertainty would make it extremely difficult to measure the EFT coefficient $\alpha$ from matching the simulation power spectrum to a theory power spectrum on large scales.

This uncertainty can be significantly reduced by matching ratios of power spectra from the same simulation to theory predictions. This can be understood as follows. The uncertainty on a ratio of power spectra can be computed from Eq.~\eqref{eq:sample_cov_pab} as:
\beq
\frac{\sigma^2(P_{ab}/P_{cd})}{(P_{ab}/P_{cd})}
=
\frac{1}{N} \left[
2 + \frac{1}{r_{ab}^2} + \frac{1}{r_{cd}^2} - 2\frac{P_{ac}P_{bd} + P_{ad}P_{bc}}{P_{ab}P_{cd}}
\right]
+\frac{1}{V} \left[ 
\frac{T_{aabb}}{P_{ab}^2} + \frac{T_{ccdd}}{P_{cd}^2} - 2 \frac{T_{ancd}}{P_{ab} P_{cd}}
\right].
\eeq
Fig.~\ref{fig:errorP} shows that considering ratios of power spectra reduces the cosmic variance from $10\%$ to $0.05\%$ at $k\sim 5\times 10^{-3} \ihMpc$. Such an uncertainty would be marginally sufficient to allow our measurement of the EFT coefficient $\alpha$.

Another consequence of Eq.~\eqref{eq:sample_var_pab} is that the relative uncertainty on $P_{\text{nl} i}$ is much greater when $i$ is even, since $r_{\text{nl} i }$ is much lower for even values of $i$. Indeed, $r_{\text{nl} i } = \frac{P_{\text{nl}i}}{\sqrt{P_{\text{nl}\text{nl}} P_{ii}}} \simeq r_{1i}$ for $i$ odd, while $r_{\text{nl} i } \simeq \sqrt{\frac{P_{22}}{P_{11}}} r_{2i} \ll r_{2i}$ for $i$ even. This is shown in Fig.~\ref{fig:snr_simu_cross_pt}, and explains the scatter in the measurements seen in Fig.~\ref{fig:simu_cross_pt}.

%%%%%%%%%%%%%%%%%%%%%%%%%%%%%%%%%%%%%%%%%%%%%%%%%%%%%%%%%%%%%%%
%%%%%%%%%%%%%%%%%%%%%%%%%%%%%%%%%%%%%%%%%%%%%%%%%%%%%%%%%%%%%%%
\section{Transients}
\label{sec:transients}
%%%%%%%%%%%%%%%%%%%%%%%%%%%%%%%%%%%%%%%%%%%%%%%%%%%%%%%%%%%%%%%
%%%%%%%%%%%%%%%%%%%%%%%%%%%%%%%%%%%%%%%%%%%%%%%%%%%%%%%%%%%%%%%
In this Appendix, we consider the imprint of a finite starting redshift of the simulation on its late times results. These so called transients were studied in perturbation theory \cite{Scoccimarro:1998} and simulations \cite{Crocce:2006ve} at the level of 1LPT initial conditions. We will review the derivation and extend it to higher orders.\\
The equation of motion can be written in vector notation for $\chi=(\psi_{i,i},\eta_{i,i})=(\psi_{i,i},\derd \psi_{i,i}/\derd \ln a)$
\begin{align}
\frac{\derd \psi^{(n)}_{i,i}}{\derd \ln a}-\eta^{(n)}_{i,i}=&0\\
\frac{\derd \eta^{(n)}_{i,i}}{\derd \ln a}-\frac32\psi^{(n)}_{i,i}+\frac{1}{2}\eta^{(n)}_{i,i}=&S^{(n)}(a)
\end{align}
Here we already brought the $\mathcal{K}^{(n)}=\psi_{i,i}^{(n)}$ term arising from the determinant Eq.~\eqref{eq:determinant} to the left hand side. The source term is thus composed of the $\mathcal{L}^{(n)}$ and $\mathcal{M}^{(n)}$ terms in the determinant and the derivatives of the lower order terms from the left hand side.\\
The above equations can be summarized in vector notation as
\beq
\frac{\derd \chi_i}{\derd \ln a}+\Omega_{ij}\chi_j=S_i\; ,
\eeq
with the solution (see e.g. \cite{Scoccimarro:1998})
\beq
\chi_i(a)=g_{ij}(a,a_\text{i})\chi_j(a_\text{i})+\int_{\ln a_\text{i}}^{\ln a} \derd \ln a'\ g_{ij}(a,a') S_j(a') \; ,
\eeq
where 
\beq
g_{ij}(a,a')=\frac{a}{a'}\frac{1}{5}\begin{pmatrix}
 3& 2\\
3& 2
\end{pmatrix}+
\left(\frac{a}{a'}\right)^{-3/2}\frac{1}{5}\begin{pmatrix}
 2& -2\\
-3& 3
\end{pmatrix}
\eeq
Let us assume that we use Zel'dovich initial conditions, i.e., the 1LPT part is correct at all times, but the second and higher order solutions are zero until some initial time $a_\text{i}$. Then the second order source term is given by
\beq
S^{(2)}(a)=-\frac{3}{2}a^2 \mathcal{L}^{(2)}=-\frac{3}{4}a^2\left[\psi_{i,i}^{(1)}\psi_{j,j}^{(1)}-\psi_{i,j}^{(1)}\psi_{j,i}^{(1)}\right].
\eeq
Integrating over time with the initial condition $\chi^{(2)}(a_\text{i})=0$ we get
\begin{align}
\psi^{(2)}_{i,i}=&-\frac{3}{14}a^2\left(1 - \frac{7}{5}\frac{a_\text{i}}{a} + \frac{2}{5}\left(\frac{a_\text{i}}{a}\right)^{7/2}\right)\left[\psi_{i,i}^{(1)}\psi_{j,j}^{(1)}-\psi_{i,j}^{(1)}\psi_{j,i}^{(1)}\right]\\
=&\left(-\frac{3}{14} a^2 +\frac{3}{10} a a_\text{i} - \frac{3}{35} \frac{a_\text{i}^{7/2}}{a^{3/2}}\right)\left[\psi_{i,i}^{(1)}\psi_{j,j}^{(1)}-\psi_{i,j}^{(1)}\psi_{j,i}^{(1)}\right]
\end{align}
The first term in the brackets gives the fastest growing solution, whereas the second and third terms lead to corrections that decay as with increasing $a$.
For a starting redshift $z_\text{i}=99$, i.e. $a_\text{i}=0.01$ this corresponds to a $1.4\%$ correction to the second order displacement field at redshift $z=0$.

Let us now consider the case where we use 2LPT to set up initial conditions at $a_\text{i}$, i.e., first and second order displacement fields are correct at all times and there are transient effects on the third and higher orders. This is the case that is relevant for our suite of simulations. The third order source term is correctly given by the growing modes
\beq
S^{(3)}(a)=a^3\bigl[-3 \mathcal{M}^{1,1,1}-10\mathcal{L}^{1,2}\bigr]\; ,
\eeq
which using $\psi_{i}^{(3)}(a<a_\text{i})=0$ readily leads to the following solution for $a>a_\text{i}$
\begin{align}
\psi_{i,i}^{(3)}=&\frac{a^3}{9}\left(1 - \frac{9}{5} \left(\frac{a_\text{i}}{a}\right)^2 + \frac{4}{5}\left(\frac{a_\text{i}}{a}\right)^{9/2}\right)\bigl[-3 \mathcal{M}^{1,1,1}-10\mathcal{L}^{1,2}\bigr]\; .
\label{eq:thirdordertrans}
\end{align}
Here the first term in brackets gives the fastest growing mode and the corrections amount to $2 \times 10^{-4}$ at $z=0$ for $z_\text{i}=99$.\\
The fourth order source term has parts that are correctly predicted by LPT, but also decaying mode corrections that lead to corrections in the source term. It can be readily obtained by using the solution for the third order field including transients Eq.~\eqref{eq:thirdordertrans} in Eq.~\eqref{eq:eomderivs}
\beq
S^{(4)}(a)=a^4\biggl[-\frac{39}{2} \mathcal{M}^{1,1,2}-\frac{17}{2}\mathcal{L}^{2,2}-21\mathcal{L}^{1,3}\biggr]-2a^4\left( - \frac{9}{5} \frac{3}{2}\left(\frac{a_\text{i}}{a}\right)^2 + \frac{4}{5}\frac{3}{2}\left(\frac{a_\text{i}}{a}\right)^{9/2}\right)\mathcal{L}^{1,3}\; .
\eeq
Thus we have for the fourth order solution including transients
\beq
\begin{split}
\psi_{i,i}^{(4)}=&a^4\left[1-\frac{11}{5}\left(\frac{a_\text{i}}{a}\right)^3+\frac{6}{5}\left(\frac{a_\text{i}}{a}\right)^{11/2}\right]\biggl[-\frac{13}{11} \mathcal{M}^{1,1,2}-\frac{17}{33}\mathcal{L}^{2,2}-2\frac{7}{11}\mathcal{L}^{1,3}\biggr]\\
&-2a^4\left[-\frac{27}{35}\left(\frac{a_\text{i}}{a}\right)^2+\frac{7}{25}\left(\frac{a_\text{i}}{a}\right)^3-\frac{4}{5}\left(\frac{a_\text{i}}{a}\right)^{9/2}+\frac{6}{35}\left(\frac{a_\text{i}}{a}\right)^{11/2}\right]\mathcal{L}^{1,3}\; .
\end{split}
\eeq
The relative error for $\mathcal{L}^{2,2}$ and $\mathcal{M}^{1,1,2}$ is $2\times 10^{-6}$ and the relative error on $\mathcal{L}^{1,3}$ is $10^{-4}$ at $z=0$ for $z_\text{i}=99$. The enhanced error on $\mathcal{L}^{1,3}$ is due to the presence of transients in the third order contributions to the source term. This term decays slower than the transients on the other components of the fourth order displacement field by one power of the expansion factor $a$.

\section{Solution with Source Terms}\label{app:ctr}
Let us explicitly consider the time dependence of the EFT counterterms that arises from a given time dependence of the source term on the right hand side of the equations of motion (EoM). For convenience, we factor out $3/2\Omega_\text{m}\mathcal{H}^2$ from the source terms, such that they appear similar to the gravitational potential and such that that the EFT parameter has units of length$^2$ even at the EoM level
\beq
\ddot{\phi}^{(\tilde 1)}(\vec k)+\mathcal{H}\dot{\phi}^{(\tilde 1)} (\vec k) -\frac{3}{2}\Omega_\text{m}\mathcal{H}^2\phi^{(\tilde 1)}(\vec k)=\frac{3}{2}\Omega_\text{m}\mathcal{H}^2 \beta k^2 \phi^{(1)}(\vec k)\; .
\eeq
From now on we will work in Einstein-de-Sitter, where $\Omega_\text{m}=1$ and postulate a power law time dependence for the counterterms in the equation of motion, for instance $\beta=\beta_0 a^\gamma$. It is now convenient to rewrite the equations using $\ln a$ as the time parameter:
\beq
\mathcal{D}\phi^{(\tilde 1)}(\vec k,a)=\frac{3}{2} \beta_0 a^\gamma k^2 \phi^{(1)}(\vec k,a)
\text{  with } \mathcal{D}\equiv\biggl[\frac{\derd^2 }{\derd \ln a^2}+\frac12 \frac{\derd}{\derd \ln a}-\frac{3}{2}\biggr]\, .
\eeq
Note that here we redefined the operator $\mathcal{D}$, to also contain the full homogeneous part.
The Green's function associated with $\mathcal{D}$ is given by
\beq
G(a,\tilde a)=\frac{2}{5}\Biggl[\frac{a}{\tilde a}-\left(\frac{\tilde a}{a}\right)^{3/2}\Biggr]\, .
\eeq
This yields for the integration of a power law source
\beq
\int_0^a \derd \ln \tilde a\; G(a,\tilde a)\tilde{a}^w=\frac{2}{(w-1)(3+2w)}a^w
\eeq
and finally allows us to relate the EFT coefficient $\alpha$ of the solution to the one of the equation of motion
\beq
\phi^{(\tilde 1)}(\vec k,a)=\alpha(a) k^2 \phi^{(1)}(\vec k,a)\, \text{  with } \alpha(a) \equiv  \frac{3}{\gamma(5+2\gamma)}\beta_0a^\gamma .
\eeq
Let us now solve the equation at next-to-leading order in the counterterms, but neglecting the explicit second order source terms $E_{2,i}$ for the time being. 
\beq
\mathcal{D}\phi^{(\tilde 2)}(\vec k,a)=\frac{3}{2} \beta_0 a^\gamma k^2 \phi^{(2)}(\vec k,a)
+\frac{2}{k^2} \biggl( (\gamma+1)^2 + \frac{\gamma+1}{2} \biggr)\mathcal{L}^{1,\tilde{1}}
\eeq
The first term arises from considering the second order LPT field in the leading EFT counterterm and the second source arises from coupling the linear field and the leading order EFT solution using the the quadratic kernel. More explicitly we can write
\beq
\begin{split}
\mathcal{D}\phi^{(\tilde 2)}(\vec k,a)=\int_{\vp_1}\int_{\vp_2}\biggl[-\frac{9}{28} \beta_0 a^\gamma (\vp_1+\vp_2)^2 -\frac{3}{2}\frac{(\gamma+1)(2\gamma+3)}{\gamma(5+2\gamma)}\beta_0a^\gamma (p_1^2+p_2^2)\biggr]\\
\frac{\kappa_2(\vp_1,\vp_2)}{(\vec p_1+\vec p_2)^2}(2\pi)^3\delta^\text{(D)}(\vk-\vp_1-\vp_2)
\delta^{(1)}(\vp_1,a)\delta^{(1)}(\vp_2,a)\\
\end{split}
\eeq
Integrating the source with the Greens function, we finally have
\beq
\begin{split}
\phi^{(\tilde 2)}(\vec k,a)=&\int_{\vp_1}\int_{\vp_2}\biggl[-\frac{9}{28} \frac{2}{(1+\gamma)(7+2\gamma)}\beta_0 a^\gamma (\vp_1+\vp_2)^2 
-\frac{3}{2}\frac{2(2\gamma+3)}{\gamma(5+2\gamma)(7+2\gamma)}\beta_0a^\gamma (p_1^2+p_2^2)\biggr]\\
&\frac{\kappa_2(\vp_1,\vp_2)}{(\vec p_1+\vec p_2)^2}(2\pi)^3\delta^\text{(D)}(\vk-\vp_1-\vp_2)
\delta^{(1)}(\vp_1,a)\delta^{(1)}(\vp_2,a)\, ,\\
=&-\alpha (a)\; \int_{\vp_1}\int_{\vp_2}\biggl[\frac{3}{14} \frac{\gamma(5+2\gamma)}{(1+\gamma)(7+2\gamma)} E_{2,2}(\vec p_1,\vec p_2) 
+\frac{(2\gamma+3)}{(7+2\gamma)} E_{2,3}(\vec p_1,\vec p_2)\biggr]\\
&(2\pi)^3\delta^\text{(D)}(\vk-\vp_1-\vp_2)
\delta_{(1)}(\vp_1,a)\delta^{(1)}(\vp_2,a)\, .
\end{split}
\eeq
We thus see that expressions in the form of the $E_{2,2}$ and $E_{2,3}$ counterterms are generated by the time integration of the leading order counterterm, but not $E_{2,1}$. 
Part of the counterterm $\phi^{(\tilde{2})}$ appears in $P_{2\tilde{2}}$ to cancel the UV mistake in $P_{24}$. For this part of the counterterm, the time-dependences of $P_{2\tilde{2}}$ and $P_{24}$ have to match match, which implies $\gamma=2$.

\end{document}